\documentclass[twocolumn]{article}
\usepackage{arxiv}
\usepackage[backend=biber,
style=numeric,
sorting=none,
maxcitenames=2,
maxbibnames=100, 
language=english,
isbn=false,
url=false,
firstinits=true,
uniquename=false
]{biblatex} 
\DeclareNameAlias{default}{last-first}
\DeclareLanguageMapping{english}{english-apa} 
\usepackage{subcaption}
\usepackage{placeins}
\usepackage{mathtools}
\usepackage{amssymb}
\usepackage{amsmath}
\usepackage{amsfonts}   
\usepackage{fancyhdr}
\usepackage{times}
\usepackage{amsthm}
\usepackage{siunitx}
\usepackage{nicefrac}

\usepackage{todonotes}
\usepackage{algorithm}
\usepackage{algorithmic}

\newcommand{\given}{\,|\,}

\usepackage{float}

\usepackage{accents}

\usepackage{enumitem}

\usepackage{xcolor}
\definecolor{darkblue}{RGB}{0,0,128}
\definecolor{darkgreen}{RGB}{0, 128, 0}
\definecolor{darkred}{RGB}{128, 0, 0}
\definecolor{black}{RGB}{0, 0, 0}
\definecolor{errorcolor}{HTML}{481567}
\definecolor{viridisgreen}{HTML}{55C667}

\usepackage{hyperref}
\usepackage{url}
\usepackage{changes}
\hypersetup{
	colorlinks=true,
	linkcolor=darkblue,
	filecolor=darkblue,
	urlcolor=darkblue,
	citecolor=darkblue,
	pdfauthor={L. Schumacher, P.-C. B\"{u}rkner, A. Voss, U. K\"{o}the, S. T. Radev},
	pdftitle={Neural Superstatistics}
}

\title{Neural Superstatistics for Bayesian Estimation of Dynamic Cognitive Models}
\author{
    Lukas Schumacher*\\
    Institute of Psychology\\
    Heidelberg University\\
    \href{mailto:lukas.schumacher@psychologie.uni-heidelberg.de}{lukas.schumacher@psychologie.uni-heidelberg.de}
    \And
    Paul-Christian Bürkner\\
    Department of Statistics\\
    TU Dortmund University\\
    \And
    Andreas Voss\\
    Institute of Psychology\\
    Heidelberg University\\
    \And
    Ullrich Köthe\\
    Computer Vision and Learning Lab\\
    Heidelberg University\\
    \And
    Stefan T.~Radev\\
    Cluster of Excellence STRUCTURES\\
    Heidelberg University
}

\begin{document}
\twocolumn[{%
  \begin{@twocolumnfalse}
    \maketitle
  \end{@twocolumnfalse}
}]
\begin{abstract}
Mathematical models of cognition are often memoryless and ignore potential fluctuations of their parameters.
However, human cognition is inherently dynamic.
Thus, we propose to augment mechanistic cognitive models with a temporal dimension and estimate the resulting dynamics from a superstatistics perspective.
Such a model entails a hierarchy between a low-level observation model and a high-level transition model. The observation model describes the local behavior of a system, and the transition model specifies how the parameters of the observation model evolve over time.
To overcome the estimation challenges resulting from the complexity of superstatistical models, we develop and validate a simulation-based deep learning method for Bayesian inference, which can recover both time-varying and time-invariant parameters.
We first benchmark our method against two existing frameworks capable of estimating time-varying parameters.
We then apply our method to fit a dynamic version of the diffusion decision model to long time series of human response times data.
Our results show that the deep learning approach is very efficient in capturing the temporal dynamics of the model. Furthermore, we show that the erroneous assumption of static or homogeneous parameters will hide important temporal information.
\end{abstract}

\section*{Introduction}
Mathematical models are important tools for conceptualizing human cognition and predicting observable behavior. 
Such models aim to provide a mathematical formalization of cognitive processes by mapping latent cognitive constructs to model parameters and specifying how these generate manifest data \autocite{farrell2018}.
The surge of cognitive model applications has made it possible to test precise mechanistic hypotheses and to predict performance in various domains, such as decision-making \autocite{voss2013, ratcliff2016}, learning \autocite{eckstein2020, gershman2017}, or memory \autocite{oberauer2018, yoo2022}.

The majority of cognitive models treat human data as independent and identically distributed (IID) observations.
The IID assumption implies that these models largely ignore the temporal changes of latent cognitive constructs.
However, such constructs are inherently dynamic, regardless of a particular time scale \autocite{vanorden2003, wagenmakers2004, gilden2001, collins2018}.
For instance, there is little dispute that constructs, such as working memory capacity \autocite{brockmole2013} or mental speed \autocite{vonkrause2022}, change over the human life span. These constructs also vary on much shorter time scales, for example, within experimental sessions \autocite{riley2012, favela2020}.

In psychological experiments, cognitive affordances are influenced not only by external task demands but also by internal mental processes and brain states that change over time. 
There are many possible explanations for the resulting systematic and unsystematic fluctuations, for instance, fatigue \autocite{ratcliff2011, walsh2017}, practice \autocite{kahana2018, evans2018}, mind-wandering \autocite{mittner2016, christoff2016}, or motivational factors \autocite{brosowsky2020, kiuru2020}. 
In this article, we argue that cognitive mechanisms should be treated as complex dynamic systems and that cognitive models should account for the dynamics of their components to fully understand and capture the rich structure of empirical human data.

\begin{figure}[h]
\centering
\includegraphics[width=0.48\textwidth]{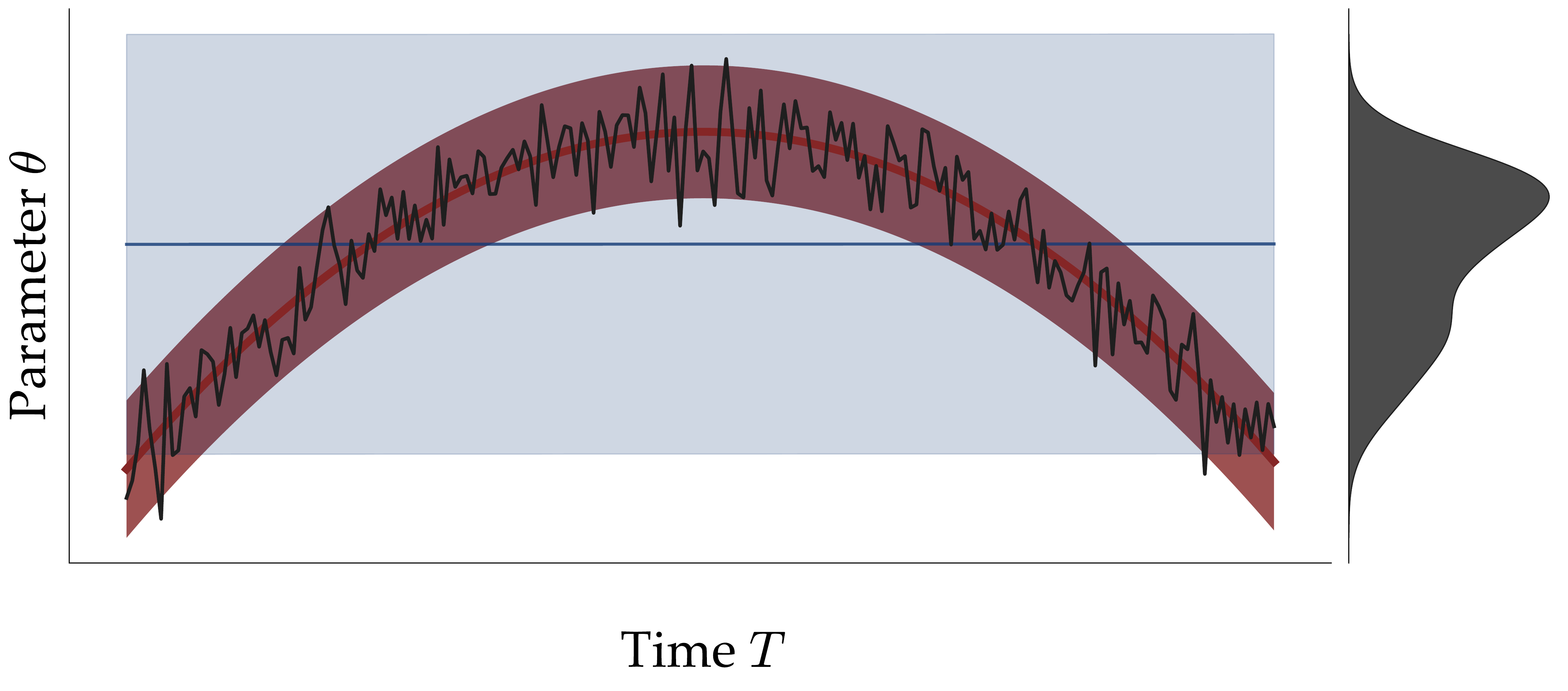}
\caption{Conceptual illustration of a hypothetical parameter $\theta$ varying over time (solid black line). 
The solid blue line and shaded blue region depict the posterior mean and the $95$\%-CI of a static model, respectively. 
The solid red line and shaded red region depict the posterior mean and $95$\%-CI of a dynamic model, respectively.
Treating the parameter as static (i.e., stationary) by marginalizing out the effects of time leads to inflated uncertainty estimates (matching the width of the marginal distribution, depicted in gray) and obscures the underlying change.}
\label{fig:concept}
\end{figure}

Ignoring temporal fluctuations and changes in cognitive parameters can have drastic consequences for the descriptive, explanatory, and predictive merits of cognitive models.
Consider a simple inverted U-shape hypothetical trajectory of a single parameter, as depicted in \autoref{fig:concept}.
Typical cognitive models assuming IID observations \autocite{voss2013, oberauer2018} would estimate a flat trajectory (depicted in blue) whose uncertainty would match the width of the marginal parameter distribution (depicted in gray). 
Differently, dynamic models would account for temporal change and achieve a much greater information gain (depicted in red).
Indeed, this is not just a hypothetical scenario, and we subsequently demonstrate its consequences in a real data application (cf. \autoref{fig:average_param_dynamics}).

One way to mathematically formalize dynamic systems is by treating them as stochastic generative processes that produce data with temporal dependencies (i.e., time series data). 
As most complex systems are inherently non-linear, these time series often do not exhibit simple fluctuations around a stable mean with a fixed variance, but resemble a heterogeneous random walk \autocite{mark2018}. 
\textcite{beck2003} coined the term \textit{superstatistics}, which refers to a superposition of multiple stochastic processes on different temporal scales that can describe heterogeneous temporal dynamics \autocite{hanel2011}. 
Thus, instead of assuming static model parameters, a superstatistics modeling approach introduces a hierarchy of at least two models: A low-level (i.e., observation or microscopic) model that formalizes the local behavior of a system and a high-level (i.e., transition or macroscopic) model that describes the parameter dynamics of the low-level model. Note that there is no absolute time scale for low- and high-level processes. The meaning of these terms is relative and always depends on the scale relevant to the research question.

A viable approach for modeling parameter transitions is offered by hidden Markov models (HMMs). 
For instance, \textcite{kucharsky2021} accounted for different response states during a decision-making task by combining a HMM with an evidence accumulation model of decision-making.
This model combination allows for discontinuous changes on longer time scales and continuous changes on shorter time scales.
Similarly, \textcite{gunawan2022} extended a hierarchical version of the same decision-making model with a HMM and applied it to three existing long time series of response time and choice data. 
Both studies demonstrate that the HMM approach can reveal plausible fluctuations of decision model parameters in cognitive tasks.

However, the superstatistics framework is far more general and flexible in representing macroscopic fluctuations. 
First, it does not require modelers to pre-define a small set of possible modes (i.e., distinct system behaviors).
Further, models within the superstatistics framework can be agnostic about the concrete dynamics of the model parameters -- the most plausible dynamic can be directly estimated in a data-driven fashion.
For example, using a superstatistics framework, \textcite{metzner2021} demonstrated that the transition between different sleep stages is less abrupt than previously suggested.

The superstatistics framework has been utilized in physics \autocite{yalcin2016, rabassa2015, williams2020}, the life-sciences \autocite{bogachev2017} and economics \autocite{vanderstraeten2009, denys2016}, but it has not yet been disseminated in the cognitive sciences.
Under the assumption that cognitive processes are dynamic and complex, it seems natural to equip existing cognitive models with superstatistical aspects.
However, to our knowledge, no previous study besides \textcite{metzner2021} has employed superstatistical methods for studying the dynamic aspects of cognitive parameters.
Existing dynamic models of cognition fit stationary time series models (e.g., autoregressive models) to the observed behavior \autocite{wagenmakers2004} but do not incorporate a low-level mechanistic model that formalizes the underlying cognitive process(es). 
Thus, these time series models describe how behavior changes over time but do not explain how behavior occurs at a specific point in time.
On the other hand, popular mechanistic models tailored to describe local behavior, such as diffusion decision models \autocite[DDM,][]{voss2007, voss2013, ratcliff1978}, either ignore the dynamic aspects of their parameters entirely or represent parameters as deterministic functions of time \autocite{vonkrause2021, diederich2006, urai2019a, vanrooij2013, gasimova2014}. 

In this work, we argue that the superstatistics framework can reveal a more nuanced picture of cognitive dynamics and behavioral fluctuations. 
This is possible because we formalize the dynamic aspect of the low-level parameters as a higher-order stochastic process.
Consequently, we estimate the low-level parameters at each time step directly from the data.
Thus, their temporal evolution is only constrained by the modeler's choice of prior distributions and by the high-level transition model.
Nevertheless, superstatistical models can be rigorously validated in the same way as their static counterparts, using standard model criticism methods, such as simulation-based calibration (SBC) to assess computational faithfulness, parameter recovery for inferential calibration, posterior re-simulation checks for assessing model adequacy, as well as cross-validation for assessing predictive performance \autocite{gelman2020, van2021bayesian}.
Superstatistical models allow us to address questions about how cognitive systems undergo distinct transitions in various settings \autocite{kucharsky2021}.
Further, one can examine which model parameters explain behavioral fluctuations without predefined equations that fix the hypothesized temporal evolution of specific parameters.

Superstatistical models can be quite challenging to estimate and compare for a number of reasons, especially in a Bayesian framework for principled uncertainty quantification \autocite{mark2018}.
First, both the high-level and low-level models are stochastic, so there is considerable uncertainty about the values of all model parameters (i.e., static and dynamic) given a finite number of observations.
Second, the low-level models might be complex and non-linear so that there is not always a closed-form analytic expression relating model parameters to data (i.e., the likelihood function is \textit{intractable}), or the likelihood might be computationally very expensive to evaluate.
Finally, even for stationary low-level models, the computational cost might become insurmountable when these models are applied to multiple data sets, since standard Bayesian methods are not amortized and thus need to be re-run sequentially (unless massively parallelized) and from scratch for each data set \autocite{radev2020, mestdagh2019prepaid}.

Indeed, estimation challenges may be the main reason for the underrepresentation of superstatistical models in psychology and the cognitive sciences. 
However, we argue that recent advances in (amortized) simulation-based inference \autocite[SBI,][]{cranmer2020, radev2020, burkner2022some} render estimation challenges secondary and allow researchers to create and test high-fidelity models of cognition without worrying about analytic tractability.
SBI encompasses methods that use synthetic data to approximate intractable posterior distributions of unknown parameters.
Moreover, amortized SBI with neural networks represents a particularly efficient way to perform posterior estimation on multiple data sets by investing the primary computational effort in a relatively expensive training phase \autocite{cranmer2020, burkner2022some}. Once simulation-based training has converged, the trained networks can be applied to any number of observations or set of observations consistent with the model's structure. 

\begin{figure}
\centering
\includegraphics[width=0.48\textwidth]{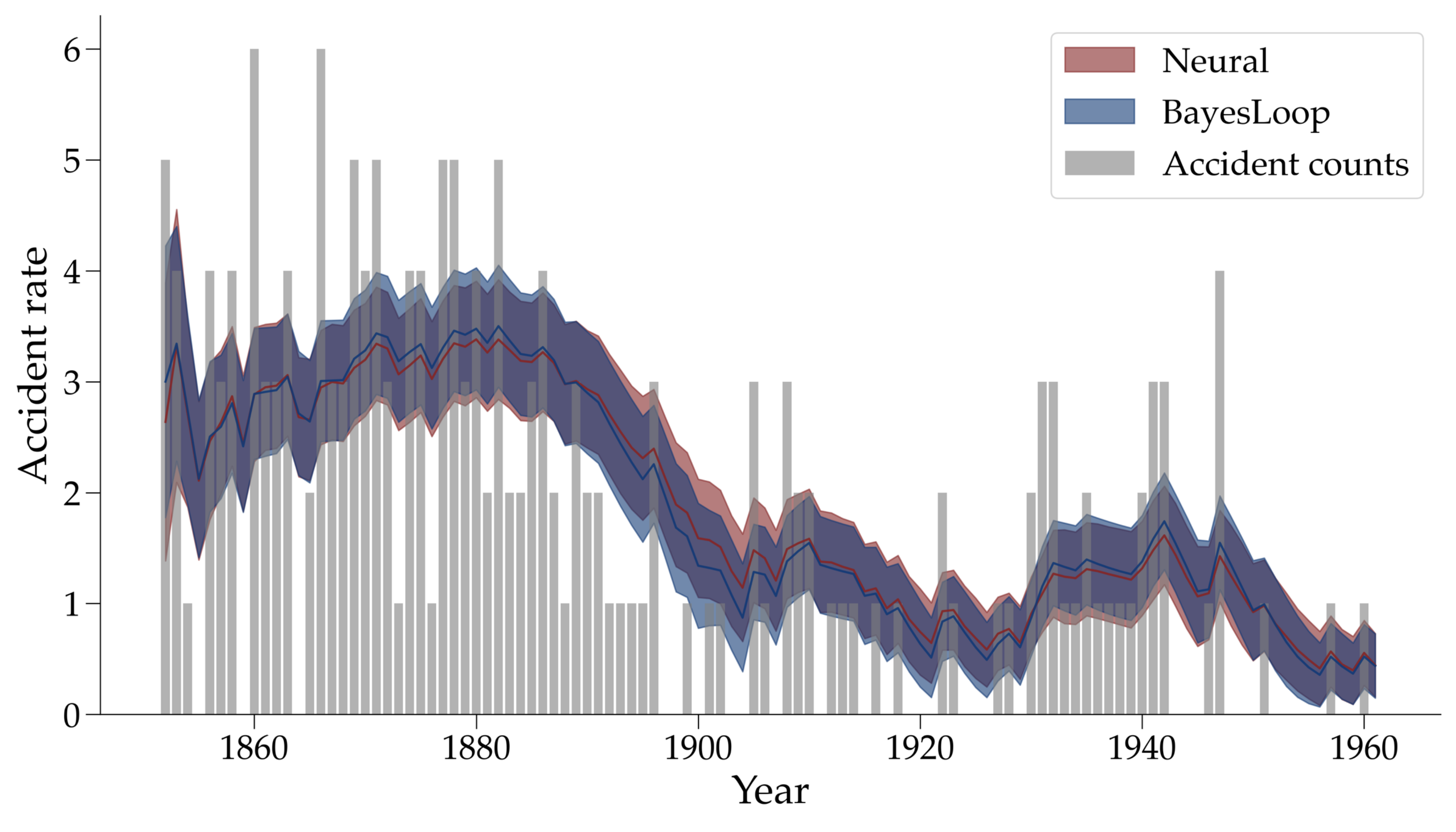}
\caption{Coal mining disasters in the United Kingdoms between 1852
and 1962. The annual reported accident counts are depicted using gray bars. The mean posterior of the rate parameter $\lambda$ of a Poisson process with Gaussian fluctuation is shown with solid lines for both estimation methods separately. The shaded area represents $\pm1$ posterior standard deviation.}
\label{fig:coal_mining}
\end{figure}

The main purpose of this article is two-fold.
First, we demonstrate and validate the use of superstatistics in cognitive modeling via an out-of-the-box extension of a popular mechanistic cognitive model, namely, the DDM.
Second, we develop and validate a novel Bayesian estimation method grounded in the \texttt{BayesFlow} framework for amortized neural SBI \autocite{radev2020}. 
To this end, we first perform benchmark comparisons with existing frameworks on simulated data. 
We then specify a non-stationary DDM and fit it to long time series of response times obtained from human participants.
Moreover, with this application, we empirically demonstrate how stationary models assuming IID observations can hide a number of interesting dynamic patterns and fluctuations present in behavioral data.

\section*{Results}

\subsection*{Benchmark Studies}
To ensure the trustworthiness of our method, we first benchmark its performance against two existing Bayesian frameworks which use different estimation algorithms: \texttt{bayesloop} \autocite{mark2018} and \texttt{Stan} \autocite{carpenter2017stan}.
The former employs grid approximation for low-dimensional problems, whereas the latter relies on Hamiltonian Monte Carlo \autocite[HMC,][]{neal_mcmc_2011} sampling.
Both frameworks operate in a non-amortized way and can only estimate superstatistical models with closed-form likelihoods.

\subparagraph*{\textit{Coal Mining Accidents}}
Currently, \texttt{bayesloop} cannot fit low-level models as complex as the DDM, nor high-level models such as the Gaussian process.
Therefore, we compare the estimation performance of our method on a simpler example based on the coal mining accident data (freely available from \autocite{mark2018}).
These data comprise counts of coal mining accidents in the United Kingdom between 1852 and 1962.
The low-level model is a simple Poisson distribution with a parameter $\lambda$ that corresponds to the accident rate.
One can assume that the accident rate in coal mines was not constant during this more than a century-long period.
Therefore, the accident rate $\lambda$ is allowed to fluctuate over time according to the Gaussian random walk transition model (cf. equation \eqref{GaussianTransition}).
Both estimation methods use the same informative prior distribution for the low-level parameter $\lambda_0 \sim$ Exp$(0.5)$ and high-level parameter $\sigma \sim$ Beta$(1, 25)$.

Using the \texttt{bayesloop} software, we approximated a grid with $4000$ equally spaced points ranging from $0$ to $15$ for $\lambda$ and from $0$ to $1$ for $\sigma$, respectively. 
This calculation lasted approximately $38$ minutes on a standard desktop computer. Training the neural network for $20$ epochs took approx. $18$ minutes, and obtaining $4000$ posterior samples took less than a second.
Thus, in this case, the training effort amortizes even after \textit{a single} data set.

\autoref{fig:coal_mining} shows the annual count of coal mining accidents overlaid with the estimated dynamic accident rate $\lambda$ (posterior mean and $\pm1$ standard deviation).
Both methods estimate an almost identical latent trajectory for the low-level model parameter $\lambda$.
Between the years $1880$ and $1900$, we observe a decrease in coal mining accidents followed by two temporary increases around the years $1905$ and $1930$. 
The estimated parameter dynamic closely follows these data patterns.
Thus, we conclude that our neural method can estimate a plausible parameter dynamic for a simple low-level model and performs equally well compared to \texttt{bayesloop}.

\subparagraph*{\textit{Static Diffusion Decision Model}}
As a second benchmark, we compare our neural method to \texttt{Stan} in terms of Bayesian updating, assuming a ``true'' DDM with time-invariant parameters.
This benchmark serves two goals. 
Firstly, it aims to compare the estimation performance of our method with that of \texttt{Stan}, which is regarded as the gold standard for sampling-based Bayesian inference.
Secondly, it aims to assure that our method can correctly identify stationary parameters when fitting a dynamic model on data generated from a stationary process (i.e., it does not estimate ``pseudo-dynamics'').

\begin{figure}[h]   
\centering
\includegraphics[width=0.48\textwidth]{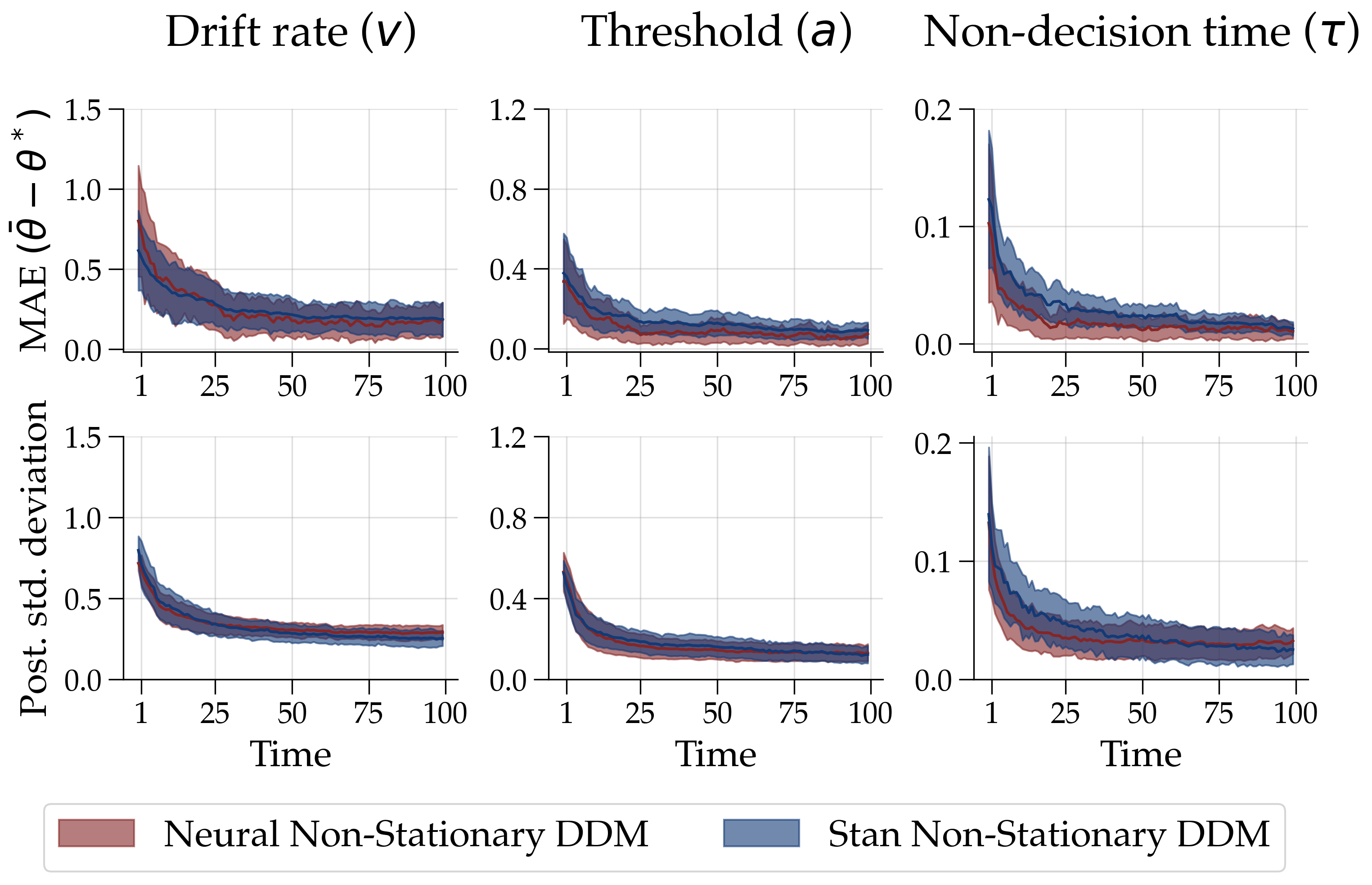}
\caption{Comparison between the neural and Stan estimation method. \textbf{First row}: Median absolute error (MAE) between the ground truth data-generating parameters and the estimated posterior means across the $100$ simulations over time. \textbf{Second row}: Posterior standard deviation aggregated across the $100$ simulated data sets over time (solid lines). The shaded area depicts the median absolute deviation (MAD).}
\label{fig:plot_stan_benchmark}
\end{figure}

To this end, we simulated $100$ data sets with $100$ observations, each using a static DDM with $3$ free parameters (see \textbf{A Non-Stationary Diffusion Decision Model} section) without parameter fluctuation over time.
Then, we fit a non-stationary DDM with a Gaussian random walk transition model (cf. equation \eqref{GaussianTransition}) to all $100$ data sets using both estimation methods.
Again, we use the same prior distributions (see Appendix) to ensure comparability.
We compared the two methods based on the following two performance metrics:
i) the median absolute error (MAE) between the estimated posterior means and the data generating stationary parameters averaged across all $100$ simulations, and ii) the average posterior standard deviation over time.
These two metrics are common indicators for inferential model calibration, which aims to analyze the global behavior of the posterior distribution given possible observations from the prior predictive distribution \autocite{betancourt2018}.
The former metric informs us how well the posterior recovers the true model configurations (analogous to posterior \textit{z}-scores).
The latter metric indicates how much the posterior is informed by the data beyond the prior knowledge that was encoded in the prior distribution (analogous to posterior contraction) \autocite{betancourt2018}.

The upper panel of \autoref{fig:plot_stan_benchmark} depicts the absolute difference between the true data generating parameters and the dynamically estimated posterior means over time, averaged over all $100$ simulations for both methods separately.
On average, the posterior means show a relatively large deviation from the true data generating parameters on early trials of the data.
This difference then quickly decreases and flattens after approximately $25$ trials.
The performance of both methods concerning this metric is almost indistinguishable.

The lower panel of \autoref{fig:plot_stan_benchmark} displays the median posterior contraction measured as posterior standard deviation over time for all $3$ parameters separately.
We observe considerable posterior contraction within the first $25$ time points.
Again, the performance of both methods is nearly identical.
However, there is a large difference in estimation time between the two methods.
As we are interested in the filtering posterior distributions, the \texttt{Stan} model has to be refit with every additional observation of a time series. 
Hence, we fit the \texttt{Stan} model to each simulated $x_{1:t}$, $t = 1,\dots,T$, which amounted to $T = 100$ re-fits per simulated data set.
Fitting the model to all $100$ synthetic data sets resulted in $100$ x $100$ model fits.
This procedure took over $1$ week of non-stop computing on a standard desktop computer -- whereas training the neural network lasted approximately $8$ hours with almost instantaneous fit to the $100$ data sets thereafter.
This is a non-negligible difference that will grow with longer time series, more data sets, or increased complexity until reaching a point where models can no longer be estimated with \texttt{Stan} due to limited processing resources or time constraints (see next section).

In summary, our method closely approximates the results obtained from \texttt{bayesloop} and \texttt{Stan} on the considered benchmark examples.
Note, however, that our method is primarily designed for models where the above frameworks cannot be applied -- higher dimensional models, possibly lacking a closed-form likelihood (i.e., available only as stochastic simulators), or many data sets consisting of long time series. 
The next application we present could be tackled with our neural approximators, but not with the above two frameworks.

\begin{figure}[h]
    \centering
    \includegraphics[width=0.45\textwidth]{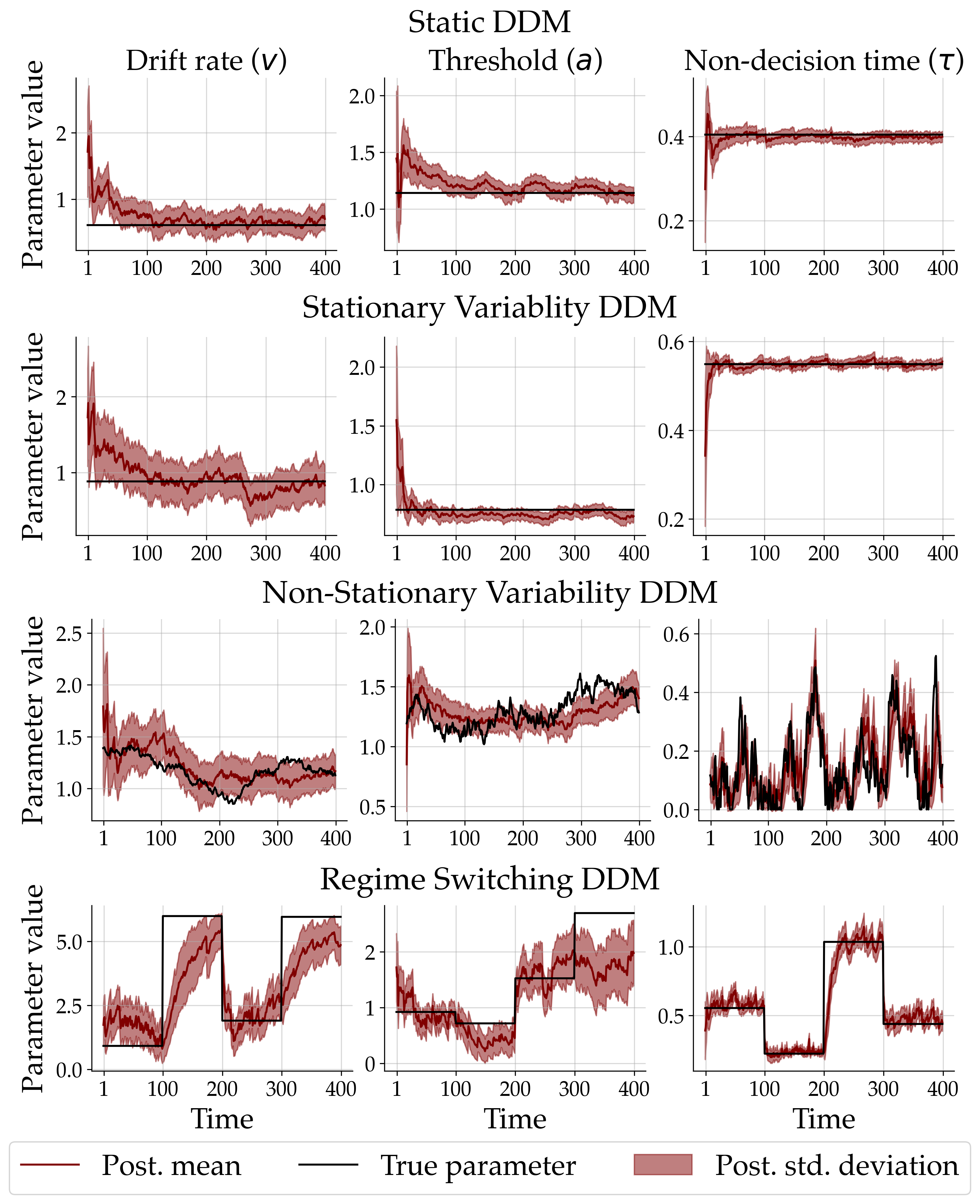}
    \caption{
      Example time-varying parameters estimated by our neural method in each scenario of the simulation study. Each row depicts the posterior estimates obtained from a single simulated person. The third row corresponds to the dynamic model used for training the network (i.e., well-specified case). The first, second, and fourth rows correspond to model variants not seen during training (i.e., misspecified cases).
    }
    \label{fig:sim_study_exemplar}
\end{figure}

\subsection*{Simulation Study}
Next, we probe the parameter recoverability of a non-stationary DDM under different induced misspecifications (i.e., models that differ from the one used for training the network). 
To this end, we performed an extensive study for which we simulated data sets consisting of $T = 400$ time points in four different scenarios: i) A static DDM with constant parameters; ii) a DDM with stationary variability (commonly referred to as ``inter-trial variability'') where the 3 DDM parameter fluctuate randomly around a constant value; iii) a non-stationary DDM with a Gaussian random walk transition model; iv) and a DDM with constant parameters that jump abruptly and uniformly at three predefined time points (i.e., a regime switching model).
Crucially, we trained the neural approximator only with simulations from the non-stationary model. 
However, during amortized inference, we applied the network to $200$ data sets from each of the four scenarios. 
Thus, we could investigate the network's response in the \textit{open world} setting where the true data generator may differ from the reference model used during the training phase.

\begin{figure}[h]
    \includegraphics[width=0.48\textwidth]{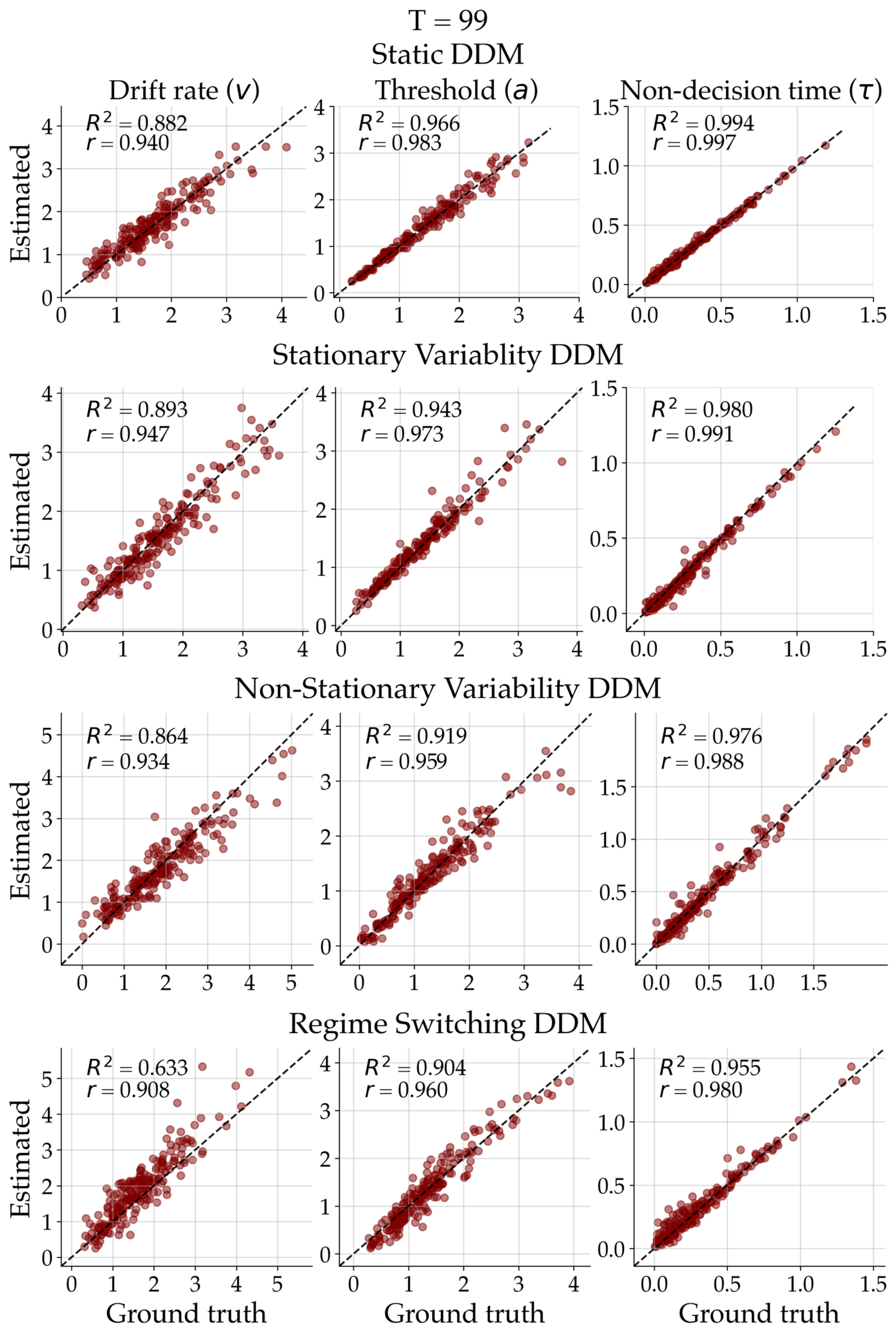}
    \caption{
        Ground truth-data generating parameters plotted against posterior means for all 3 parameters and simulation scenarios separately at time point $T = 99$ (just before the change of regime of the regime switching DDM). 
    }
    \label{fig:true_estimated_98}
\end{figure}

\autoref{fig:sim_study_exemplar} shows an exemplar fit of the non-stationary DDM with a random walk transition model to data sets from each of the four simulation scenarios.
In the top row, we see that the estimated parameter trajectories converge to the constant ground-truth parameters. 
A similar pattern emerges when the ground-truth parameters randomly fluctuate around a constant value (second row), yet we observe less uncertainty reduction.
The third row depicts the posterior estimates based on a data set simulated from the reference non-stationary DDM (i.e., the well-specified case).
Besides some local deviations from the ground-truth parameter trajectory, the model is able recover the overall trend of the dynamics.
In the fourth row, we can inspect the posterior estimates from a data set simulated from the regime switching DDM which allows the parameters to ``jump'' uniformly at three time points to any value within the parameter bounds.
Despite the severe misspecification, the random walk DDM is able to recover the discontinuous trajectories surprisingly well; still, the gradual change implied by the random walk transition does not allow for a rapid adaptation and exhibits a notable lag after each switch.

\autoref{fig:true_estimated_98} depicts the true data generating and the estimated posterior means at time point $T = 99$ (right before the first jump of the regime switching transition model). We observe excellent recovery performance for all $3$ parameters in all $4$ simulation scenarios at the selected time point. The recovery performance at other time points as well as further details and analyses (i.e., MAE over time) can be found in the Appendix.

\subsection*{Human Data Applications}
Following our benchmarking and simulation studies, we applied non-stationary versions of the DDM to two separate data sets collected from response time (RT) experiments: i) A standard random-dot motion task (a maximum of $T = 1320$ trials per participant), and ii) very long time series (a maximum of $T = 3200$ trials per participant) from a lexical decision task.
The first application serves as a starting point with data stemming from a popular task in experimental psychology.
The second application showcases the utility of our method to estimate a complex non-stationary DDM with a Gaussian process (GP) transition model and multiple drift rate parameters for different difficulty conditions.
Before fitting a model to empirical data, it is imperative to assess the faithfulness of the approximation method \autocite{gelman2020, schad2021}.
To this end, we perform simulation-based calibration \autocite[SBC;][]{talts2018, sailynoja2021}.
These analyses suggest that our neural Bayesian method exhibits reasonable calibration, with slightly miscalibrated posteriors for the non-decision time parameter (see Appendix for more details on calibration).

\subparagraph*{\textit{Random-Dot Motion Task}}
First, we fit a non-stationary DDM with a Gaussian random walk transition model to a data set retrieved from the experimental study of \textcite{evans2017}.
We chose this data set because the purpose of the original study was to investigate the decline of the threshold parameter over time. 
The experiment had a $3$ (\textit{Low}, \textit{Medium}, and \textit{High} feedback) by $2$ (\textit{Time} and \textit{Trial} condition) factorial between-subject design.
Differently from our approach, \textcite{evans2017} subdivided the time series into trial bins and fitted a stationary hierarchical Bayesian DDM to each bin separately.
Therefore, we can compare the parameter trajectories recovered by our neural superstatistics method with the estimates obtained by the original authors using Markov chain Monte Carlo (MCMC).

\begin{figure}[h]
    \includegraphics[width=0.48\textwidth]{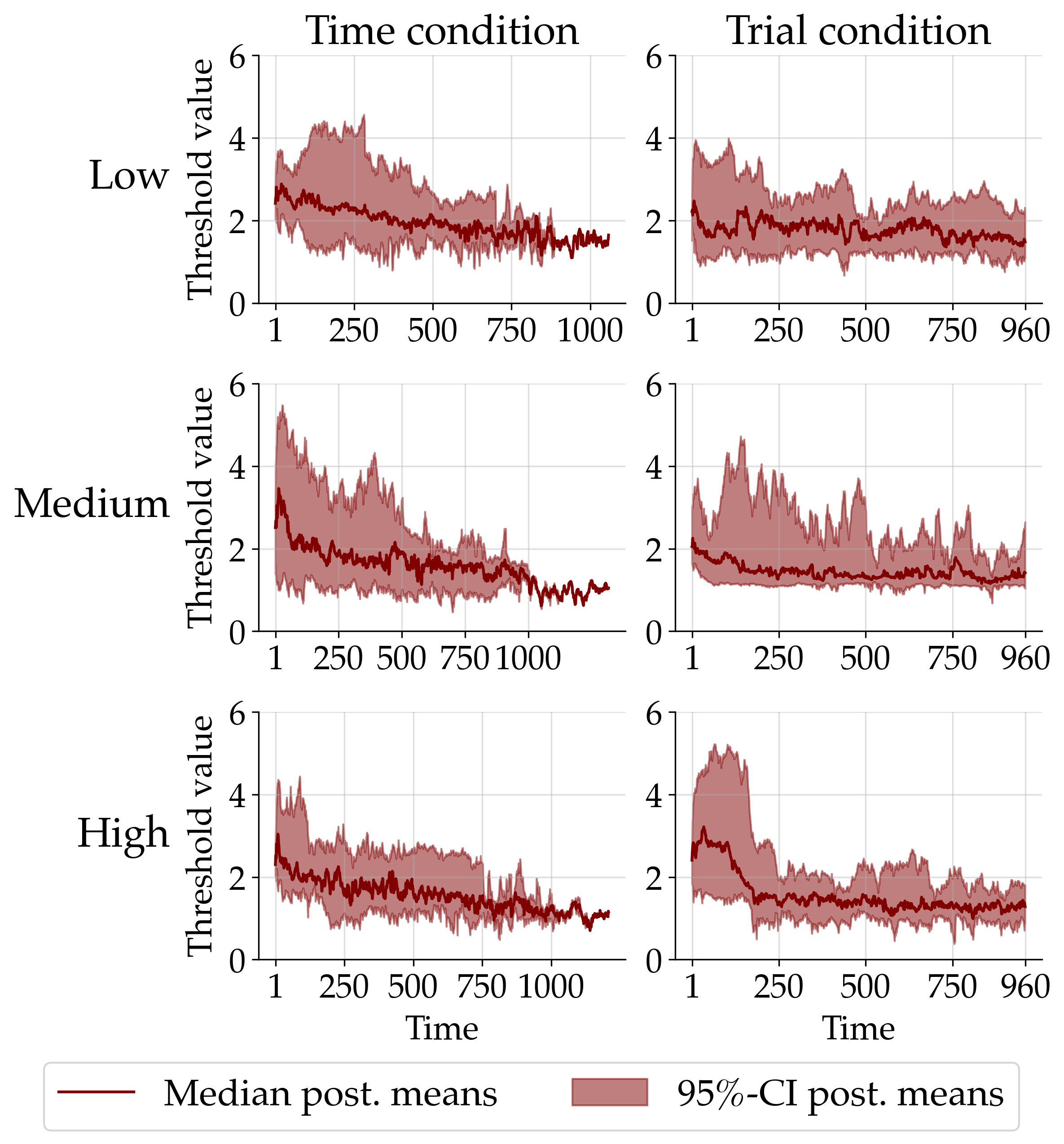}
    \caption{
        Estimated trajectories of the DDM threshold parameter aggregated across all individuals for each between-subject experimental condition. The first column corresponds to the \textit{Time} and the second to the \textit{Trial} condition. The rows correspond to the three feedback conditions, \textit{Low}, \textit{Medium}, and \textit{High}, respectively. The red solid lines depict the median of the individual posterior means and the red shaded area the 95\% credibility interval of these posterior means.
    }
    \label{fig:optimal_policy_plot}
\end{figure}

\autoref{fig:optimal_policy_plot} depicts the trajectory of the threshold parameter aggregated across all individuals in a separate panel for each experimental condition.
Note, that in the \textit{Time} condition participants had a fixed amount of time they could spend on the task resulting in different  time intervals. 
When we compare our estimates to those obtained by \textcite{evans2017}, it becomes evident that both approaches yield similar qualitative and quantitative patterns. 
This result complements our promising results ``in silico'' and points to the convergent validity of our superstatistics approach in applications with real data. 

\begin{figure*}[h]
\includegraphics[width=0.99\textwidth]{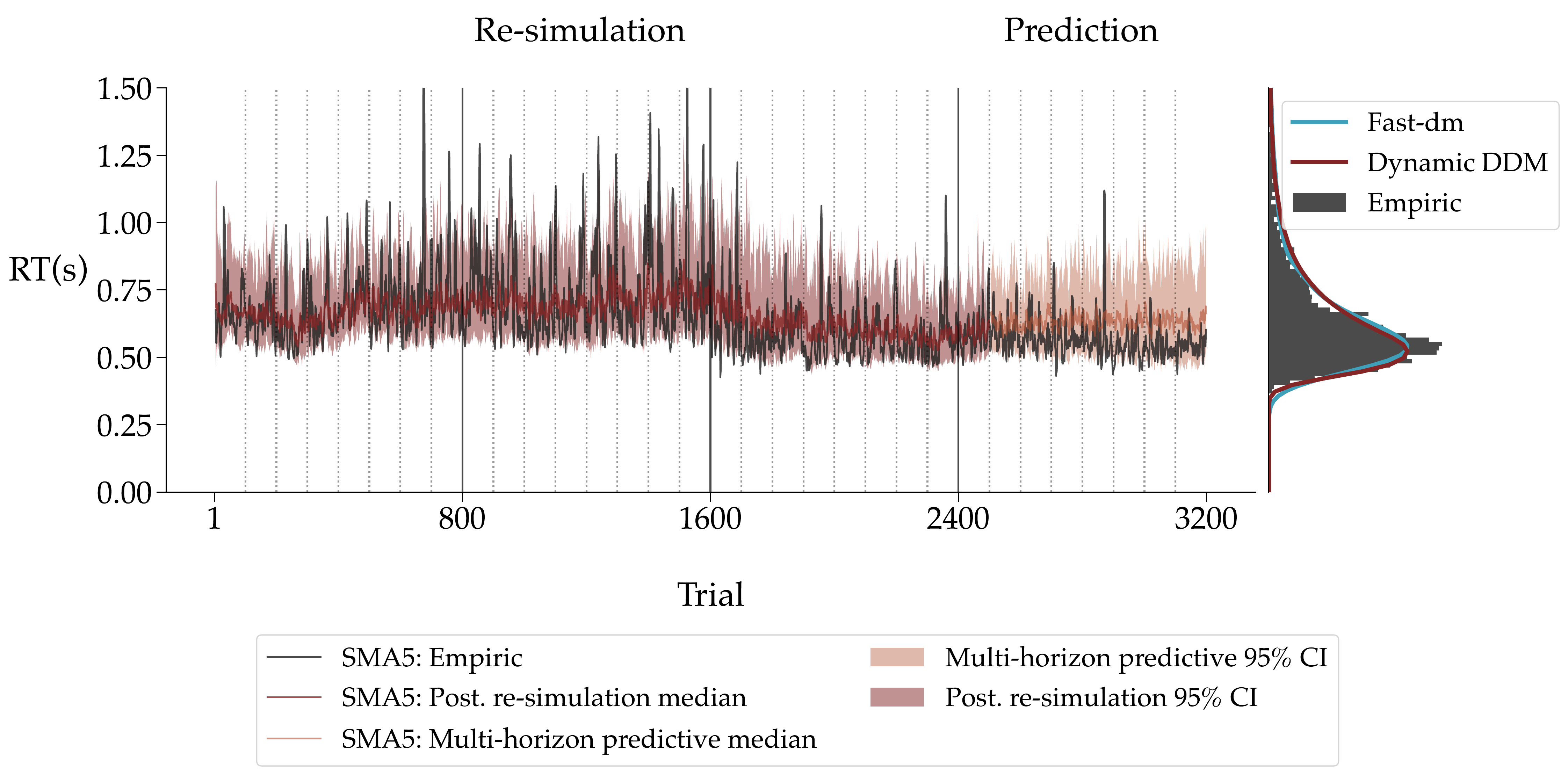}
\caption{Model fit to human data. \textbf{Left panel} The empirical RT time series of a single individual in black. From trial 1 to 2500, the median posterior re-simulation (aka \textit{retrodictive check}) using the non-stationary DDM is shown in red. The models' multi-horizon prediction is depicted for the remaining trials in orange. The shaded areas for the posterior re-simulation and multi-horizon prediction correspond to 95\% credibility intervals. All the time series were smoothed via a simple moving average (SMA) with a period of 5. The dotted vertical lines indicate the end of an experimental block, and the solid vertical lines the end of an experimental session. \textbf{Right panel} The raw RT distribution is plotted as a histogram in black. The re-simulated RT distributions from the non-stationary DDM and reference re-simulations from the static DDM using \texttt{fast-dm} are shown as kernel density estimates (KDEs) in red and blue, respectively.}
\label{fig:time_series}
\end{figure*}

\subparagraph*{\textit{Lexical Decision Task}} 
We fit the non-stationary DDM with a GP transition model (cf.~equation \eqref{GaussianProcess}) to human behavioral data originating from a lexical decision-making task.
The data consist of long RT and choice time series from four experimental conditions.
For this application, we use four separate drift rates -- one for each experimental condition.
The length of these time series made it impossible to estimate the model with \texttt{Stan} (due to memory limitations and infeasible compute time). 
Thus, to increase the trustworthiness of the results obtained with our neural method, we resort to the established \texttt{fast-dm} software \autocite{voss2007} as a benchmark, which is capable of estimating homogeneous (block) trial-by-trial fluctuations (i.e., inter-trial variabilities).
We then compare the goodness of absolute fit in terms of re-simulation accuracy between both estimation methods and investigate the multi-horizon predictive performance of our method.
Further, we analyze the main advantage of the non-stationary DDM, that is, the inferred trial-by-trial parameter dynamics, and compare those to the static \texttt{fast-dm} parameter estimates.
Note that \texttt{fast-dm} is not a Bayesian method and is thus not included in our previous benchmark studies.

\begin{figure*}[h]
\includegraphics[width=1\textwidth]{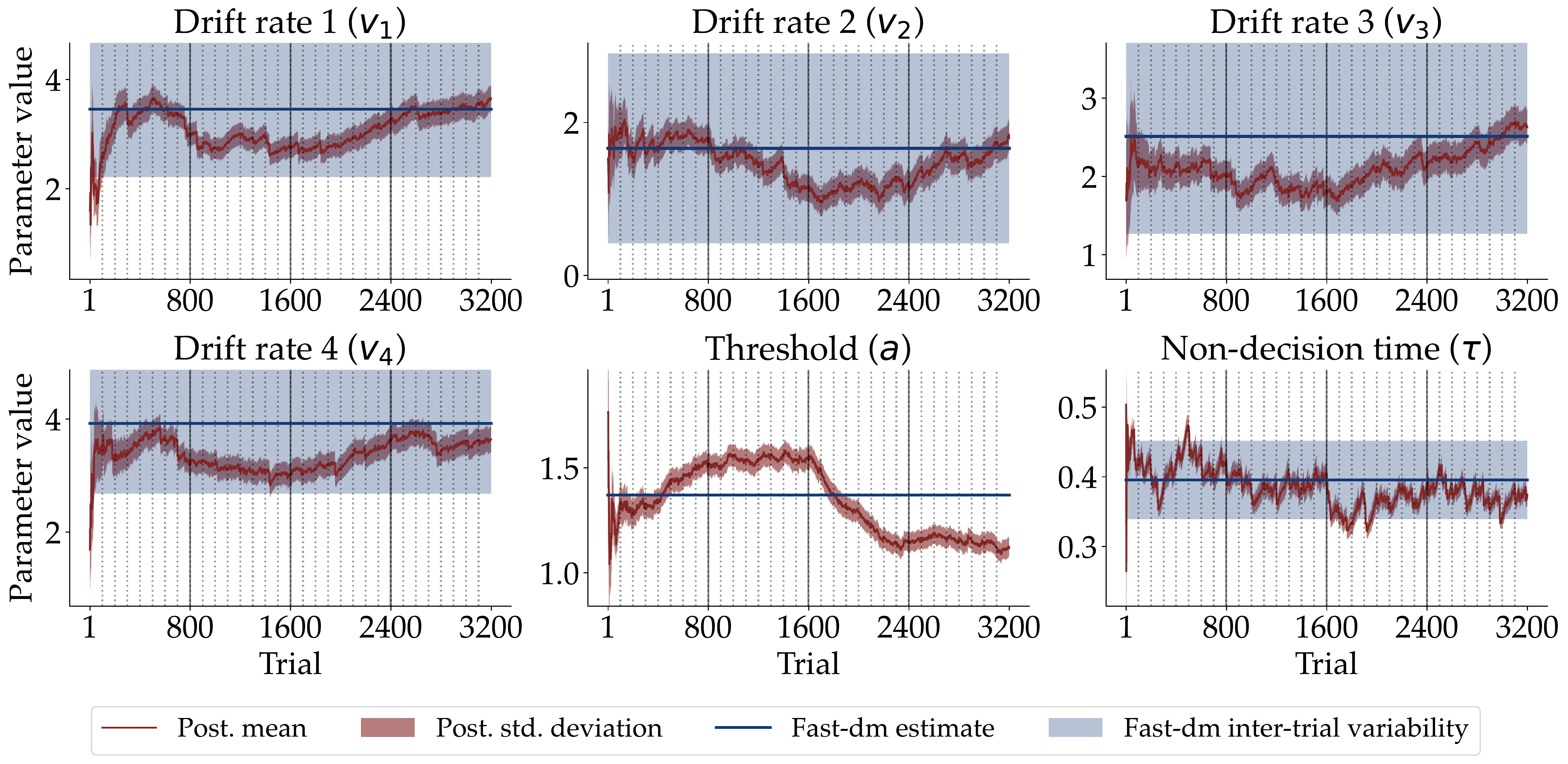}
\caption{Estimated parameter dynamics. The trial-wise posterior mean and $\pm1$ standard deviation for all six parameters, namely the four drift rates $v_1$ - $v_4$ (one for each experimental condition), the threshold $a$, and the non-decision time $\tau$ of an individual participant. The point estimates of the stationary DDM parameters and the corresponding inter-trial variabilities (except for the threshold $a$) are shown in solid blue lines and blue shaded areas, respectively.}
\label{fig:average_param_dynamics}
\end{figure*}

The left panel of \autoref{fig:time_series} depicts the empirical RT time series data of an individual participant in black (Figures for the remaining participants are available in the Appendix).
To evaluate whether the non-stationary DDM is capable of capturing the empirical data, we perform posterior re-simulations on the first $3$ blocks of the experiment (trials $1 - 2500$).
To this end, we draw $100$ samples from the posterior distributions over $\theta_{0:2499}$ to simulate $100$ posterior-re-simulated data sets.
The resulting RT time series are then summarized with the median and the $95\%$ credibility interval (CI) across simulations and depicted in red color.
We smooth the trial-by-trial empirical data and model outputs via a simple moving average (SMA) with a period of $5$ to ease visual inspection of potential trends. 
Note, that the re-simulation from the \texttt{fast-dm} model is only shown in the marginal RT distribution on the right panel to avoid visual clutter.

The overall time series show that the individual's RTs decrease over time.
Furthermore, the variability of the RTs, which is most pronounced in the first session, decreases considerably over time.
The non-stationary DDM not only captures both of these overall trends, but also represents the shorter time oscillations within the empirical RT time series.
The data also exhibits various sudden ``jumps'' in RTs, probably due to fluctuations in non-decisional processes, such as inattention. 
Unsurprisingly, these jumps are not fully accounted for by our non-stationary DDM since the high-level model (GP with squared exponential kernel) does not allow for sudden large changes in the low-level parameters.

We purposefully leave out the remaining $700$ trials from the posterior re-simulation analysis to also test the predictive capabilities of the non-stationary DDM against held-out empirical data \autocite{yarkoni2017, buerkner2020}.
To this end, we generate $100$ new parameter dynamics according to equation \eqref{GaussianProcess} with randomly drawn posterior samples of $\theta_{2499}$ as initial parameter values and posterior samples of the high-level Gaussian process parameters $\eta$.
Then, we simulate $100$ novel RT time series for the remaining $700$ trials using the simulated parameter trajectories. 
The resulting RT time series are summarized in the same manner as before (median, $95\%$ CI) and again smoothed with an SMA. 
The corresponding multi-horizon posterior predictions are depicted in \autoref{fig:time_series} with an orange color.
The dynamic model yields accurate predictions on the held-out data and thus does not overfit the training data.
Moreover, the held-out time series remain in the 95\% CI of the multi-horizon prediction, which is the case for all individual data sets (see Appendix).

The right panel of \autoref{fig:time_series} depicts the empirical RT distributions (black) along with the data generated by the non-stationary DDM (red) and the static DDM (blue).
Note that the three empirical RT distributions show a substantial overlap. 
Since the \texttt{fast-dm} re-simulations serve as a benchmark for the non-stationary DDM, it is essential to quantify if there are pronounced deviations between the re-simulated and the empirical RT distributions.
To this end, we estimate the pairwise maximum mean discrepancy (MMD) between the three distributions for each individual separately and then average the resulting values across participants.
MMD is a kernel-based statistical metric of equality between distributions \autocite{gretton12a}.

Accordingly, our analysis reveals no pronounced differences between the three distributions. 
The average MMD between the empirical RT distributions and the ones predicted by the non-stationary DDM ($\overline{MMD} = 0.026, SD = 0.008$) is lower than between the empirical and the ones predicted by the \texttt{fast-dm} model ($\overline{MMD} = 0.042, SD = 0.027$). 
The SDs of the average MMD values indicate that data generated with the non-stationary DDM are not only slightly more accurate on average but also more consistent compared to data generated from the standard DDM.
For the sake of completeness, we also compare both re-simulated RT distributions ($\overline{MMD} = 0.035, SD = 0.019$). 
This comparison reveals that the re-simulated RT distributions of the static DDM are more similar to the one obtained by the non-stationary DDM than to the empirical RT distribution.
Altogether, both models can reproduce the empirical RT distributions with high fidelity, but the non-stationary DDM fits the data slightly better than the static DDM estimated with \texttt{fast-dm}.

In summary, our non-stationary DDM can closely re-simulate and predict the temporal trajectory of empirical RT time series as well as corresponding raw RT distributions from all individuals (see Appendix). 
Even though the standard DDM also accounts for the marginal RT distribution, it cannot generate the observed heterogeneous RT time series data (cf. \autoref{fig:time_series}).

However, the most decisive advantage of our non-stationary DDM over its stationary counterpart is that it can recover parameter dynamics directly from the empirical data. 
As the static parameters of \texttt{fast-dm} can only vary homogeneously around their mean, we cannot detect any systematic changes in the parameters over time.
However, the dynamic parameters estimated with our neural method strongly suggest such systematic changes.
\autoref{fig:average_param_dynamics} depicts the dynamics of the estimated trial-by-trial posterior means and $\pm1$ standard deviation for all DDM parameters of the same participant as above in red (see Appendix for the parameter dynamics of the remaining participants as well as the average parameter dynamic).
The corresponding point estimates (solid line) and inter-trial variabilities (shaded area) obtained with \texttt{fast-dm} are shown in blue.

All parameters of the non-stationary DDM seem to exhibit considerable fluctuations and notable oscillations throughout the experiment.
Due to the assumption of homogeneous variation, the inter-trial variabilities inferred with \texttt{fast-dm} vastly overestimate the uncertainty in parameter estimates (cf. \autoref{fig:average_param_dynamics}).
The dynamic drift rates fluctuate roughly within the uncertainty corridors spanned by the homogeneous inter-trial variabilities, but exhibit much tighter error bars.
As a consequence, local drift rates are much less uncertain than the homogeneous variability parameters indicate.
On the other hand, the dynamic non-decision time $\tau$ fluctuates more than the corresponding flat inter-trial variability.
Note that \texttt{fast-dm} does not support estimating inter-trial variability of the threshold $a$, so we only report the estimates of our neural method, suggesting a substantial decrease of the threshold parameter throughout the experiment.
Notably, we observe a considerable mismatch between heterogeneous and homogeneous dynamics in almost all individuals (see Appendix).

\section*{Discussion}
In this work, we explored the merits of superstatistics for representing non-stationary dynamics in cognitive processes, along with the utility of a neural Bayesian method for estimating superstatistical models.
We verified the computational faithfulness and adequacy of our method using simulations and two benchmark studies.
We then applied our method to a dynamic, non-stationary diffusion decision model and estimated the temporal trajectories of its key parameters, namely, drift rates, decision threshold, and non-decision time from the data of two experiments.
We showed that such a non-stationary model i) can indeed be fit to long time series of human data with high fidelity and ii) that the inferred heterogeneous dynamics reveal patterns that would have remained hidden by traditional stationary models \autocite{voss2013, oberauer2018}.
To our knowledge, this is the first attempt to augment a stationary cognitive model by employing a superstatistics framework.

Previous research has suggested that response times often exhibit heterogeneous dynamics \autocite{gilden2001, wagenmakers2004}. 
It has also been shown that even the history of past choices can influence specific parameters of the DDM \autocite{urai2019a}.
Hence, it seems plausible that the cognitive processes represented by the DDM parameters vary over time even within an experimental session due to internal psychological factors.
This is exactly what was implied by the individual parameter dynamics inferred from the lexical decision task data set.
However, as the data originates from an experiment that was not designed explicitly to test dynamic modeling, we need to be wary of any \textit{ad hoc} interpretations concerning the estimated parameter dynamics.

Nevertheless, some of the recovered patterns may suggest interpretable underlying changes.
For instance, the threshold parameter seemed to decrease within an experimental session for many individuals.
This indicates that participants generally responded less cautiously toward the end of an experimental session. 
A plausible explanation for this change in response caution might be that participants became increasingly bored during a session and started to decrease their ambitions.
Note that current DDM modeling approaches rarely account for such variation in the threshold parameter.
Further, the drift rates generally tended to increase over time, suggesting that participants' increased their information processing speed over time.
A change in the average rate of information uptake typically results in shorter RTs, which is precisely what we observed in most individual data sets (cf. Appendix).
These increases in drift rates over time could imply the occurrence of learning effects.
An important next step will be to tailor experiments with systematic manipulations from which we expect specific changes in some cognitive process and test whether the estimated parameter dynamics exhibit these changes.

Notwithstanding, our neural method has certain limitations.
As can be seen in \autoref{fig:plot_stan_benchmark}, the values for most parameters change strongly at the beginning of the time series.
One could be tempted to (falsely) claim that the psychological constructs mapped to these parameters drastically change at the beginning of the first session of the experiment.
However, these early parameter trajectories should be interpreted with great caution as they can be quite dependent on the initial prior. As a result, we cannot easily differentiate between initially large Bayesian updates to move away from the prior or actual changes in the underlying process.  
As is the case for any dynamic process, our modeling approach may also not be sensible for data sets with few observations.
In the context of psychological experiments, a possible remedy could be to use burn-in trials at the beginning of an experiment that only serves the purpose of having some data points to inform the plausible parameter values.
At the same time, these could serve as practice trials during which participants get accustomed to the task.

Furthermore, the simulation study has demonstrated that the non-stationary DDM exhibits a good performance in recovering parameters across various scenarios.
However, it is essential to acknowledge that there still exists an error between the true and estimated parameters.
Especially for the drift rate parameter errors around $0.25$ have been observed frequently.
Consequently, interpreting small local changes in parameter values requires caution.
Despite this limitation, we firmly believe that the proposed method excels particularly in scenarios where moderately large changes in parameters are expected to occur over the course of a couple of time steps.

Another limitation concerns the implementation of the low-level mechanistic model, that is, the DDM itself.
We assumed four different drift rates -- one for each stimulus type -- which is the standard procedure used in the application of stationary DDMs \autocite{voss2013}.
This parameter is usually regarded as a proxy for average information uptake speed. 
However, in theory, there should only be one drift rate per participant \autocite{ratcliff2016} that changes over time, for instance, due to experimental manipulation.
Thus, a non-stationary DDM could also incorporate only one drift rate parameter. 
In our experiment, the manipulation (i.e., four conditions) was randomized throughout the experiment. 
This implies that besides fluctuation stemming from other sources, the drift rate would ``jump'' from trial to trial based on this change in task difficulty.
To account for these jumps, we would need a different high-level transition model whose changes can be bigger than
what a smooth Gaussian process or Gaussian random walk allows.
In order to keep the content of this article manageable, we decided against proposing a novel transition model.

Finally, there are numerous degrees of freedom when implementing a computational model -- not only with respect to the low-level observation model, but also regarding the high-level transition model.
Exploring different model specifications and then deciding which is the most sensible for the type of task and data at hand requires Bayesian model comparison.
Concerning dynamic cognitive models, it would be of particular interest to test which high-level transition model specification is most plausible for a given setting \autocite{mark2018}. 
Since Bayesian model comparison is a topic in its own right, future studies should investigate the utility of simulation-based methods \autocite{radev2021, schmitt2022} for comparing competing superstatistical models.

We acknowledge that our study may not provide a definitive argument for when and why a non-stationary DDM is superior to a static DDM.
The primary objective of this article is to showcase the implementation of non-stationary parameters within a superstatistics framework.
However, we believe that the superstatistics framework, coupled with powerful neural approximators, gives rise to many new modeling opportunities and makes it possible to augment virtually any computational model with time-varying parameters.
We think that there are many interesting research questions out there that could be investigated with the approach we propose in this work.
Future studies can use our method to estimate even more challenging cognitive models than the DDM explored in this work and further extend its scope beyond cognitive science and psychology.

\section*{Methods}

\subsection*{Experimental Tasks}
\subparagraph*{\textit{Random-Dot Motion Task}}
The data set used in this study was adopted from \textcite{evans2017}.
It includes data from 58 individuals, after excluding participants with a response accuracy below $70\%$. Each individual was randomly assigned to one of six groups, which were formed by two factors: the \textit{time} vs. \textit{trial} condition and three levels of feedback details.
During the experiment, participants solved a total of $24$ blocks of the task.
In the \textit{trial} condition, each block comprised $40$ trials, whereas in the \textit{time} condition, each block lasted for 1 minute.
In each trial, participants were presented with a random dot kinematogram and were required to determine if some of the dots coherently moved to the top-left or top-right direction.
For more in-depth information about the experimental setup and methodology, refer to the comprehensive details provided in \textcite{evans2017}.

\subparagraph*{\textit{Lexical Decision Task}}
A total of 11 students from Heidelberg University participated in the experiment. 
Their average age was $23.81$ ($SD = 3.30$) and $10$ of the participants were female. 
All individuals gave written informed consent to the study, which was approved by the local ethics committee. The study was conducted according to the ethical declarations of Helsinki.

The participants performed a lexical decision-making task.
On each trial, they had to assess if a presented letter string was a German word.
As stimuli, we used high and low-frequency words, pseudo words that were generated by replacing vowels of existing words, and random letter strings.
These four experimental conditions were pseudo-randomly presented throughout $3200$ trials.
All participants solved their task on 4 separate days (sessions) consisting of $800$ trials each. 
The sessions were further split into $8$ blocks of $100$ trials with short breaks between them.
On each trial, participants' choice (German word; non-German word) and response time was recorded.

\subsection*{Model Family}
Following \textcite{mark2018}, we consider dynamic models that entail a low-level model with time-dependent parameters $\theta_t$, which vary according to a high-level model with static parameters $\eta$.
The low-level model is defined by a likelihood function $\mathcal{L}$, and the high-level model consists of a transition function $\mathcal{T}$.

In this work, we aim to tackle general superstatistical models for which the low-level model likelihood $\mathcal{L}$ may not be available in closed-form.
Such models are implemented as randomized stateful simulators that generate observable trajectories $\{x_t\}_{t=1}^T$ via the following (very general) recurrent system:
\begin{align}\label{eq:gen}
    \theta_t &= \mathcal{T}(\theta_{0:t-1}, \eta, \xi_t) \quad\, \text{with}\quad \xi_t \sim p(\xi \given \eta)\\
    x_t &= \mathcal{G}(x_{1:t-1}, \theta_t, z_t) \quad \text{with}\quad z_t \sim p(z \given \theta_t).
\end{align}
In the above equation, $\mathcal{T}$ is an arbitrary high-level transition function parameterized by $\eta$, $\mathcal{G}$ stands for an arbitrary (non-linear) transformation which encodes the functional assumptions of the low-level model. $\xi_t \sim p(\xi)$ and $z_t \sim p(z)$ are sources of random noise. 
The initial parameter configuration $\theta_0$ follows a prior distribution $\theta_0 \sim p(\theta)$ which encodes available information about plausible parameter values.

One example of a transition model $\mathcal{T}$ is a convolution with a Gaussian distribution, which implies a gradual change in the low-level model's parameters resembling a random walk:
\begin{align}
    \mathcal{T}(\theta_{t-1}, \eta, \xi_t) = \theta_{t-1} + \eta\,\xi_t \quad\text{with}\quad
    \xi_t \sim \mathcal{N}(0, 1).
\label{GaussianTransition}
\end{align}
Another similar example would be a convolution with a fat-tailed distribution, allowing for abrupt changes in the parameter space. 
Furthermore, since our simulation-based setting is not limited to transition models with a Markov property, we can also test more complex transitions, such as a vector autoregression \autocite[VAR,][]{toda1994vector}:
\begin{equation}
    \mathcal{T}(\theta_{t-p:t-1}, \eta, \xi_t) = c + A_{1} \theta_{t-1} + \cdots +A_{p} \theta_{{t-p}} + \xi_{t},
\end{equation}
where $p$ is the order of the VAR model (i.e., its look-back period), $\xi_t \sim \mathcal{N}(0, \sigma)$, and $\eta = \{c, A_1,\dots,A_p, \sigma\}$ are the high-level parameters of the model.

We can even test transition models which depend on the entire history of the process, such as a Gaussian process \autocite[GP,][]{rasmussen2003gaussian}
\begin{equation}
    \theta_{1:T} \sim \mathcal{GP}(\mu_{\theta}, K_{\theta})
    \label{GaussianProcess}
\end{equation}
with mean function $\mu_{\theta}$ and covariance function $K_{\theta}$ defined through the vector of time indices.
The high-level parameters $\eta$ in this case would be the free kernel parameters, such as the amplitude $\sigma$ or the length-scale $l$ of a Gaussian kernel
\begin{equation}
    k(\theta_t, \theta_{t'}) = \sigma^2 \exp\left(\frac{||\theta_t - \theta_{t'} ||^2}{2l^2}\right).
    \label{GaussianKernel}
\end{equation}
A typical task in Bayesian analysis of dynamic systems is to recover both the entire trajectory of dynamic parameters $\{\theta_t\}^{T}_{t=1}$ as well as the vector of static parameters $\eta$.
Since for many discrete dynamic systems, the current data point $x_t$ depends on the current parameter configuration $\theta_t$ as well as on the observable history of the system $x_{1:t-1}$, we can write the (implicit) point-wise likelihood as 
\begin{equation}\label{eq:pointlik}
    \mathcal{L}_t = p(x_t \given x_{1:t-1}, \theta_t).
\end{equation}
The point-wise likelihood describes the probability of each data point, given the parameter values of the same time step and all past data points \autocite{mark2018}.
Notably, we do not require this likelihood to be available in closed-form; we only need the ability to generate random draws through the forward-time generative process specified by equation \eqref{eq:gen}.

Assuming the above factorization of the likelihood is possible, we aim to estimate the joint \textit{filtering} posterior distribution of $\theta_t$ and $\eta$ up to each discrete time-step $t$
\begin{equation}\label{eq:jointpost}
    p(\theta_t, \eta \given x_{1:t}) \propto \mathcal{L}_t\,p(\theta_t \given x_{1:t-1}, \eta)\, p(\eta \given x_{1:t-1}). 
\end{equation}
This posterior encodes the reduction in uncertainty regarding the dynamic states evolving over time and the static parameter values being increasingly constrained by the data. 
From this joint distribution, we can derive the corresponding marginal posteriors as follows:
\begin{align}
    p(\theta_t \given x_{1:t}) &= \int p(\theta_t, \eta \given x_{1:t}) \, d\eta, \\
    p(\eta \given x_{1:t}) &= \int p(\theta_t, \eta \given x_{1:t}) \,d \theta_t. 
\end{align}
These distributions describe the average parameter dynamics over all possible high-level parameters and the best estimate for the high-level parameters up to discrete time-step $t$, respectively.
Thus, learning both distributions amounts to standard Bayesian updating with an additional uncertainty factor due to the high-level transition model $\mathcal{T}$.
Thus, posterior contraction over time will strongly depend on the form of the transition model and may even increase in some cases, such as models allowing for sudden ``jumps'' in their parameters (i.e., regime switching behavior).

\subsection*{Neural Bayesian Estimation}

Various methods for estimating dynamic models have been proposed in the literature.
Markov chain Monte Carlo (MCMC) methods offer a viable but computationally demanding approach based on random draws from the posterior \autocite{blei2017variational}. 
Variational inference (VI) methods approximate the true target posterior with simple, tractable densities and thus are a faster alternative to MCMC at the cost of a potential loss of posterior accuracy \autocite{blei2017variational}. 
A recent promising approach for low-dimensional problems is the grid-based method of \textcite{mark2018}, which represents parameter distribution on discrete lattices and enables efficient approximation of model evidence.

\begin{figure}[h]
\centering
\includegraphics[width=0.42\textwidth]{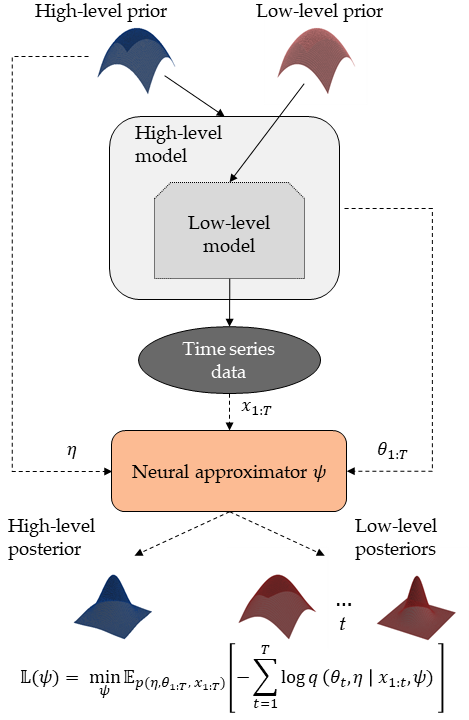}
\caption{A graphical illustration of our neural inference method.
A recurrent neural approximator updates the posterior of the low-level model parameters $\theta_t$ each time step $t$ and yields the posterior over the high-level model parameters $\eta$ considering all available data.
The low-level prior constrains the initial dynamic parameter values $\theta_0$, which then get passed to the high-level transition model.
Together, the two priors and the two models comprise a stochastic simulator that trains the neural approximator to perform amortized Bayesian updating.}
\label{fig:conceptual}
\end{figure}

However, the above methods all depend on the ability to evaluate the likelihood function $\mathcal{L}_t$ at each time point explicitly. 
This restriction makes it impossible to efficiently test the growing number of simulator-based or non-analytic models of cognition to observed data \autocite{radev2020, van_rooij2019}.
Furthermore, MCMC and standard variational methods cannot leverage experience and require the same repeated computational effort for every new data set.
For instance, when multiple participants complete a cognitive task, the same estimation procedures need to be repeated for each participant from scratch.
Differently, hierarchical Bayesian models can be employed to jointly estimate group- and participant-level parameters, but they come with high computational costs and also rely on a closed-form likelihood function.

In contrast, \textit{amortized inference} refers to methods with a ``pre-paid'' computational cost - after an expensive optimization or training phase, the same procedure can be instantly applied to any data set whose structure is compatible with the model \autocite{radev2020, mestdagh2019prepaid}.
As a useful ``side effect'', amortization allows us to easily perform extensive checks of computational faithfulness and parameter recoverability ``in silico'', since we can obtain posterior samples from hundreds or even thousands of simulated data sets by applying the same pre-trained network.
Amortized Bayesian inference is typically realized by specialized neural networks, which are trained to become estimation experts from repeated model simulations \autocite{greenberg2019automatic, radev2020}.
The architecture of these networks can easily encode the probabilistic symmetry of the data, for instance, recurrent networks for temporal data \autocite{gers2000learning} or permutation-invariant networks for IID data \autocite{bloem2020probabilistic}.

Crucially, dynamic models with time-varying parameters present a challenge to existing neural architectures since they induce a new joint posterior at each time-step $p(\theta_t, \eta \given x_{1:t})$.
However, most previous architectures can only estimate a single set of parameters with no temporal information \autocite{radev2020, greenberg2019automatic, cranmer2020}.
Thus, we propose to use a recurrent probabilistic neural architecture that estimates the joint posterior over all static and dynamic parameters for all discrete time points in a single forward pass.

\subsection*{Recurrent Estimation Method}

Our proposed architecture consists of several neural components. 
First, a recurrent neural network (RNN) with learnable parameters $\psi^{(r)}$ embodying long short-term memory (LSTM) consumes the observed data sequentially:
\begin{align}
    h_t = \text{LSTM}(x_{t}, h_{t-1}; \psi^{(r)}),
\end{align}
where the hidden state $h_t$ at each time point $t$ represents the internal memory of the network over arbitrary temporal intervals. 
Thus, we can treat $h_t$ as a compact representation of the observable history up to time point $t$.
We employ a standard LSTM network, which consists of three gates: an input gate, an output gate, and a forget gate.
These gates are responsible for weighing and integrating old and new information.
Importantly, LSTM networks can naturally deal with sequences of varying length, which enables them to process streams of ``online'' data \autocite{gers2000learning}.

In order to recover the time-varying parameters $\theta_t$ of the low-level model as well as the static high-level parameters $\eta$, we use the hidden state $h_t$ as a conditioning vector for a generative neural network with trainable weights $\psi^{(g)}$.
This network can be implemented as a conditional variant of any popular generative architecture for inference, such as coupling networks \autocite{ardizzone2019}, autoregressive flows \autocite{papamakarios2017masked}, or standard neural networks with probabilistic outputs \autocite{kendall2017uncertainties}.
The generative network is responsible for approximating the current joint posterior up to time step $t$ given the outputs of the recurrent summary network: $q(\theta_t, \eta \given x_{1:t}, \psi) \equiv q(\theta_t, \eta \given h_t, \psi)$.
To reduce notational clutter, we set $\psi = (\psi^{(r)}, \psi^{(g)})$ and assume that $h_t$ is expressive enough to encode all information contained in the data for correctly updating the prior (i.e., $h_t$ is a maximally informative \textit{summary statistic} of $x_{1:t}$).

Alternatively, we can also directly target one of the two equivalent factorizations of the joint posterior, namely:
\begin{align}
    p(\theta_t, \eta \given x_{1:t}) &= p(\theta_t \given x_{1:t}, \eta)\,p(\eta \given x_{1:t}) \label{eq:marg_1}\\ 
    &= p(\eta \given x_{1:t}, \theta_t)\,p(\theta_t \given x_{1:t}) \label{eq:marg_2}.
\end{align}
While being mathematically equal, these factorizations imply different neural architectures and corresponding ancestral sampling schemes.
The former factorization (equation \eqref{eq:marg_1}) requires a generative network for first sampling the high-level parameters from $p(\eta \given x_{1:t})$ and then sampling the low-level parameters from $p(\theta_t \given x_{1:t}, \eta)$, conditional on the sampled high-level parameters.
On the other hand, the latter factorization (equation \eqref{eq:marg_2}) requires a generative network for first sampling the low-level parameters from $p(\theta_t \given x_{1:t})$ and then sampling the high-level parameters from $p(\eta \given x_{1:t}, \theta_t)$, conditional on the sampled low-level parameters.

In the current work, we consistently target the factorization in equation \eqref{eq:marg_1}, but we were able to obtain comparable filtering results with either ancestral sampling strategy.
In practice, we can either assume a multivariate Gaussian posterior for $q(\theta_t \given x_{1:t}, \eta, \psi)$ and $q(\eta \given x_{1:t}, \psi)$ as a dynamic extension of the basic method in \autocite{radev2020towards} or estimate free-form posteriors as a dynamic extension of the BayesFlow method \autocite{radev2020}.
We use the former approach for the toy \textbf{Coal Mining} benchmark and the latter approach for all other experiments in this work.

\subsection*{Simulation-Based Training}
\autoref{fig:conceptual} graphically illustrates the rationale of our simulation-based inference approach.
To train the networks, we treat the forward-time generative model as a simulator and employ equation \eqref{eq:gen} to generate multiple sets of simulated parameters and trajectories $(\eta, \theta_{1:T}, x_{1:T})$.
We then minimize the Monte Carlo estimate of the following criterion
\begin{align}
    \mathbb{L}(\psi) = \min_{\psi} \,\mathbb{E}_{p(\eta, \theta_{1:T}, x_{1:T})}\left[- \sum_{t=1}^T\log q(\theta_t, \eta \given x_{1:t}, \psi)\right],
\end{align}
where $\mathbb{E}\left[\cdot\right]$ denotes an expectation over the dynamic generative model and $\psi = (\psi^{(r)}, \psi^{(g)})$ denotes the collection of all trainable neural network parameters.
This criterion ensures that the approximate posteriors match the analytic posteriors induced by the dynamic model and can be minimized either via online (i.e., generating dynamic simulations on the fly) or via offline training (i.e., using a set of pre-computed dynamic simulations). 


\subsection*{A Non-Stationary Diffusion Decision Model}
To illustrate the potential of our approach, we will re-formulate in superstatistical terms a popular cognitive model for analyzing human response times (RTs) in binary decision tasks, namely the DDM.
The standard DDM describes the microscopic dynamics of perceptual evidence accumulations via a simple stochastic ordinary differential equation (SDE).
Accordingly, the accumulated evidence $x_j$ in experimental task $j$ follows a random walk with drift and Gaussian noise:
\begin{align}
    \mathrm{d}x_j = v\mathrm{d}t_s + z \sqrt{\mathrm{d}t_s}
    \quad\text{with}\quad
    z\sim\mathcal{N}(0, 1)
    \label{eq:DDM}
\end{align}
where $t_s$ represents time on a continuous microscopic scale (i.e., during forced-choice decision making).
A core assumption of the DDM is that task-relevant information (i.e., perceptual evidence) accumulates at a constant rate ($v$).
This process runs in a corridor with two absorbing boundaries, which represent two decision alternatives.
As soon as the accumulated evidence $x_j$ reaches either a pre-defined threshold ($a$) or $0$, the model makes a categorical decision $D_j$ for the alternative favored by the collected evidence:
\begin{align}
    D_j = 
    \begin{cases}
        1, & \text{if } x_j \geq a\\
        0, & \text{if } x_j \leq 0
    \end{cases}.
\end{align}
Further, the model assumes a constant additive factor ($\tau$) accounting for non-decision processes, such as encoding or motor responses.
Thus, the standard (static) DDM has three key parameters $\theta = (v, a, \tau)$. The starting point of the decision process is either estimated as an additional parameter or fixed at $a/2$.

The typical assumption of the standard DDM is that the parameters $\theta$ remain stationary for the duration of a given cognitive task.
In order to relax this restrictive assumption, the standard DDM has been extended to incorporate so-called inter-trial-variability for the drift rate and non-decision time parameters \autocite{ratcliff1998, ratcliff2002a}. 
In this way, the extended DDM concedes that these cognitive parameters are not static but vary over time. 
However, the assumed variation is homogeneous and memoryless, and the generative model still yields IID data, that is, the transition model coincides with independent sampling and reduces to $\theta_t = \mathcal{T}(\eta, \xi_t)$.

In contrast, our superstatistical model assumes a stateful Gaussian process (GP) high-level model, which describes the trial-by-trial dynamics of the DDM parameters according to equation \eqref{GaussianProcess} and \eqref{GaussianKernel} (see Appendix for detail).

Thereby, we want to demonstrate that our estimation method can tackle very flexible transition models $\mathcal{T}$, as long as we can simulate data from the low-level model.
However, we also fit a DDM with a simpler Gaussian random walk transition model to the data described in the \textbf{Human Data Application} section.
This simpler model corroborates our findings by suggesting qualitatively similar parameter dynamics, but yields less sharp predictions on unseen data than its GP counterpart (see Appendix for more details).

\section*{Acknowledgments}
We thank Daniel Durstewitz and Lasse Elsemüller for their helpful feedback on this project. We also thank Marie Wieschen for her efforts in data collection. L.S. and A.V. were supported by the Deutsche Forschungsgemeinschaft (DFG, German Research Foundation; grant number GRK 2277 "Statistical Modeling in Psychology"). 
P-C.B. was supported by the Deutsche Forschungsgemeinschaft (DFG, German Research Foundation) under Germany’s Excellence Strategy -- EXC-2075 - 390740016 (the Stuttgart Cluster of Excellence SimTech). 
S.T.R was supported by the Deutsche Forschungsgemeinschaft (DFG, German Research Foundation) under Germany’s Excellence Strategy -- EXC-2181 - 390900948 (the Heidelberg Cluster of Excellence STRUCTURES).
For the publication fee we acknowledge financial support by Deutsche Forschungsgemeinschaft within the funding programme "Open Access Publikationskosten" as well as by Heidelberg University.

\section*{Author contributions}
L.S., S.T.R., designed the research, created and applied the models, and wrote the initial draft of the manuscript. P.C.B. significantly contributed with his statistical and scientific expertise to the methods and the written content. A.V. and U.K. supervised the project and contributed to all stages.

\section*{Data and Code Availability}
All models, data, and scripts for reproducing the results of this work are publicly available in the project's repository \url{https://github.com/bayesflow-org/Neural-Superstatistics}.
Our methods are implemented in the \texttt{BayesFlow} Python library for amortized Bayesian workflows \autocite{radev2023bayesflow}.

\section*{Competing interests}
The authors declare no competing interest.

\newpage
\printbibliography

@article{evans2017,
  title={People adopt optimal policies in simple decision-making, after practice and guidance},
  author={Evans, Nathan J and Brown, Scott D},
  journal={Psychonomic Bulletin \& Review},
  volume={24},
  pages={597--606},
  year={2017},
  publisher={Springer}
}

@article{gunawan2022,
  title={Time-evolving psychological processes over repeated decisions.},
  author={Gunawan, David and Hawkins, Guy E and Kohn, Robert and Tran, Minh-Ngoc and Brown, Scott D},
  journal={Psychological review},
  volume={129},
  number={3},
  pages={438},
  year={2022},
  publisher={American Psychological Association},
  doi = {10.1037/rev0000351}
}

@article{vonKrause2021,
  title={Stability and change in diffusion model parameters over two years},
  author={von Krause, Mischa and Radev, Stefan T and Voss, Andreas and Quintus, Martin and Egloff, Boris and Wrzus, Cornelia},
  journal={Journal of Intelligence},
  volume={9},
  number={2},
  pages={26},
  year={2021},
  publisher={MDPI}
}

@article{kahana2018,
  title={The variability puzzle in human memory.},
  author={Kahana, Michael J and Aggarwal, Eash V and Phan, Tung D},
  journal={Journal of Experimental Psychology: Learning, Memory, and Cognition},
  volume={44},
  number={12},
  pages={1857},
  year={2018},
  publisher={American Psychological Association},
  doi = {10.1037/xlm0000553}
}

@article{evans2018,
  title={Refining the law of practice.},
  author={Evans, Nathan J and Brown, Scott D and Mewhort, Douglas JK and Heathcote, Andrew},
  journal={Psychological review},
  volume={125},
  number={4},
  pages={592},
  year={2018},
  publisher={American Psychological Association},
  doi = {10.1037/rev0000105}
}

@article{christoff2016,
  title={Mind-wandering as spontaneous thought: a dynamic framework},
  author={Christoff, Kalina and Irving, Zachary C and Fox, Kieran CR and Spreng, R Nathan and Andrews-Hanna, Jessica R},
  journal={Nature Reviews Neuroscience},
  volume={17},
  number={11},
  pages={718--731},
  year={2016},
  publisher={Nature Publishing Group},
  doi = {10.1038/nrn.2016.113.}
}

@article{ratcliff2011,
  title={Diffusion model for one-choice reaction-time tasks and the cognitive effects of sleep deprivation},
  author={Ratcliff, Roger and Van Dongen, Hans PA},
  journal={Proceedings of the National Academy of Sciences},
  volume={108},
  number={27},
  pages={11285--11290},
  year={2011},
  publisher={National Acad Sciences},
  doi = {10.1073/pnas.1100483108}
}

@article{walsh2017,
  title={Computational cognitive modeling of the temporal dynamics of fatigue from sleep loss},
  author={Walsh, Matthew M and Gunzelmann, Glenn and Van Dongen, Hans},
  journal={Psychonomic bulletin \& review},
  volume={24},
  number={6},
  pages={1785--1807},
  year={2017},
  publisher={Springer},
  doi = {10.3758/s13423-017-1243-6 }
}

@article{sailynoja2021,
  doi = {10.48550/ARXIV.2103.10522},
  url = {https://arxiv.org/abs/2103.10522},
  author = {Säilynoja, Teemu and Bürkner, Paul-Christian and Vehtari, Aki},
  title = {Graphical Test for Discrete Uniformity and its Applications in Goodness of Fit Evaluation and Multiple Sample Comparison},
  publisher = {arXiv},
  year = {2021},
  copyright = {arXiv.org perpetual, non-exclusive license}
}

@inbook{van_rooij2019,
    place={Cambridge},
    title={Applications. In Cognition and Intractability: A Guide to Classical and Parameterized Complexity Analysis},
    publisher={Cambridge University Press},
    author={van Rooij, Iris and Blokpoel, Mark and Kwisthout, Johan and Wareham, Todd},
    year={2019},
    pages={215–216},
    doi = {10.1017/9781107358331}
    }

@article{talts2018,
  doi = {10.48550/ARXIV.1804.06788},
  url = {https://arxiv.org/abs/1804.06788},
  author = {Talts, Sean and Betancourt, Michael and Simpson, Daniel and Vehtari, Aki and Gelman, Andrew},
  title = {Validating Bayesian Inference Algorithms with Simulation-Based Calibration},
  publisher = {arXiv},
  year = {2018},
  copyright = {arXiv.org perpetual, non-exclusive license}
}

@article{schmitt2022,
  doi = {10.48550/ARXIV.2210.07278},
  url = {https://arxiv.org/abs/2210.07278},
  author = {Schmitt, Marvin and Radev, Stefan T. and Bürkner, Paul-Christian},
  title = {Meta-Uncertainty in Bayesian Model Comparison},
  publisher = {arXiv},
  year = {2022},
  copyright = {Creative Commons Attribution Non Commercial No Derivatives 4.0 International}
}

@article{radev2021,
  author={Radev, Stefan T. and D'Alessandro, Marco and Mertens, Ulf K. and Voss, Andreas and Köthe, Ullrich and Bürkner, Paul-Christian},
  journal={IEEE Transactions on Neural Networks and Learning Systems},
  title={Amortized Bayesian Model Comparison With Evidential Deep Learning},
  year={2021},
  volume={},
  number={},
  pages={1-15},
  doi={10.1109/TNNLS.2021.3124052}
  }

@article{eckstein2020,
  title = {Computational Evidence for Hierarchically Structured Reinforcement Learning in Humans},
  author = {Eckstein, Maria K. and Collins, Anne G. E.},
  year = {2020},
  journal = {Proceedings of the National Academy of Sciences},
  volume = {117},
  number = {47},
  pages = {29381--29389},
  publisher = {{Proceedings of the National Academy of Sciences}},
  doi = {10.1073/pnas.1912330117}
}

@article{schad2021,
  title = {Toward a Principled {{Bayesian}} Workflow in Cognitive Science},
  author = {Schad, Daniel J. and Betancourt, Michael and Vasishth, Shravan},
  year = {2021},
  journal = {Psychological Methods},
  volume = {26},
  pages = {103--126},
  publisher = {{American Psychological Association}},
  address = {{US}},
  issn = {1939-1463},
  doi = {10.1037/met0000275}
}

@article{betancourt2018,
  doi = {10.48550/ARXIV.1803.08393},
  author = {Betancourt, Michael},
  title = {Calibrating Model-Based Inferences and Decisions},
  publisher = {arXiv},
  year = {2018}
}

@article{yarkoni2017,
    title = {Choosing {Prediction} {Over} {Explanation} in {Psychology}: {Lessons} {From} {Machine} {Learning}},
    volume = {12},
    issn = {1745-6924},
    shorttitle = {Choosing {Prediction} {Over} {Explanation} in {Psychology}},
    doi = {10.1177/1745691617693393},
    number = {6},
    journal = {Perspectives on Psychological Science: A Journal of the Association for Psychological Science},
    author = {Yarkoni, Tal and Westfall, Jacob},
    year = {2017},
    pmid = {28841086},
    pmcid = {PMC6603289},
    pages = {1100--1122}
}

@article{buerkner2020,
	doi = {10.1080/00949655.2020.1783262},
	year = 2020,
	publisher = {Informa {UK} Limited},
	volume = {90},
	number = {14},
	pages = {2499--2523},
	author = {Paul-Christian Bürkner and Jonah Gabry and Aki Vehtari},
	title = {Approximate leave-future-out cross-validation for Bayesian time series models},
	journal = {Journal of Statistical Computation and Simulation}
}

@article{brockmole2013,
  title = {Age-Related Change in Visual Working Memory: A Study of 55,753 Participants Aged 8-75},
  shorttitle = {Age-Related Change in Visual Working Memory},
  author = {Brockmole, James R. and Logie, Robert H.},
  year = {2013},
  journal = {Frontiers in Psychology},
  volume = {4},
  pages = {12},
  issn = {1664-1078},
  doi = {10.3389/fpsyg.2013.00012},
  pmcid = {PMC3557412},
  pmid = {23372556}
}

@article{gretton12a,
  title = {A Kernel Two-Sample Test},
  author = {Gretton, Arthur and Borgwardt, Karsten M. and Rasch, Malte J. and Sch{\"o}lkopf, Bernhard and Smola, Alexander},
  year = {2012},
  journal = {Journal of Machine Learning Research},
  volume = {13},
  number = {25},
  pages = {723--773}
}

@article{voss2007,
  title = {Fast-Dm: {{A}} Free Program for Efficient Diffusion Model Analysis},
  shorttitle = {Fast-Dm},
  author = {Voss, Andreas and Voss, Jochen},
  year = {2007},
  journal = {Behavior Research Methods},
  volume = {39},
  number = {4},
  pages = {767--775},
  issn = {1554-3528},
  doi = {10.3758/BF03192967}
}

@misc{gelman2020,
  doi = {10.48550/ARXIV.2011.01808},
  author = {Gelman, Andrew and Vehtari, Aki and Simpson, Daniel and Margossian, Charles C. and Carpenter, Bob and Yao, Yuling and Kennedy, Lauren and Gabry, Jonah and Bürkner, Paul-Christian and Modrák, Martin},
  title = {Bayesian Workflow},
  publisher = {arXiv},
  year = {2020}
}

@article{beck2003,
  title = {Superstatistics},
  author = {Beck, C. and Cohen, E. G. D.},
  year = {2003},
  journal = {Physica A: Statistical Mechanics and its Applications},
  volume = {322},
  pages = {267--275},
  issn = {0378-4371},
  doi = {10.1016/S0378-4371(03)00019-0},
  langid = {english}
}

@article{bogachev2017,
  title = {Superstatistical Model of Bacterial {{DNA}} Architecture},
  author = {Bogachev, Mikhail I. and Markelov, Oleg A. and Kayumov, Airat R. and Bunde, Armin},
  year = {2017},
  journal = {Scientific Reports},
  volume = {7},
  number = {1},
  pages = {43034},
  publisher = {{Nature Publishing Group}},
  issn = {2045-2322},
  doi = {10.1038/srep43034},
  copyright = {2017 The Author(s)},
  langid = {english}
}

@article{brosowsky2020,
  title = {Mind Wandering, Motivation, and Task Performance over Time: {{Evidence}} That Motivation Insulates People from the Negative Effects of Mind Wandering},
  shorttitle = {Mind Wandering, Motivation, and Task Performance over Time},
  author = {Brosowsky, Nicholaus P. and DeGutis, Joseph and Esterman, Mike and Smilek, Daniel and Seli, Paul},
  year = {2020},
  journal = {Psychology of Consciousness: Theory, Research, and Practice},
  pages = {No Pagination Specified-No Pagination Specified},
  publisher = {{Educational Publishing Foundation}},
  address = {{US}},
  issn = {2326-5531},
  doi = {10.1037/cns0000263}
}

@article{collins2018,
  title = {Within- and across-Trial Dynamics of Human {{EEG}} Reveal Cooperative Interplay between Reinforcement Learning and Working Memory},
  author = {Collins, Anne G. E. and Frank, Michael J.},
  year = {2018},
  journal = {Proceedings of the National Academy of Sciences},
  volume = {115},
  number = {10},
  pages = {2502--2507},
  publisher = {{Proceedings of the National Academy of Sciences}},
  doi = {10.1073/pnas.1720963115}
}

@article{cranmer2020,
  title = {The Frontier of Simulation-Based Inference},
  author = {Cranmer, Kyle and Brehmer, Johann and Louppe, Gilles},
  year = {2020},
  journal = {Proceedings of the National Academy of Sciences},
  volume = {117},
  number = {48},
  pages = {30055--30062},
  publisher = {{Proceedings of the National Academy of Sciences}},
  doi = {10.1073/pnas.1912789117}
}

@article{denys2016,
  title = {Universality of Market Superstatistics},
  author = {Denys, Mateusz and Gubiec, Tomasz and Kutner, Ryszard and Jagielski, Maciej and Stanley, H. Eugene},
  year = {2016},
  journal = {Physical Review E},
  volume = {94},
  number = {4},
  pages = {042305},
  publisher = {{American Physical Society}},
  doi = {10.1103/PhysRevE.94.042305}
}

@article{diederich2006,
  title = {Modeling the Effects of Payoff on Response Bias in a Perceptual Discrimination Task: {{Bound-change}}, Drift-Rate-Change, or Two-Stage-Processing Hypothesis},
  shorttitle = {Modeling the Effects of Payoff on Response Bias in a Perceptual Discrimination Task},
  author = {Diederich, Adele and Busemeyer, Jerome R.},
  year = {2006},
  journal = {Perception \& Psychophysics},
  volume = {68},
  number = {2},
  pages = {194--207},
  issn = {1532-5962},
  doi = {10.3758/BF03193669},
  langid = {english}
}

@book{farrell2018,
  title = {Computational {{Modeling}} of {{Cognition}} and {{Behavior}}},
  author = {Farrell, Simon and Lewandowsky, Stephan},
  year = {2018},
  publisher = {{Cambridge University Press}},
  address = {{Cambridge}},
  doi = {10.1017/CBO9781316272503},
  isbn = {978-1-107-10999-5}
}

@article{favela2020,
  title = {Cognitive Science as Complexity Science},
  author = {Favela, Luis H.},
  year = {2020},
  journal = {WIREs Cognitive Science},
  volume = {11},
  number = {4},
  pages = {e1525},
  issn = {1939-5086},
  doi = {10.1002/wcs.1525},
  langid = {english}
}

@article{gasimova2014,
  title = {Dynamical Systems Analysis Applied to Working Memory Data},
  author = {Gasimova, Fidan and Robitzsch, Alexander and Wilhelm, Oliver and Boker, Steven M. and Hu, Yueqin and H{\"u}l{\"u}r, Gizem},
  year = {2014},
  journal = {Frontiers in Psychology},
  volume = {5},
  issn = {1664-1078},
  doi = {10.3389/fpsyg.2014.00687}
}

@inproceedings{rasmussen2003gaussian,
  title={Gaussian processes in machine learning},
  author={Rasmussen, Carl Edward},
  booktitle={Summer school on machine learning},
  pages={63--71},
  year={2003},
  organization={Springer},
  doi = {10.1007/978-3-540-28650-9_4}
}

@article{radev2020towards,
  title={Towards end-to-end likelihood-free inference with convolutional neural networks},
  author={Radev, Stefan T and Mertens, Ulf K and Voss, Andreas and K{\"o}the, Ullrich},
  journal={British Journal of Mathematical and Statistical Psychology},
  volume={73},
  number={1},
  pages={23--43},
  year={2020},
  publisher={Wiley Online Library},
  doi={10.1111/bmsp.12159}
}

@article{neal_mcmc_2011,
	title = {{MCMC} using {Hamiltonian} dynamics},
	volume = {2},
	number = {11},
	journal = {Handbook of markov chain monte carlo},
	author = {Neal, Radford M and {others}},
	year = {2011},
	note = {Publisher: Chapman and Hall/CRC},
	pages = {2},
    doi = {10.1201/b10905-7}
 
}

@article{gershman2017,
  title = {Reinforcement Learning and Episodic Memory in Humans and Animals: An Integrative Framework},
  shorttitle = {Reinforcement Learning and Episodic Memory in Humans and Animals},
  author = {Gershman, Samuel J. and Daw, Nathaniel D.},
  year = {2017},
  journal = {Annual review of psychology},
  volume = {68},
  pages = {101--128},
  issn = {0066-4308},
  doi = {10.1146/annurev-psych-122414-033625},
  pmcid = {PMC5953519},
  pmid = {27618944}
}

@article{gilden2001,
  title = {Cognitive Emissions of 1/f Noise},
  author = {Gilden, D. L.},
  year = {2001},
  journal = {Psychological Review},
  volume = {108},
  number = {1},
  pages = {33--56},
  issn = {0033-295X},
  doi = {10.1037/0033-295x.108.1.33},
  langid = {english},
  pmid = {11212631}
}

@article{hanel2011,
  title = {Generalized Entropies and the Transformation Group of Superstatistics},
  author = {Hanel, Rudolf and Thurner, Stefan and {Gell-Mann}, Murray},
  year = {2011},
  journal = {Proceedings of the National Academy of Sciences},
  volume = {108},
  number = {16},
  pages = {6390--6394},
  publisher = {{Proceedings of the National Academy of Sciences}},
  doi = {10.1073/pnas.1103539108}
}

@article{toda1994vector,
  title={Vector autoregression and causality: a theoretical overview and simulation study},
  author={Toda, Hiro Y and Phillips, Peter CB},
  journal={Econometric reviews},
  volume={13},
  number={2},
  pages={259--285},
  year={1994},
  publisher={Taylor \& Francis},
  doi = {10.1080/07474939408800286}
}

@article{kiuru2020,
  title = {The Dynamics of Motivation, Emotion, and Task Performance in Simulated Achievement Situations},
  author = {Kiuru, Noona and Spinath, Birgit and Clem, Anna-Leena and Eklund, Kenneth and Ahonen, Timo and Hirvonen, Riikka},
  year = {2020},
  journal = {Learning and Individual Differences},
  volume = {80},
  pages = {101873},
  issn = {1041-6080},
  doi = {10.1016/j.lindif.2020.101873},
  langid = {english}
}

@article{kucharsky2021,
  title = {Hidden {{Markov Models}} of {{Evidence Accumulation}} in {{Speeded Decision Tasks}}},
  author = {Kucharsk{\'y}, {\v S}imon and Tran, N.-Han and Veldkamp, Karel and Raijmakers, Maartje and Visser, Ingmar},
  year = {2021},
  journal = {Computational Brain \& Behavior},
  volume = {4},
  number = {4},
  pages = {416--441},
  issn = {2522-087X},
  doi = {10.1007/s42113-021-00115-0},
  langid = {english}
}

@article{mark2018,
  title = {Bayesian Model Selection for Complex Dynamic Systems},
  author = {Mark, Christoph and Metzner, Claus and Lautscham, Lena and Strissel, Pamela L. and Strick, Reiner and Fabry, Ben},
  year = {2018},
  journal = {Nature Communications},
  volume = {9},
  number = {1},
  pages = {1803},
  publisher = {{Nature Publishing Group}},
  issn = {2041-1723},
  doi = {10.1038/s41467-018-04241-5},
  copyright = {2018 The Author(s)},
  langid = {english}
}

@misc{ardizzone2019,
  title = {Guided {{Image Generation}} with {{Conditional Invertible Neural Networks}}},
  author = {Ardizzone, Lynton and L{\"u}th, Carsten and Kruse, Jakob and Rother, Carsten and K{\"o}the, Ullrich},
  year = {2019},
  month = jul,
  number = {arXiv:1907.02392},
  publisher = {{arXiv}},
  doi = {10.48550/arXiv.1907.02392},
  archiveprefix = {arXiv}
}

@article{papamakarios2017masked,
  title={Masked autoregressive flow for density estimation},
  author={Papamakarios, George and Pavlakou, Theo and Murray, Iain},
  journal={Advances in neural information processing systems},
  volume={30},
  year={2017}
}

@article{kendall2017uncertainties,
  title={What uncertainties do we need in bayesian deep learning for computer vision?},
  author={Kendall, Alex and Gal, Yarin},
  journal={Advances in neural information processing systems},
  volume={30},
  year={2017}
}

@article{burkner2022some,
  title={Some models are useful, but how do we know which ones? Towards a unified Bayesian model taxonomy},
  author={B{\"u}rkner, Paul-Christian and Scholz, Maximilian and Radev, Stefan},
  journal={arXiv preprint arXiv:2209.02439},
  year={2022},
  doi={10.48550/arXiv.2209.02439}
}

@article{mestdagh2019prepaid,
  title={Prepaid parameter estimation without likelihoods},
  author={Mestdagh, Merijn and Verdonck, Stijn and Meers, Kristof and Loossens, Tim and Tuerlinckx, Francis},
  journal={PLoS computational biology},
  volume={15},
  number={9},
  pages={e1007181},
  year={2019},
  publisher={Public Library of Science San Francisco, CA USA},
  doi = {10.1371/journal.pcbi.1007181}
}

@article{carpenter2017stan,
  title={Stan: A probabilistic programming language},
  author={Carpenter, Bob and Gelman, Andrew and Hoffman, Matthew D and Lee, Daniel and Goodrich, Ben and Betancourt, Michael and Brubaker, Marcus and Guo, Jiqiang and Li, Peter and Riddell, Allen},
  journal={Journal of statistical software},
  volume={76},
  number={1},
  year={2017},
  publisher={Columbia Univ., New York, NY (United States); Harvard Univ., Cambridge, MA~…},
  doi = {10.18637/jss.v076.i01}
}

@article{blei2017variational,
  title={Variational inference: A review for statisticians},
  author={Blei, David M and Kucukelbir, Alp and McAuliffe, Jon D},
  journal={Journal of the American statistical Association},
  volume={112},
  number={518},
  pages={859--877},
  year={2017},
  publisher={Taylor \& Francis},
  doi = {10.1080/01621459.2017.1285773}
}

@article{metzner2021,
  title = {Sleep as a Random Walk: A Super-Statistical Analysis of {{EEG}} Data across Sleep Stages},
  shorttitle = {Sleep as a Random Walk},
  author = {Metzner, Claus and Schilling, Achim and Traxdorf, Maximilian and Schulze, Holger and Krauss, Patrick},
  year = {2021},
  journal = {Communications Biology},
  volume = {4},
  number = {1},
  pages = {1--11},
  publisher = {{Nature Publishing Group}},
  issn = {2399-3642},
  doi = {10.1038/s42003-021-02912-6},
  copyright = {2021 The Author(s)},
  langid = {english}
}

@article{mittner2016,
  title = {A {{Neural Model}} of {{Mind Wandering}}},
  author = {Mittner, Matthias and Hawkins, Guy E. and Boekel, Wouter and Forstmann, Birte U.},
  year = {2016},
  journal = {Trends in Cognitive Sciences},
  volume = {20},
  number = {8},
  pages = {570--578},
  issn = {1364-6613},
  doi = {10.1016/j.tics.2016.06.004},
  langid = {english}
}

@article{oberauer2018,
  title = {Benchmarks for Models of Short-Term and Working Memory},
  author = {Oberauer, Klaus and Lewandowsky, Stephan and Awh, Edward and Brown, Gordon D. A. and Conway, Andrew and Cowan, Nelson and Donkin, Christopher and Farrell, Simon and Hitch, Graham J. and Hurlstone, Mark J. and Ma, Wei Ji and Morey, Candice C. and Nee, Derek Evan and Schweppe, Judith and Vergauwe, Evie and Ward, Geoff},
  year = {2018},
  journal = {Psychological Bulletin},
  volume = {144},
  number = {9},
  pages = {885--958},
  publisher = {{American Psychological Association}},
  address = {{US}},
  issn = {1939-1455},
  doi = {10.1037/bul0000153}
}

@article{van2021bayesian,
  title={Bayesian statistics and modelling},
  author={van de Schoot, Rens and Depaoli, Sarah and King, Ruth and Kramer, Bianca and M{\"a}rtens, Kaspar and Tadesse, Mahlet G and Vannucci, Marina and Gelman, Andrew and Veen, Duco and Willemsen, Joukje and others},
  journal={Nature Reviews Methods Primers},
  volume={1},
  number={1},
  pages={1--26},
  year={2021},
  publisher={Nature Publishing Group},
  doi = {10.1038/s43586-020-00001-2}
}

@inproceedings{abadi2016tensorflow,
  title={$\{$TensorFlow$\}$: a system for $\{$Large-Scale$\}$ machine learning},
  author={Abadi, Mart{\'\i}n and Barham, Paul and Chen, Jianmin and Chen, Zhifeng and Davis, Andy and Dean, Jeffrey and Devin, Matthieu and Ghemawat, Sanjay and Irving, Geoffrey and Isard, Michael and others},
  booktitle={12th USENIX symposium on operating systems design and implementation (OSDI 16)},
  pages={265--283},
  year={2016}
}

@article{radev2023bayesflow,
  title={BayesFlow: Amortized Bayesian Workflows With Neural Networks},
  author={Radev, Stefan T and Schmitt, Marvin and Schumacher, Lukas and Elsem{\"u}ller, Lasse and Pratz, Valentin and Sch{\"a}lte, Yannik and K{\"o}the, Ullrich and B{\"u}rkner, Paul-Christian},
  journal={arXiv preprint arXiv:2306.16015},
  year={2023}
}

@article{rabassa2015,
  title = {Superstatistical Analysis of Sea-Level Fluctuations},
  author = {Rabassa, P. and Beck, C.},
  year = {2015},
  doi = {10.1016/j.physa.2014.08.068}
}

@article{radev2020,
  author={S. T. {Radev} and U. K. {Mertens} and A. {Voss} and L. {Ardizzone} and U. {Köthe}},
  journal={IEEE Transactions on Neural Networks and Learning Systems}, 
  title={{BayesFlow}: Learning Complex Stochastic Models With Invertible Neural Networks}, 
  year={2020},
  volume={},
  number={},
  pages={},
  doi={10.1109/TNNLS.2020.3042395}
}

@article{ratcliff1978,
  title = {A Theory of Memory Retrieval},
  author = {Ratcliff, Roger},
  year = {1978},
  journal = {Psychological Review},
  volume = {85},
  number = {2},
  pages = {59--108},
  publisher = {{American Psychological Association}},
  address = {{US}},
  issn = {1939-1471(Electronic),0033-295X(Print)},
  doi = {10.1037/0033-295X.85.2.59}
}

@article{gers2000learning,
  title={Learning to forget: Continual prediction with LSTM},
  author={Gers, Felix A and Schmidhuber, J{\"u}rgen and Cummins, Fred},
  journal={Neural computation},
  volume={12},
  number={10},
  pages={2451--2471},
  year={2000},
  publisher={MIT Press},
  doi={10.1049/cp:19991218}
}

@article{bloem2020probabilistic,
  title={Probabilistic Symmetries and Invariant Neural Networks.},
  author={Bloem-Reddy, Benjamin and Teh, Yee Whye},
  journal={J. Mach. Learn. Res.},
  volume={21},
  pages={90--1},
  year={2020}
}

@inproceedings{greenberg2019automatic,
  title={Automatic posterior transformation for likelihood-free inference},
  author={Greenberg, David and Nonnenmacher, Marcel and Macke, Jakob},
  booktitle={International Conference on Machine Learning},
  pages={2404--2414},
  year={2019},
  organization={PMLR},
  doi = {10.48550/arXiv.1905.07488}
}

@article{ratcliff2016,
  title = {Diffusion {{Decision Model}}: {{Current Issues}} and {{History}}},
  shorttitle = {Diffusion {{Decision Model}}},
  author = {Ratcliff, Roger and Smith, Philip L. and Brown, Scott D. and McKoon, Gail},
  year = {2016},
  journal = {Trends in Cognitive Sciences},
  volume = {20},
  number = {4},
  pages = {260--281},
  issn = {1364-6613},
  doi = {10.1016/j.tics.2016.01.007},
  langid = {english}
}

@article{riley2012,
  title = {Dynamics of Cognition},
  author = {Riley, Michael A. and Holden, John G.},
  year = {2012},
  journal = {WIREs Cognitive Science},
  volume = {3},
  number = {6},
  pages = {593--606},
  issn = {1939-5086},
  doi = {10.1002/wcs.1200},
  langid = {english}
}

@article{urai2019a,
  title = {Choice History Biases Subsequent Evidence Accumulation},
  author = {Urai, Anne E and {de Gee}, Jan Willem and Tsetsos, Konstantinos and Donner, Tobias H},
  editor = {Verstynen, Timothy and {Shinn-Cunningham}, Barbara G and Verstynen, Timothy},
  year = {2019},
  journal = {eLife},
  volume = {8},
  pages = {e46331},
  publisher = {{eLife Sciences Publications, Ltd}},
  issn = {2050-084X},
  doi = {10.7554/eLife.46331}
}

@article{vanderstraeten2009,
  title = {Superstatistical Fluctuations in Time Series: {{Applications}} to Share-Price Dynamics and Turbulence},
  shorttitle = {Superstatistical Fluctuations in Time Series},
  author = {{Van der Straeten}, Erik and Beck, Christian},
  year = {2009},
  journal = {Physical Review E},
  volume = {80},
  number = {3},
  pages = {036108},
  publisher = {{American Physical Society}},
  doi = {10.1103/PhysRevE.80.036108}
}

@article{vanorden2003,
  title = {Self-Organization of Cognitive Performance},
  author = {Van Orden, Guy C. and Holden, John G. and Turvey, Michael T.},
  year = {2003},
  journal = {Journal of Experimental Psychology: General},
  volume = {132},
  number = {3},
  pages = {331--350},
  publisher = {{American Psychological Association}},
  address = {{US}},
  issn = {1939-2222},
  doi = {10.1037/0096-3445.132.3.331}
}

@article{vanrooij2013,
  title = {Modeling the {{Dynamics}} of {{Risky Choice}}},
  author = {{van Rooij}, Marieke M. J. W. and Favela, Luis H. and Malone, MaryLauren and Richardson, Michael J.},
  year = {2013},
  journal = {Ecological Psychology},
  volume = {25},
  number = {3},
  pages = {293--303},
  publisher = {{Routledge}},
  issn = {1040-7413},
  doi = {10.1080/10407413.2013.810502}
}

@article{vonkrause2022,
  title = {Mental Speed Is High until Age 60 as Revealed by Analysis of over a Million Participants},
  author = {{von Krause}, Mischa and Radev, Stefan T. and Voss, Andreas},
  year = {2022},
  journal = {Nature Human Behaviour},
  pages = {1--9},
  publisher = {{Nature Publishing Group}},
  issn = {2397-3374},
  doi = {10.1038/s41562-021-01282-7},
  copyright = {2022 The Author(s), under exclusive licence to Springer Nature Limited},
  langid = {english}
}

@article{voss2013,
  title = {Diffusion {{Models}} in {{Experimental Psychology}}},
  author = {Voss, Andreas and Nagler, Markus and Lerche, Veronika},
  year = {2013},
  journal = {Experimental Psychology},
  volume = {60},
  number = {6},
  pages = {385--402},
  publisher = {{Hogrefe Publishing}},
  issn = {1618-3169},
  doi = {10.1027/1618-3169/a000218}
}

@article{wagenmakers2004,
  title = {Estimation and Interpretation of 1/F{$\alpha$} Noise in Human Cognition},
  author = {Wagenmakers, Eric-Jan and Farrell, Simon and Ratcliff, Roger},
  year = {2004},
  journal = {Psychonomic Bulletin \& Review},
  volume = {11},
  number = {4},
  pages = {579--615},
  issn = {1531-5320},
  doi = {10.3758/BF03196615},
  langid = {english}
}

@article{williams2020,
  title = {Superstatistical Approach to Air Pollution Statistics},
  author = {Williams, Griffin and Sch{\"a}fer, Benjamin and Beck, Christian},
  year = {2020},
  journal = {Physical Review Research},
  volume = {2},
  number = {1},
  pages = {013019},
  publisher = {{American Physical Society}},
  doi = {10.1103/PhysRevResearch.2.013019}
}

@article{yalcin2016,
  title = {Extreme Event Statistics of Daily Rainfall: Dynamical Systems Approach},
  shorttitle = {Extreme Event Statistics of Daily Rainfall},
  author = {Yalcin, G. Cigdem and Rabassa, Pau and Beck, Christian},
  year = {2016},
  journal = {Journal of Physics A: Mathematical and Theoretical},
  volume = {49},
  number = {15},
  pages = {154001},
  publisher = {{IOP Publishing}},
  issn = {1751-8121},
  doi = {10.1088/1751-8113/49/15/154001},
  langid = {english}
}

@article{yoo2022,
  title = {How {{Working Memory}} and {{Reinforcement Learning Are Intertwined}}: {{A Cognitive}}, {{Neural}}, and {{Computational Perspective}}},
  shorttitle = {How {{Working Memory}} and {{Reinforcement Learning Are Intertwined}}},
  author = {Yoo, Aspen H. and Collins, Anne G. E.},
  year = {2022},
  journal = {Journal of Cognitive Neuroscience},
  volume = {34},
  number = {4},
  pages = {551--568},
  issn = {0898-929X},
  doi = {10.1162/jocn_a_01808}
}

@article{ratcliff1998,
  title = {Modeling {{Response Times}} for {{Two-Choice Decisions}}},
  author = {Ratcliff, Roger and Rouder, Jeffrey N.},
  year = {1998},
  journal = {Psychological Science},
  volume = {9},
  number = {5},
  pages = {347--356},
  publisher = {{SAGE Publications Inc}},
  issn = {0956-7976},
  doi = {10.1111/1467-9280.00067},
  langid = {english}
}

@article{ratcliff2002a,
  title = {Estimating Parameters of the Diffusion Model: {{Approaches}} to Dealing with Contaminant Reaction Times and Parameter Variability},
  shorttitle = {Estimating Parameters of the Diffusion Model},
  author = {Ratcliff, Roger and Tuerlinckx, Francis},
  year = {2002},
  journal = {Psychonomic Bulletin \& Review},
  volume = {9},
  number = {3},
  pages = {438--481},
  issn = {1531-5320},
  doi = {10.3758/BF03196302},
  langid = {english}
}
\newpage
\appendix
\renewcommand{\thefigure}{A.\arabic{figure}}
\onecolumn
\begin{center}
    \bfseries\Large Appendix
\end{center}

\section*{Implementation Details}

All experiments, neural networks, and simulation models are implemented using the BayesFlow library \url{https://github.com/stefanradev93/BayesFlow} built on top of
TensorFlow \cite{abadi2016tensorflow}.
Code and further instructions for reproducing the results from all experiments and applications in the current manuscript is available at \url{https://github.com/bayesflow-org/Neural-Superstatistics}.  

\section*{Stan Benchmark Study}

\subsection*{Data Simulation}
To simulate the $100$ data sets each consisting of $T = 100$ trials, we used the standard DDM implementation (cf.~equation \eqref{eq:DDM} in the main text). 
The diffusion constant was fixed to $1$ and the starting point parameter to $0.5$ (i.e., symmetric starting point between the two decision boundaries). 
For data simulation, we randomly sampled parameter sets from the following prior distributions:

\begin{align*}\label{app:priors}
    v &\sim \Gamma(5.0, \frac{1}{1.3})\\
    a &\sim \Gamma(4.0, \frac{1}{3})\\
    \tau &\sim \Gamma(1.5, \frac{1}{5})\\
\end{align*}

where $\Gamma(a, b)$ denotes a Gamma distributions with shape $a$ as the first and scale $b$ as the second argument.

\subsection*{Non-Stationary DDM fitting}
We fitted a separate non-stationary DDM with a Gaussian random walk transition model to all $100$ simulated data sets. 
The same implementation and likelihood was used for Stan and our neural estimation method.  
However, all 3 parameters were allowed to vary according to a Gaussian random walk (cf. equation \eqref{GaussianTransition} in the main text). The starting values were sampled from the same prior distributions as in simulation. The hyperparameters of the random walk transition model were sampled from the following distribution:

\begin{align*}
    s_v, s_a, s_\tau &\sim \mathcal{B}(1, 25)\\
\end{align*}

where $\mathcal{B}(\alpha, \beta)$ denotes a Beta distribution with $\alpha$ and $\beta$ parameters.
In order to avoid implausible parameter values, the time-varying parameters $v_t, a_t, \tau_t$ were clipped to lower bounds $[0, 0, 0]$ and upper bounds $[6, 4, 2]$, respectively.

We trained the neural approximator via online learning (i.e., simulations on the fly) for $50$ epochs with $1000$ iterations each and a batch size of $8$. 
We use an Adam optimizer with an initial learning rate of $5 \times 10^{-4}$ and a cosine learning rate decay schedule.
After training the network, we draw $4000$ posterior samples (the same as with Stan) for each of the $100$ data sets.

\newpage

\section*{Simulation Study}

In what follows, we describe the settings for the four different simulation scenarios, namely, the static DDM, the DDM with stationary variability, the DDM with non-stationary variability, and the static DDM with random uniform jumps at pre-defined time steps (i.e., regime switching DDM). 
For each scenario, we simulated $200$ data sets, each consisting of $T = 400$ time steps.

We trained the neural approximator via online learning (i.e., simulations on the fly) for $75$ epochs with $1000$ iterations each and a batch size of $8$.
We use an Adam optimizer with an initial learning rate of $5 \times 10^{-4}$ and a cosine learning rate decay schedule.
After training the network, we draw $4000$ posterior samples (the same as with Stan) for each of the $100$ data sets.

\subsection*{Static DDM}
To simulate the $200$ data sets for the static DDM scenario, we used the same prior and likelihood as in the \textbf{Stan Benchmark Study}.

\subsection*{Stationary Variability DDM}
For the stationary variability DDM, we used the same DDM implementation as in the static DDM scenario except that we used the following variability statements:

\begin{align*}
    v_t &\sim \mathcal{N}(v, v_s)\\
    a_t &\sim \mathcal{N}(a, a_s)\\
    \tau_s &\sim \mathcal{U}(\tau - \frac{\tau_s}{2}, \tau + \frac{\tau_s}{2})\\
\end{align*}

where $\mathcal{N}(\mu, \sigma)$ denotes a Normal distribution with location $\mu$ and standard deviation $\sigma$ and $\mathcal{U}(\text{lower, } \text{upper})$ denotes an Uniform distribution with a lower and a upper bound.

The newly introduced variability parameters ($v_s, a_s, \tau_s$) were sampled from the following prior distributions:

\begin{align*}
    v_s, a_s, \tau_s &\sim \mathcal{TN}_{[0, \text{inf}]}(0, 0.1)\\
\end{align*}

where $\mathcal{TN}_{[a, b]}(\mu, \sigma)$ denotes the truncated normal distribution with location $\mu$ and standard deviation $\sigma$ truncated within the interval $[a, b]$. 

To avoid implausible values the per trial parameters $v_t, a_t, \tau_t$ were bounded with lower bounds [0, 0, 0] and upper bounds [6, 4, 2] respectively.

\subsection*{Non-Stationary DDM}
We used the same non-stationary DDM implementation as described in \textbf{Stan Benchmark Study}.

\subsection*{Regime Switching DDM}
The regime switching DDM is basically the same implementation as the static DDM, but the parameter jumped uniformly at 3 specific time steps ($T = 100; T = 200; T = 300$) and stayed again constant after the jump:

\begin{align}
    \theta_t = 
    \begin{cases}
        \theta_{t-1}, & \text{if } t \notin {100, 200, 300}\\
        \mathcal{U}(\text{lower, } \text{upper}), & \text{if } t \in \{100, 200, 300\}
    \end{cases}
\end{align}

where the lower and upper bounds of the Uniform distributions are [0, 0, 0] and [6, 4, 2], respectively. The starting values of the parameters were once again sampled form the same prior distributions as in the static DDM.

\subsection*{Amortized inference}
We fitted the same non-stationary DDM with a Gaussian transition model as described above to all four scenarios. To train the networks we used $75$ epochs with $1000$ iterations each and a batch size of $16$. After training the network we fitted the model to each simulation of each scenario separately and obtained $2000$ posterior samples.

\newpage

\subsection*{True vs. Estimated Parameters}

\begin{figure}[H]
\centering
\includegraphics[width=0.8\textwidth]{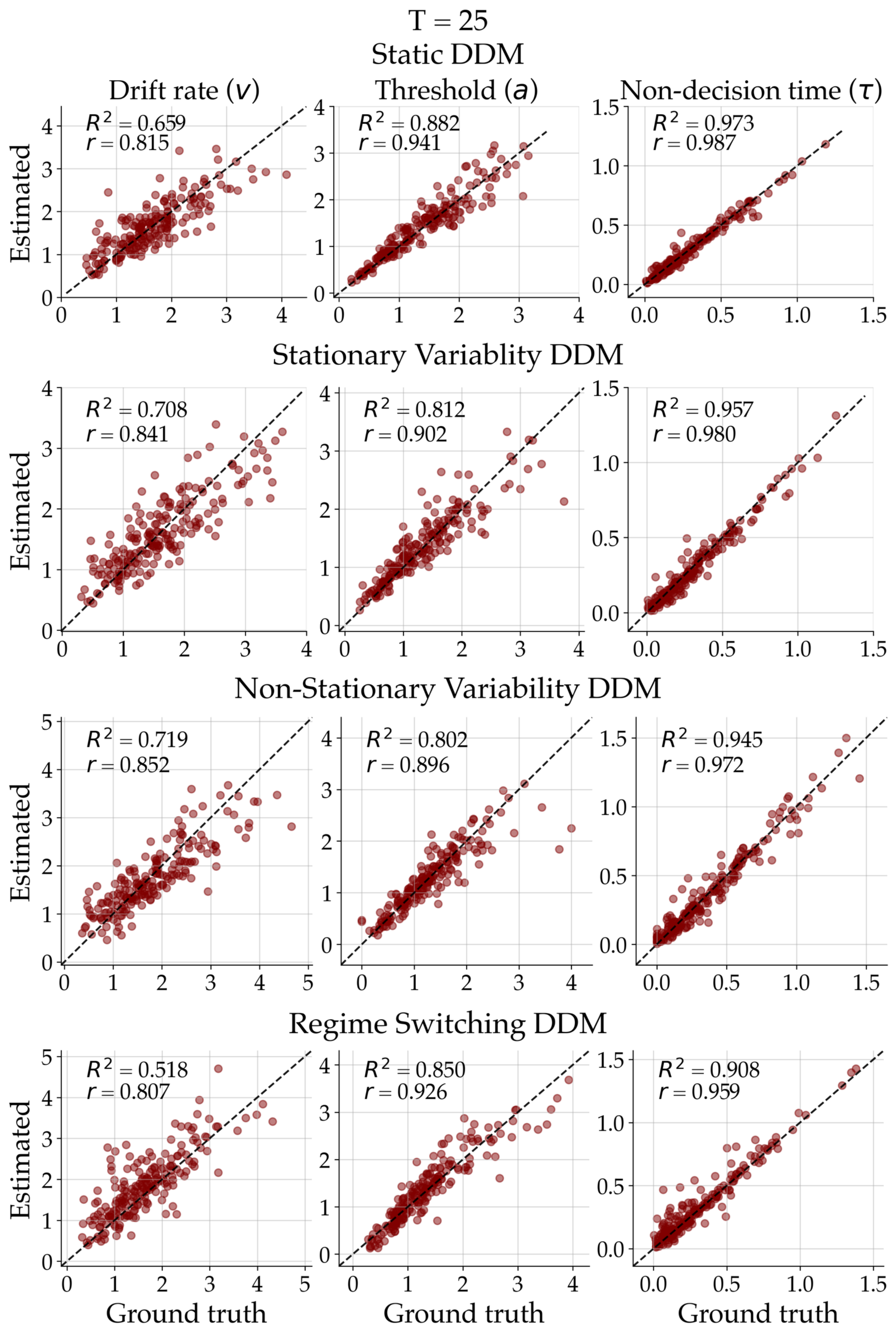}
\caption{True data generating parameters plotted against posterior means for all 3 parameters and simulation scenarios separately at time point $T = 25$.}
\label{true_estimated_25}
\end{figure}
\vfill
\hspace{0pt}
\newpage

\begin{figure}[H]
\centering
\includegraphics[width=0.8\textwidth]{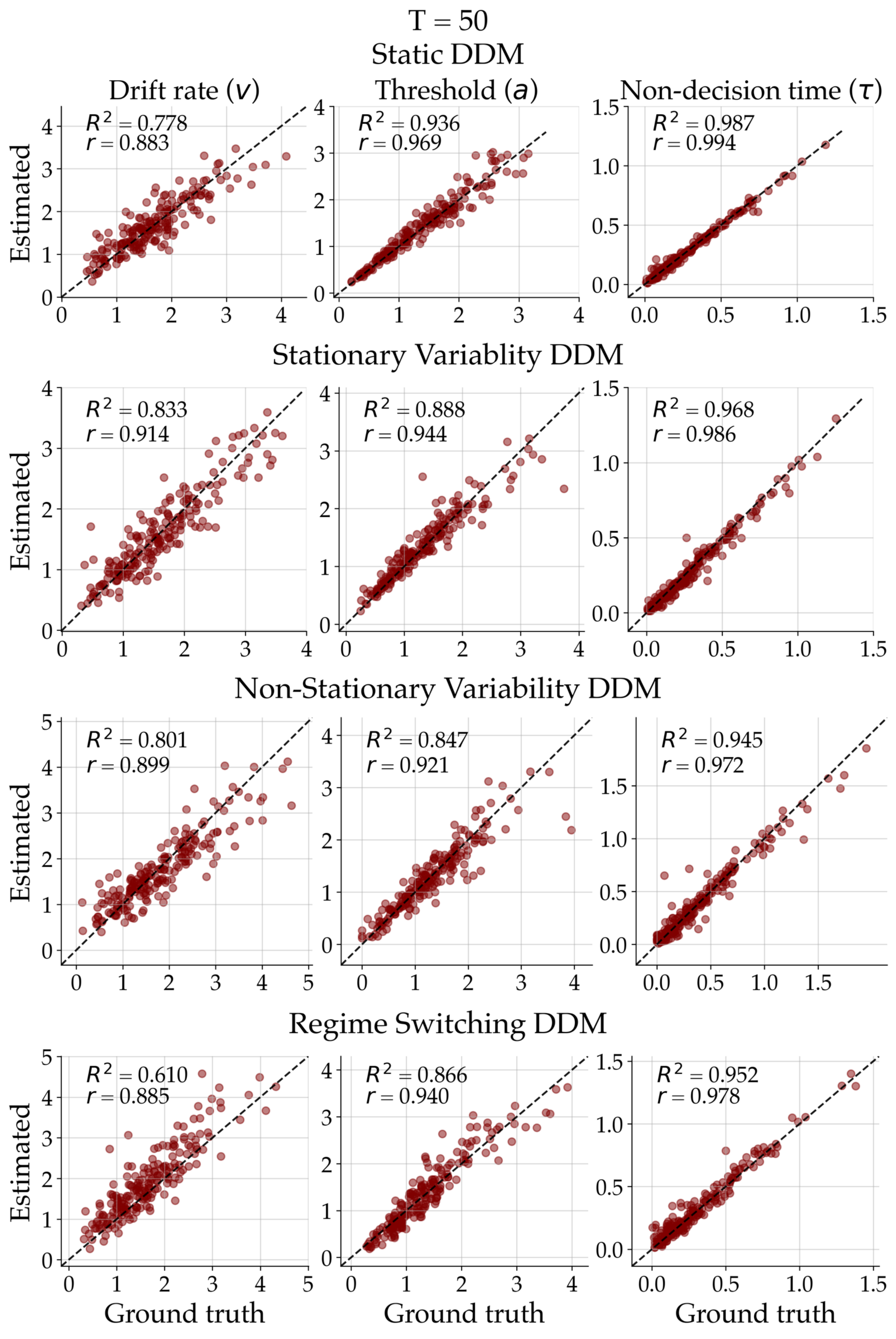}
\caption{True data generating parameters plotted against posterior means for all 3 parameters and simulation scenarios separately at time point $T = 50$.}
\label{true_estimated_25}
\end{figure}
\vfill
\hspace{0pt}
\newpage

\begin{figure}[H]
\centering
\includegraphics[width=0.8\textwidth]{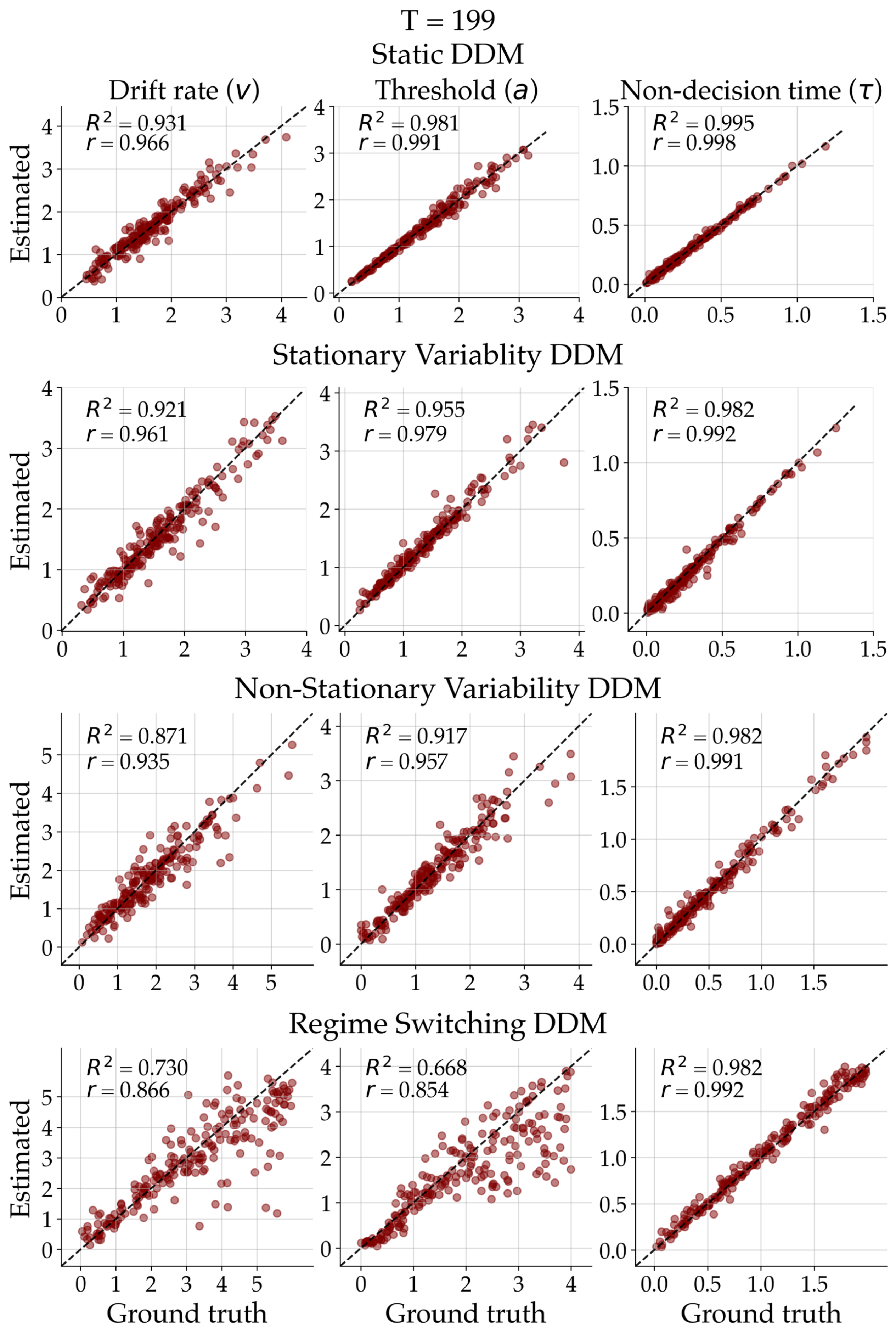}
\caption{True data generating parameters plotted against posterior means for all 3 parameters and simulation scenarios separately at time point $T = 199$.}
\label{true_estimated_25}
\end{figure}
\vfill
\hspace{0pt}
\newpage

\begin{figure}[H]
\centering
\includegraphics[width=0.8\textwidth]{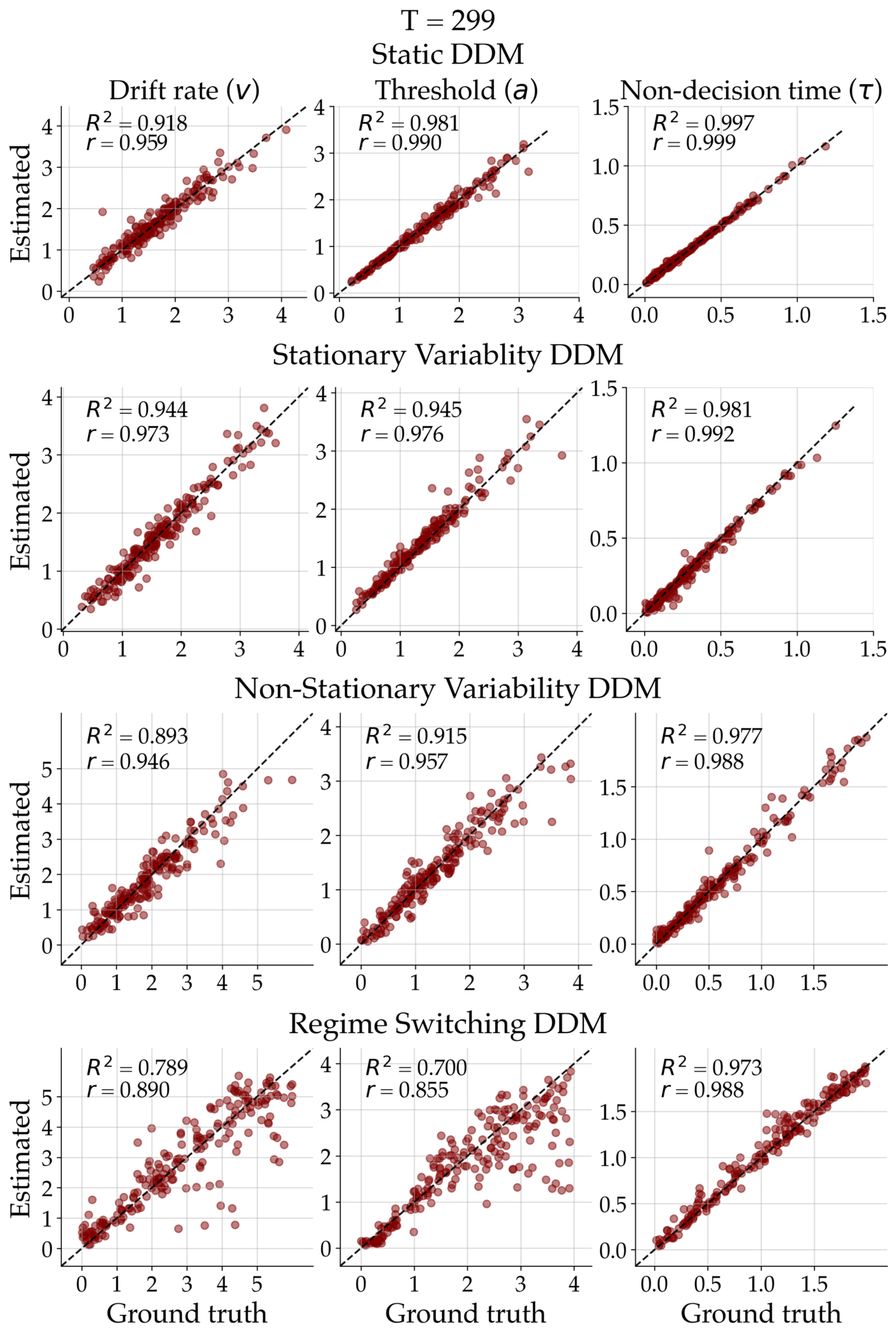}
\caption{True data generating parameters plotted against posterior means for all 3 parameters and simulation scenarios separately at time point $T = 299$.}
\label{true_estimated_25}
\end{figure}
\vfill
\hspace{0pt}
\newpage

\begin{figure}[H]
\centering
\includegraphics[width=0.8\textwidth]{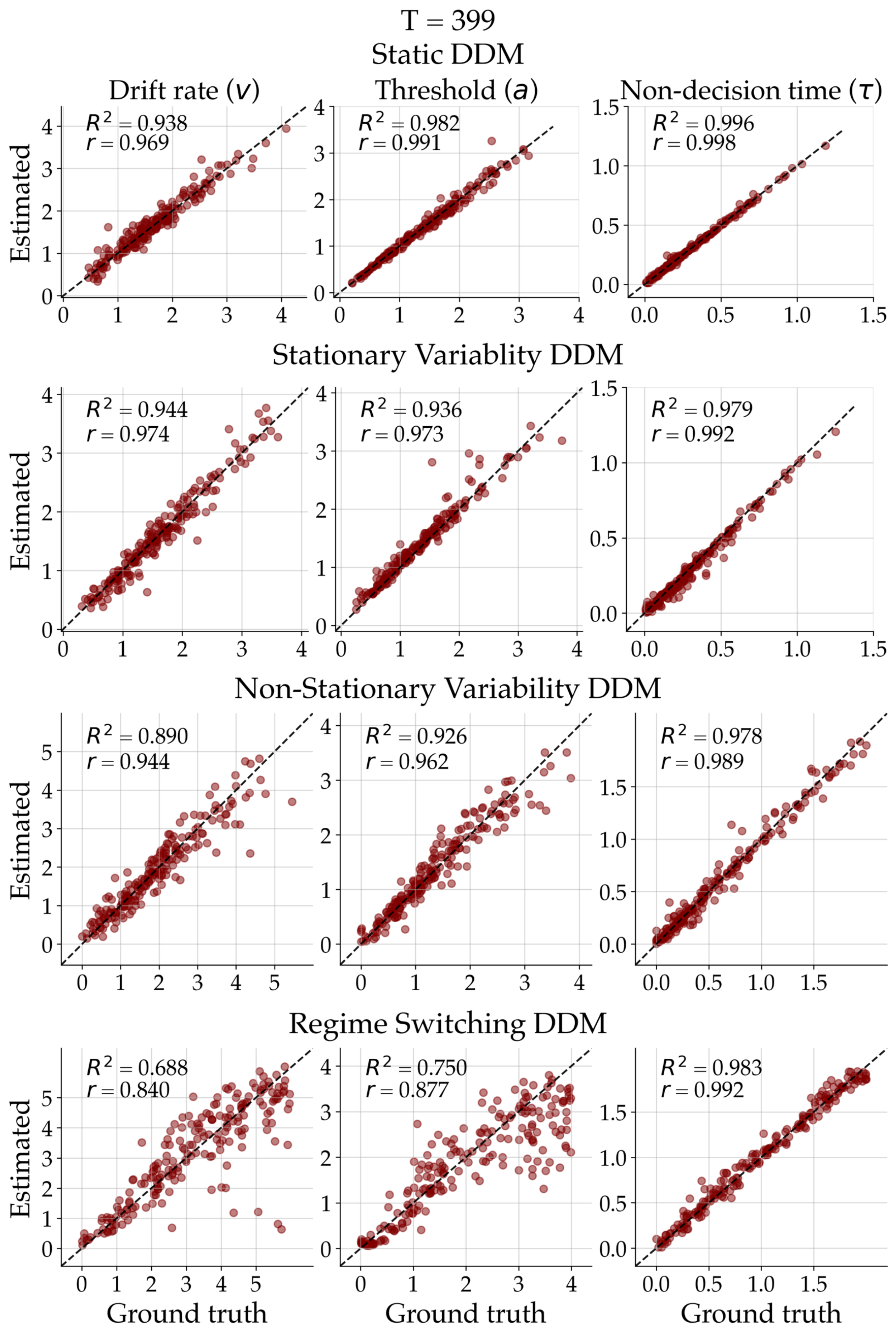}
\caption{True data generating parameters plotted against posterior means for all 3 parameters and simulation scenarios separately at time point $T = 399$.}
\label{true_estimated_25}
\end{figure}
\vfill
\hspace{0pt}
\newpage

\subsection*{Mean Absolute Error}
\begin{figure}[H]
    \centering
    \includegraphics[width=0.8\textwidth]{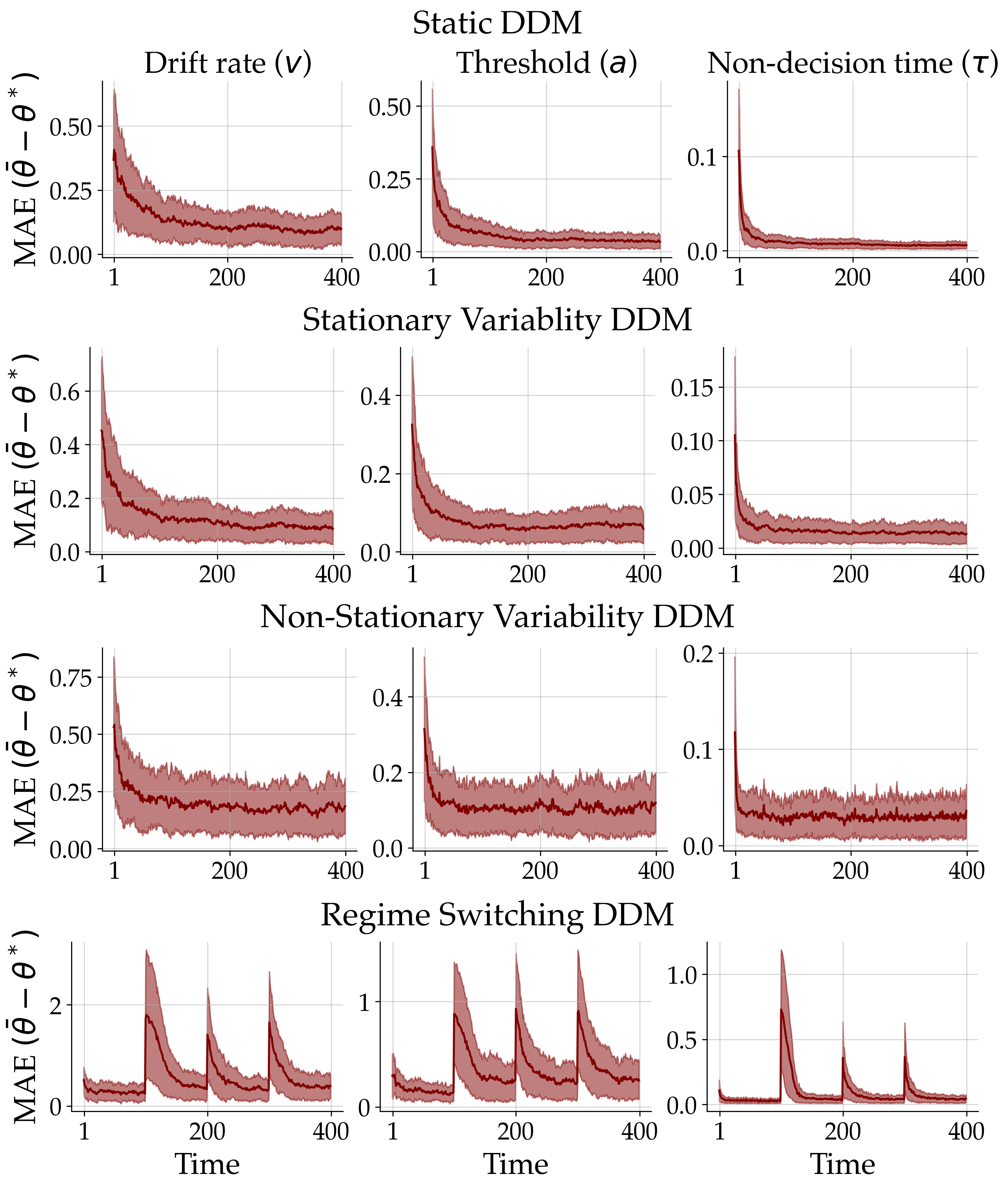}
    \caption{Median absolute error (MAE) between the data generating parameters and the estimated posterior means aggregated across the 200 simulations over time for each DDM parameter (columns) and simulation scenario (rows) separately. The red shaded areas depict the median absolute deviation of the absolute errors.}
    \label{fig:sim_study_mae}
\end{figure}

As an additional analysis of the overall parameter recovery performance of the non-stationary DDM, we computed the median absolute error (MAE) between the true data generating parameter and the posterior mean for all DDM parameters and simulation scenarios separately.
In the top row of \autoref{fig:sim_study_mae} we can see that the posterior estimates of the non-stationary DDM quickly approach the true data-generating parameter when the true parameter was constant over time.
That said, there remains some error between the true and estimated parameter even after $400$ time steps.
This error is the largest in the drift rate parameter ($\approx 0.15$).
We see similar recovery performance in the scenario, where the parameters were allowed to randomly fluctuate around a constant value (second row in \autoref{fig:sim_study_mae}).
However, we observe a larger variability in the MAE.
The third row depicts the MAE when data was simulated with the same model as we fitted to the data (i.e., the well-specified case).
Once again, the MAE quickly decreases in the beginning and then flattens out.
However, in this scenario, the MAE remains on a larger level than in the previous two scenarios.
Also, the variability of the MAE between data fits is larger.
This is not surprising because the estimation of non-stationary model parameters is more difficult than static or stationary variable parameters.
The last row in \autoref{fig:sim_study_mae} shows how the parameter estimates of the non-stationary DDM react to sudden jumps in otherwise constant parameters.
We observe that the MAE significantly increases when a jump occurred and then decreases again.

\newpage

\newpage

\section*{Human Data Application: Random-Dot Motion}
We fitted a non-stationary DDM with a Gaussian random walk transition model to each individual in the data set separately. The model implementation was the same as in the \textbf{Stan Benchmark} and the \textbf{Simulation Study}. For the training of our neural estimation method we used $75$ epoch each consisting of $1000$ iteration with a batch size of $16$. After training we obtained $2000$ posterior samples.

\section*{Human Data Application: Lexical Decision}

\subsection*{Gaussian Process DDM}
For the Gaussian Process transition model, we first create a $T$ x $T$ squared distance matrix with $T = 3200$. Based on this distance matrix we calculate the radial basis function kernel (cf.~equation \eqref{GaussianKernel} in the main text) given the two parameters, amplitude $\sigma$ and length-scale $l$, resulting in the covariance $k$ for the multivariate normal distribution of the Gaussian Process:

\begin{equation*}
    \theta_{1:T} \sim \mathcal{MVN}(\mu_{\theta}, k)
\end{equation*}

where $\mu_{\theta}$ is the mean parameter value. For these means we used the same priors we otherwise used for the starting values of the DDM parameters ($v_{0, i}$, $a_0$, $\tau_0$). In the following we present a list of the priors used by the Gaussian Process DDM simulator to generate data for the simulation study and for training the neural networks. $\Gamma(a, b)$ refers to a Gamma distribution parameterized with shape $a$ and scale $b$. 
The same prior distribution was used for all $i=4$ drift rates $v_{0, i}$. $\mathcal{U}(a, b)$ stands for a continuous uniform distribution with a lower limit $a$ and an upper limit $b$. 
$l_{j}$ denotes the length-scale parameters of the GP transition model. 
The same prior distribution was used for all $j=1,\dots,6$ length-scale parameters governing the transitions of the DDM parameters.
The amplitude parameter $\sigma$ of the Gaussian kernel is usually highly correlates with the length-scale $l$. Thus, we fixed $\sigma$ to sensible values for all low-level parameter transitions.

\begin{align*}
\label{app:priors}
v_{0, i} &\sim \Gamma(2.5, \frac{1}{1.5})\\
a_0 &\sim \Gamma(4.0, \frac{1}{3})\\
\tau_0 &\sim \Gamma(1.5, \frac{1}{5})\\
l_{j} &\sim \mathcal{U}(0.1, 10)\\
\sigma_{v_{1:4}} &= 0.15\\
\sigma_{a} &= 0.1\\
\sigma_{\tau} &= 0.05\\
\end{align*}
\newpage

\subsection*{Simulation-Based Calibration}
We validate the computational faithfulness of our Bayesian inference algorithm using simulation-based calibration, a robust method for ensuring unbiased posterior distributions.
The underlying principle is that an ensemble of posterior distributions should be indistinguishable from the prior distribution.
To accomplish this, we carry out 2000 simulations with the dynamic DDM, each generating a separate data set.
For each simulated data set, we fit the model and obtain 250 posterior samples.
These posterior distributions collectively form an ensemble.

\begin{figure}[H]
\centering
\includegraphics[width=\textwidth]{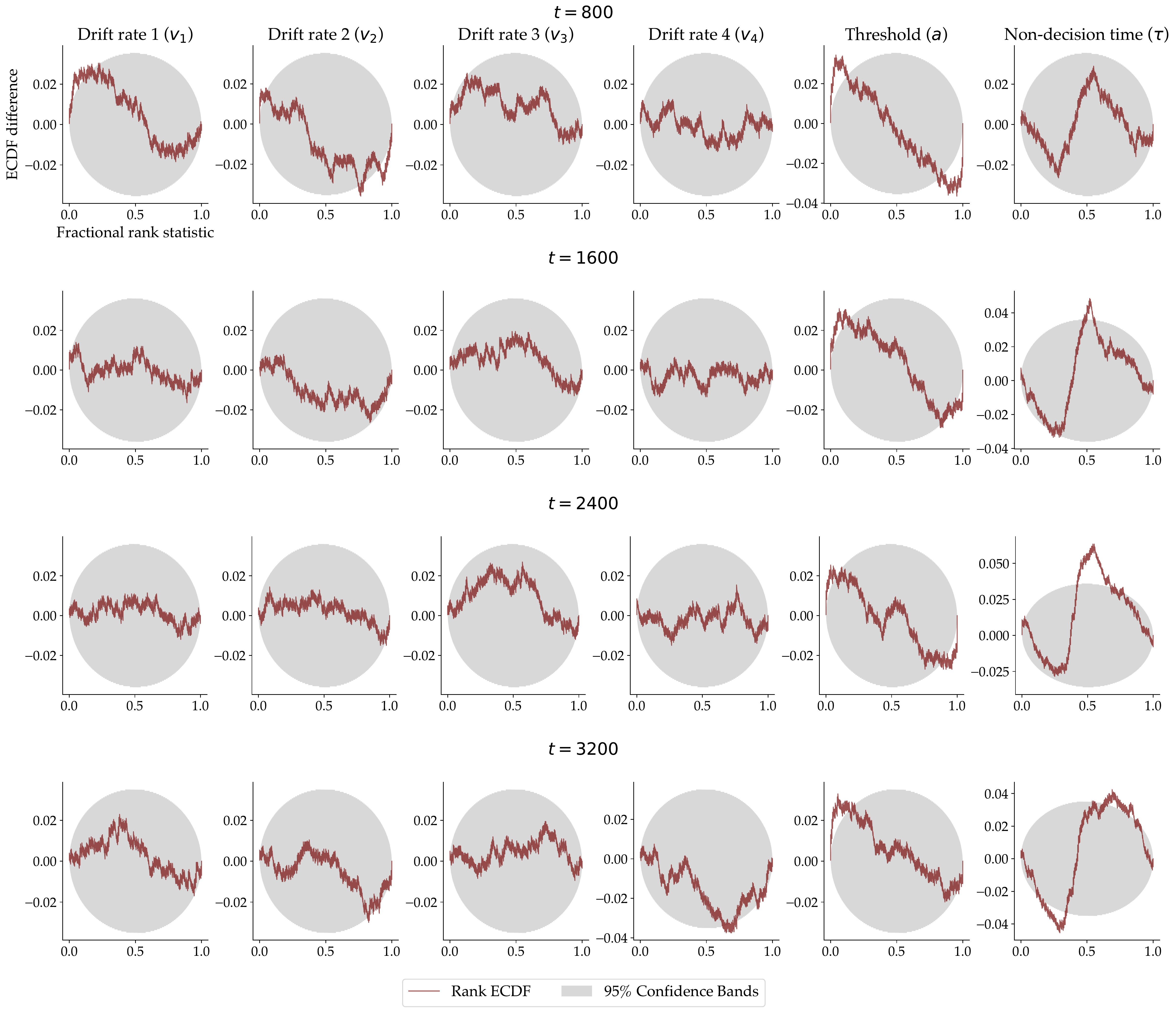}
\caption{\textbf{ECDF difference plot} 95\% simultaneous confidence bands (gray) for the empirical cumulative distribution function (ECDF; red) for all $6$ parameters at four selected time points (800, 1600, 2500, 3200) separately.}
\label{sbc_plot}
\end{figure}

When we calculate rank statistics for the ensemble relative to the prior distribution then these should be uniformly distributed.
To assess the uniformity at predefined time points, we utilize the empirical cumulative distribution function (ECDF) for each marginal rank distribution.
Comparing it with a uniform ECDF allows us to gauge how the data is distributed.
We further draw ECDF simultaneous bands using simulations from the uniform, providing an intuitive graphical test for uniformity.
For clarity, \autoref{sbc_plot} presents the ECDF difference, providing a more dynamic range for the visualization.
The red line (ECDF difference) should consistently fall within the gray shaded area (confidence band) across the entire range of fractional rank statistic values.
In the majority of cases, this criterion is met for most parameters at all selected time points.
Some slight deviations are observed for the threshold and non-decision parameters; however, these are typically small and not a major cause for concern.

\newpage

\newpage

\subsection*{Parameter Recovery Study}
A simulation study was performed to probe the dynamic DDM's capability of recovering data-generating parameter dynamics.
To this end, we simulated $1000$ data sets with the dynamic DDM and fit it to these data.
The following figures show posterior predictions of $3$ randomly selected simulated data sets and the comparison between the inferred and the true data-generating low-level parameter dynamics. The parameter recovery performance across all $1000$ data sets over all $3200$ time points for all $6$ model parameters can be inspected as a GIF in our GitHub repository (\url{https://github.com/bayesflow-org/Neural-Superstatistics})

\newpage

\begin{figure}
\centering
\includegraphics[width=\textwidth]{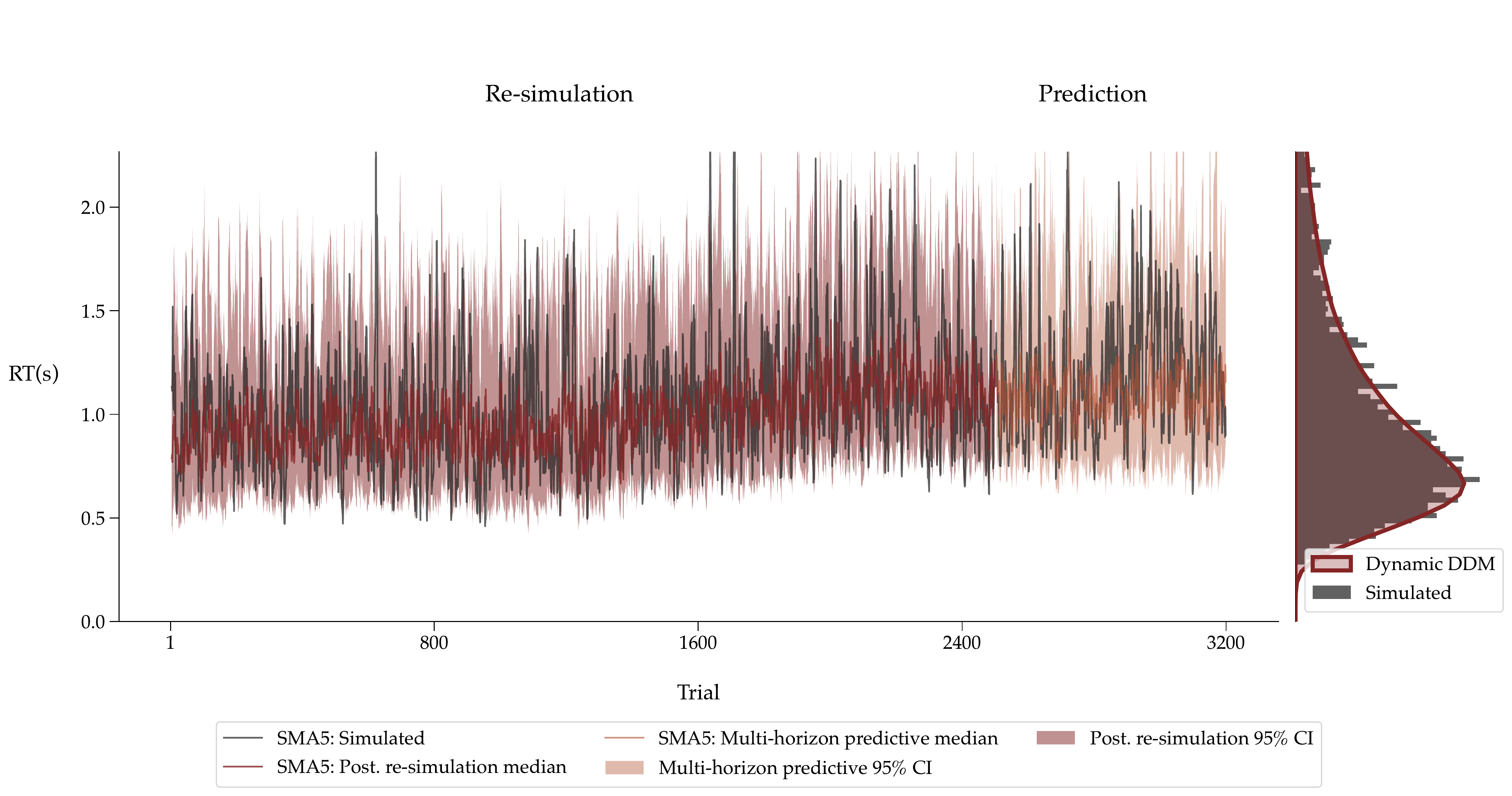}
\caption{
    \textbf{Left panel} The simulated RT time series is shown in black.
    From trial 1 to 2500, the median posterior re-simulation (aka \textit{retrodictive check}) using the dynamic DDM is shown in red.
    The models' multi-horizon prediction is depicted for the remaining trials in orange.
    The shaded areas for the posterior re-simulation and prediction correspond to the 95\% credibility interval.
    All the time series were smoothed via a simple moving average (SMA) with a period of 5.
    \textbf{Right panel} The raw simulated RT distribution is plotted as a histogram in black.
    The re-simulated RT distributions from the dynamic DDM are shown as kernel density estimates (KDEs) in red.
    }
\end{figure}
\FloatBarrier

\begin{figure}
\centering
\includegraphics[width=\textwidth]{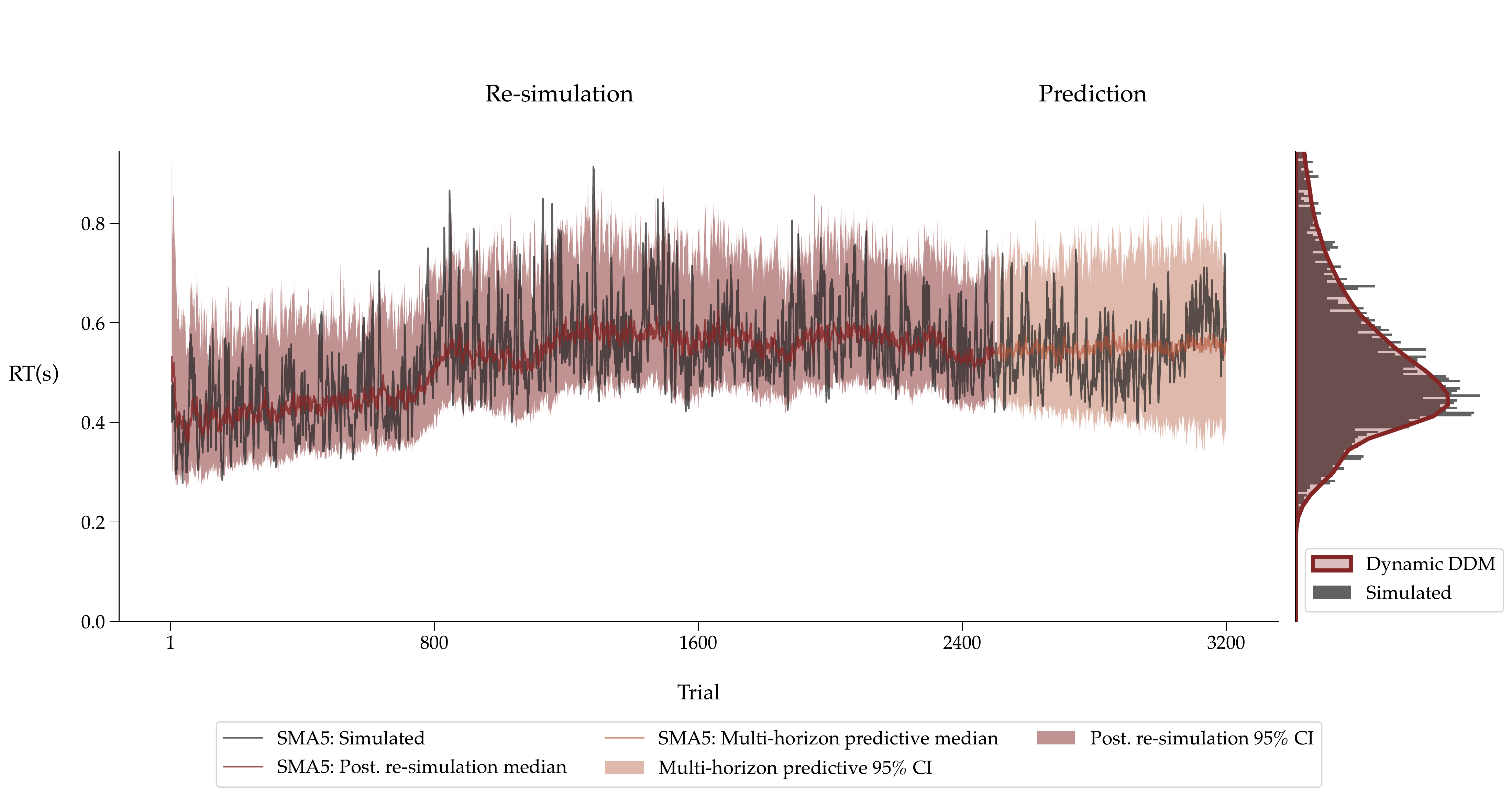}
\caption{
    \textbf{Left panel} The simulated RT time series is shown in black.
    From trial 1 to 2500, the median posterior re-simulation (aka \textit{retrodictive check}) using the dynamic DDM is shown in red.
    The models' multi-horizon prediction is depicted for the remaining trials in orange.
    The shaded areas for the posterior re-simulation and prediction correspond to the 95\% credibility interval.
    All the time series were smoothed via a simple moving average (SMA) with a period of 5.
    \textbf{Right panel} The raw simulated RT distribution is plotted as a histogram in black.
    The re-simulated RT distributions from the dynamic DDM are shown as kernel density estimates (KDEs) in red.
    }
\end{figure}
\FloatBarrier

\begin{figure}
\centering
\includegraphics[width=\textwidth]{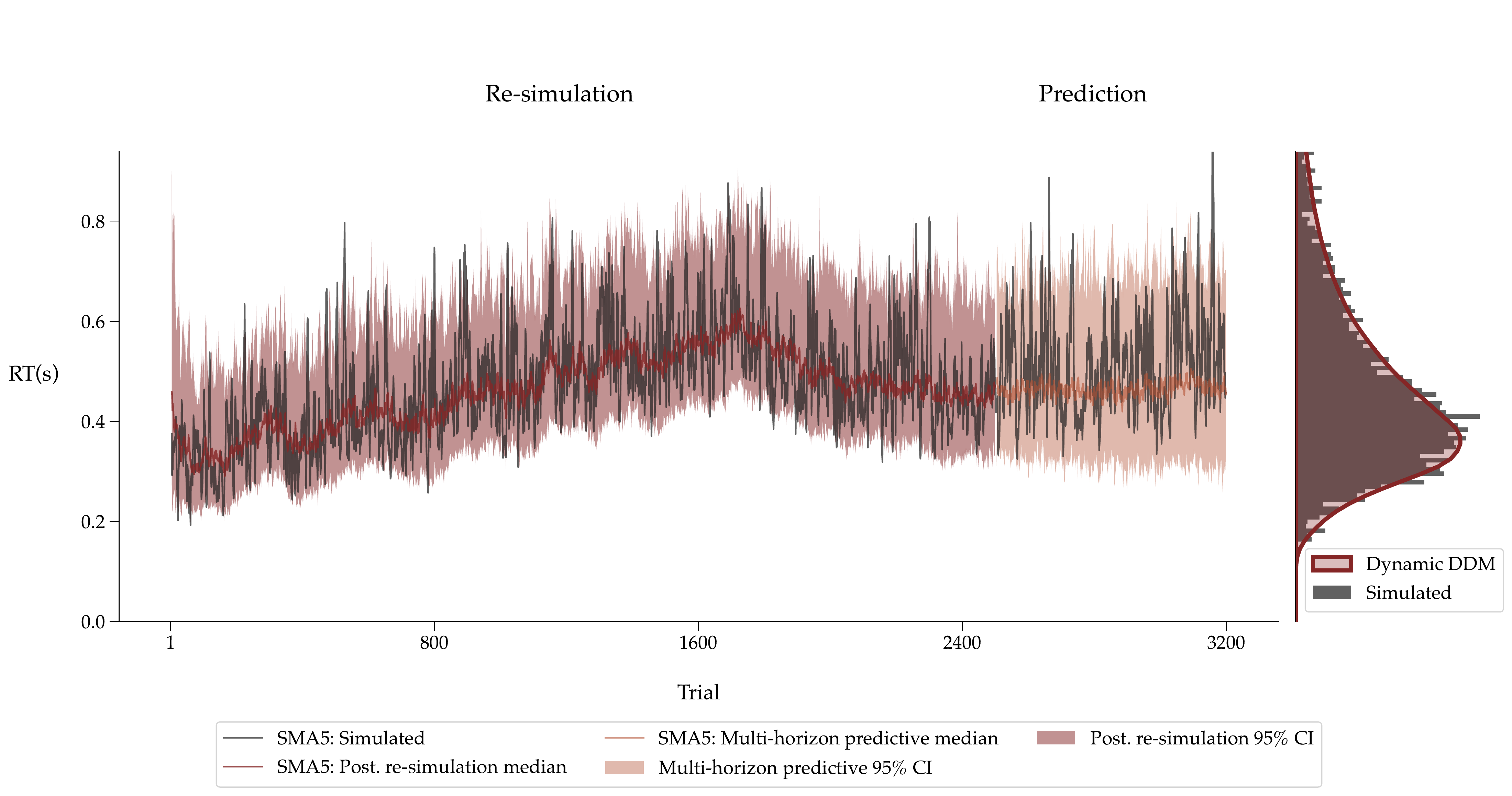}
\caption{
    \textbf{Left panel} The simulated RT time series is shown in black.
    From trial 1 to 2500, the median posterior re-simulation (aka \textit{retrodictive check}) using the dynamic DDM is shown in red.
    The models' multi-horizon prediction is depicted for the remaining trials in orange.
    The shaded areas for the posterior re-simulation and prediction correspond to the 95\% credibility interval.
    All the time series were smoothed via a simple moving average (SMA) with a period of 5.
    \textbf{Right panel} The raw simulated RT distribution is plotted as a histogram in black.
    The re-simulated RT distributions from the dynamic DDM are shown as kernel density estimates (KDEs) in red.
    }
\end{figure}
\FloatBarrier

\begin{figure}
\centering
\includegraphics[width=\textwidth]{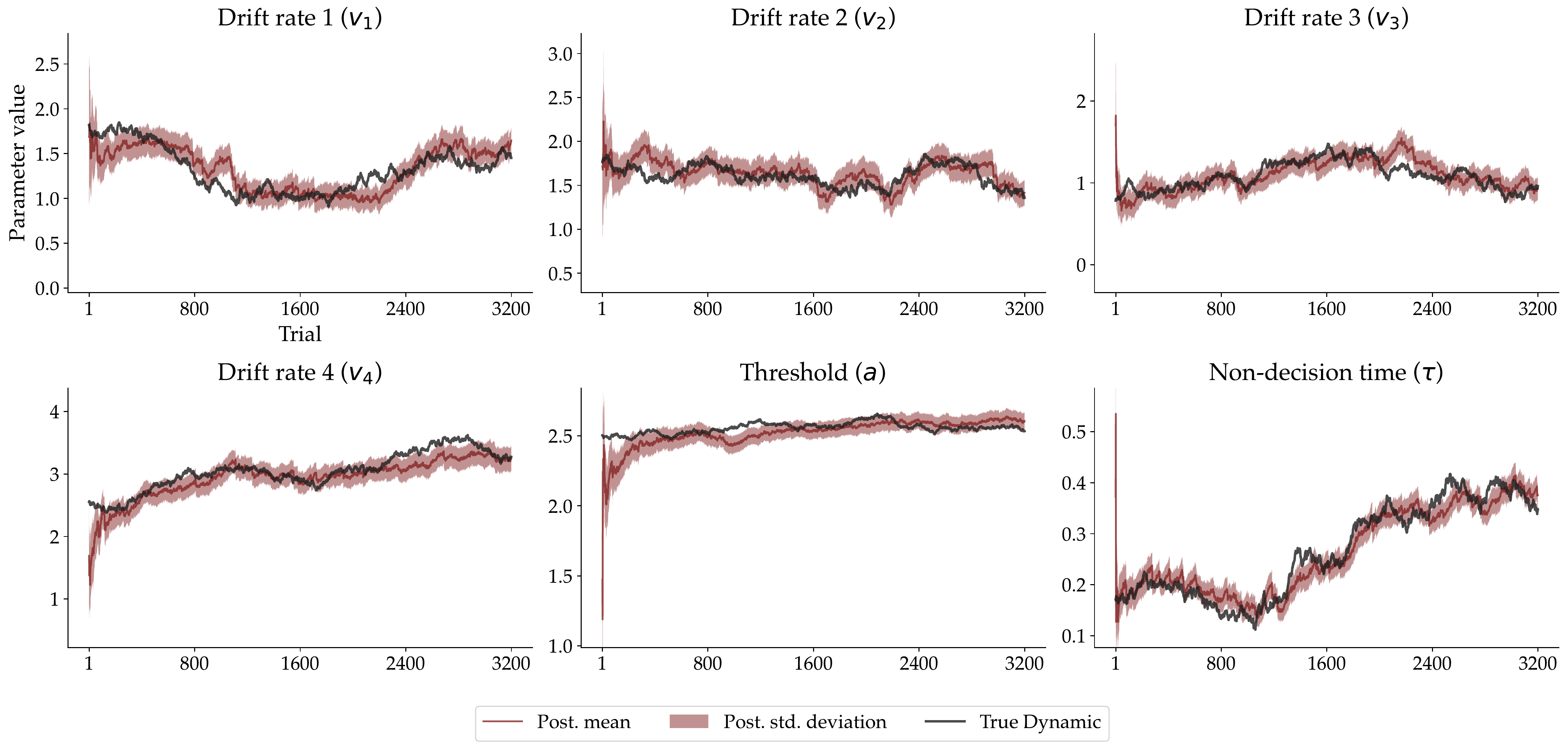}
\caption{The trial-wise posterior mean and $\pm1$ standard deviation for all six parameters, namely the four drift rates $v_1$ - $v_4$ (one for each experimental condition), the threshold $a$, and the non-decision time $\tau$ in red. The true data generating parameter dynamic in black.}
\end{figure}
\FloatBarrier

\begin{figure}
\centering
\includegraphics[width=\textwidth]{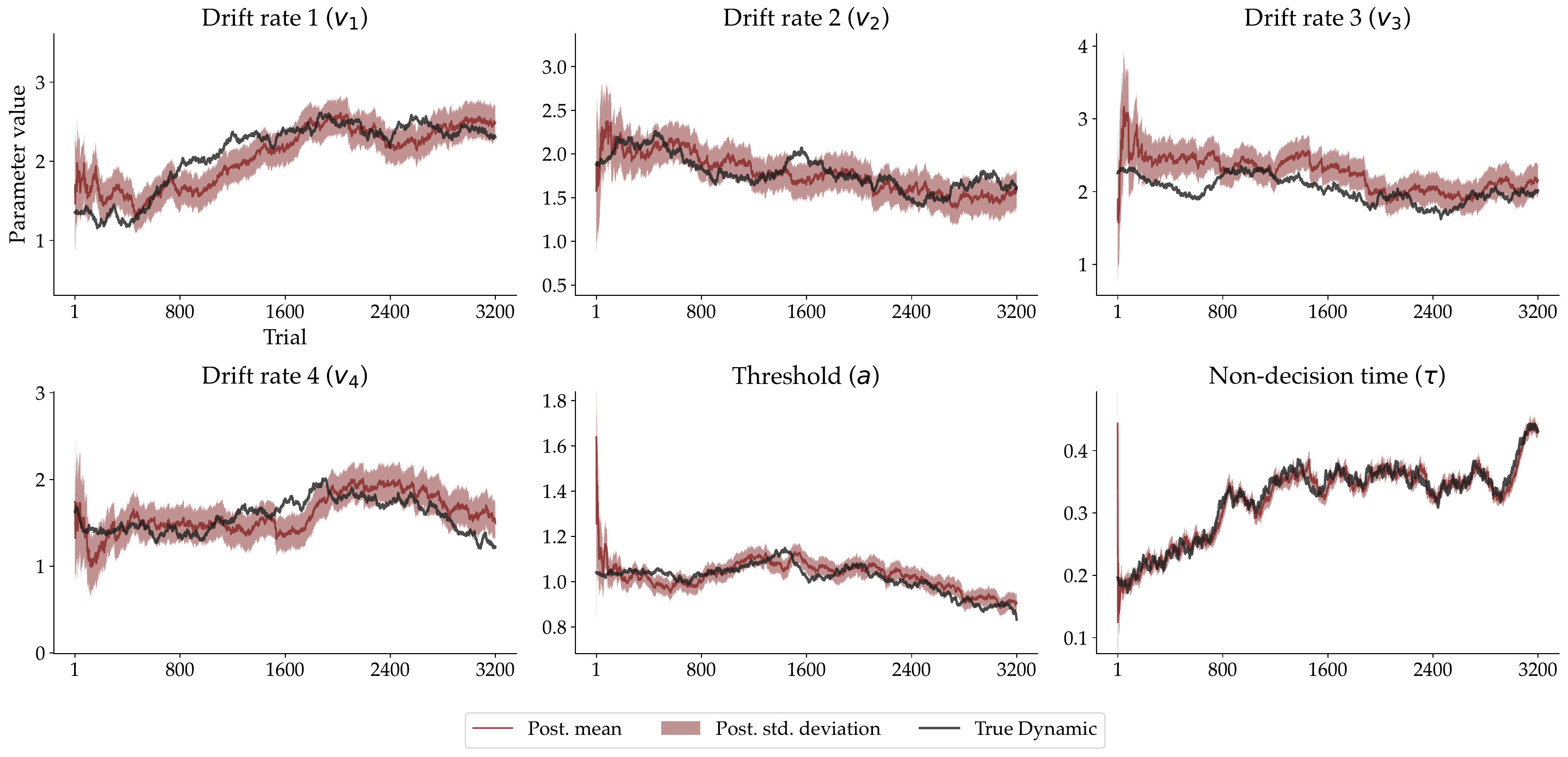}
\caption{The trial-wise posterior mean and $\pm1$ standard deviation for all six parameters, namely the four drift rates $v_1$ - $v_4$ (one for each experimental condition), the threshold $a$, and the non-decision time $\tau$ in red. The true data generating parameter dynamic in black.}
\end{figure}
\FloatBarrier

\begin{figure}
\centering
\includegraphics[width=\textwidth]{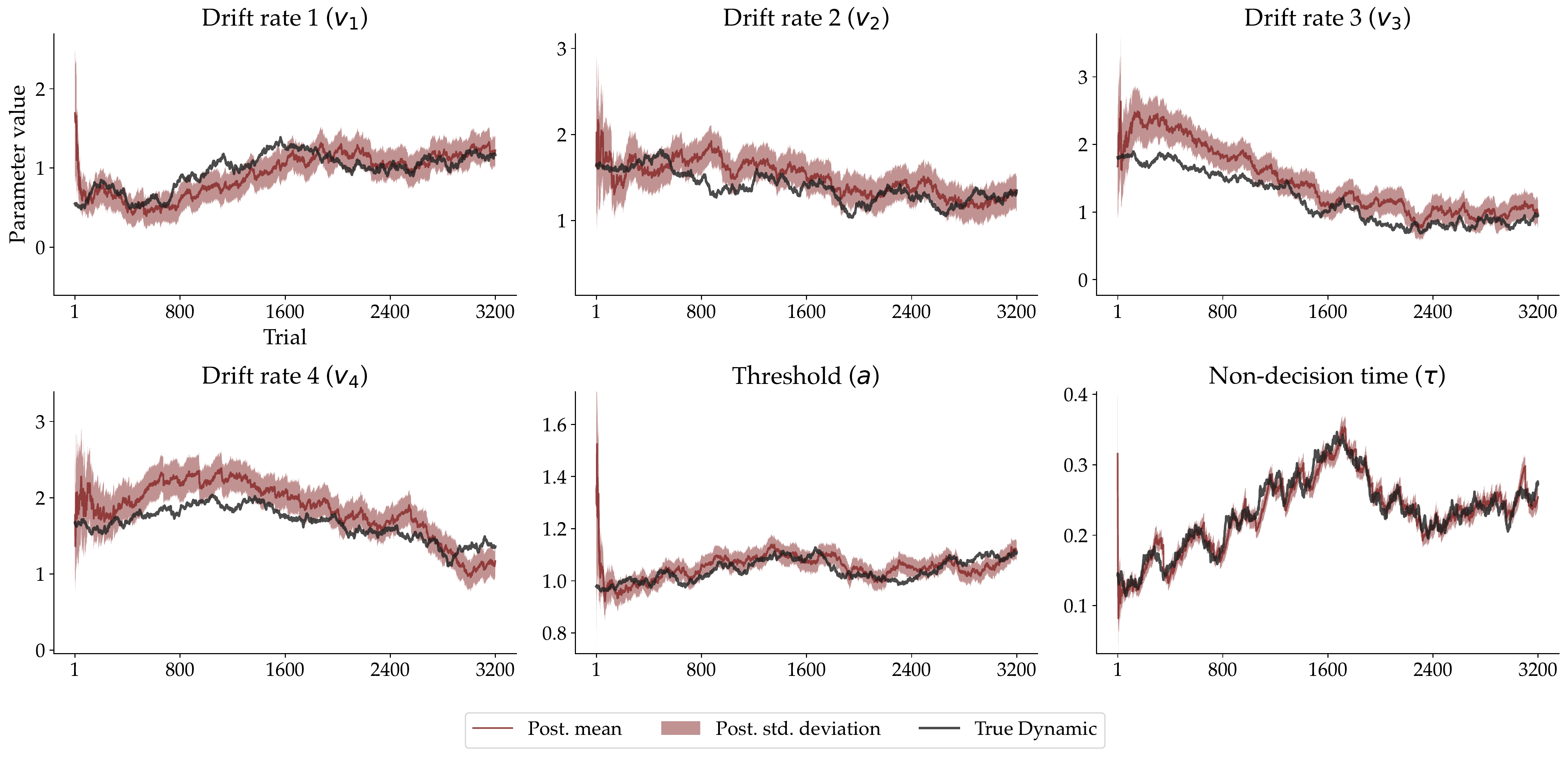}
\caption{The trial-wise posterior mean and $\pm1$ standard deviation for all six parameters, namely the four drift rates $v_1$ - $v_4$ (one for each experimental condition), the threshold $a$, and the non-decision time $\tau$ in red. The true data generating parameter dynamic in black.}
\end{figure}
\FloatBarrier

\subsection*{Individual Model Fits and Predictions}
In the following, we show the fit and multi-horizon predictions of the dynamic DDM on the individual data of the remaining $10$ participants not shown in the main text.
\hspace{0pt}
\vfill

\begin{figure}[H]
\centering
\includegraphics[width=\textwidth]{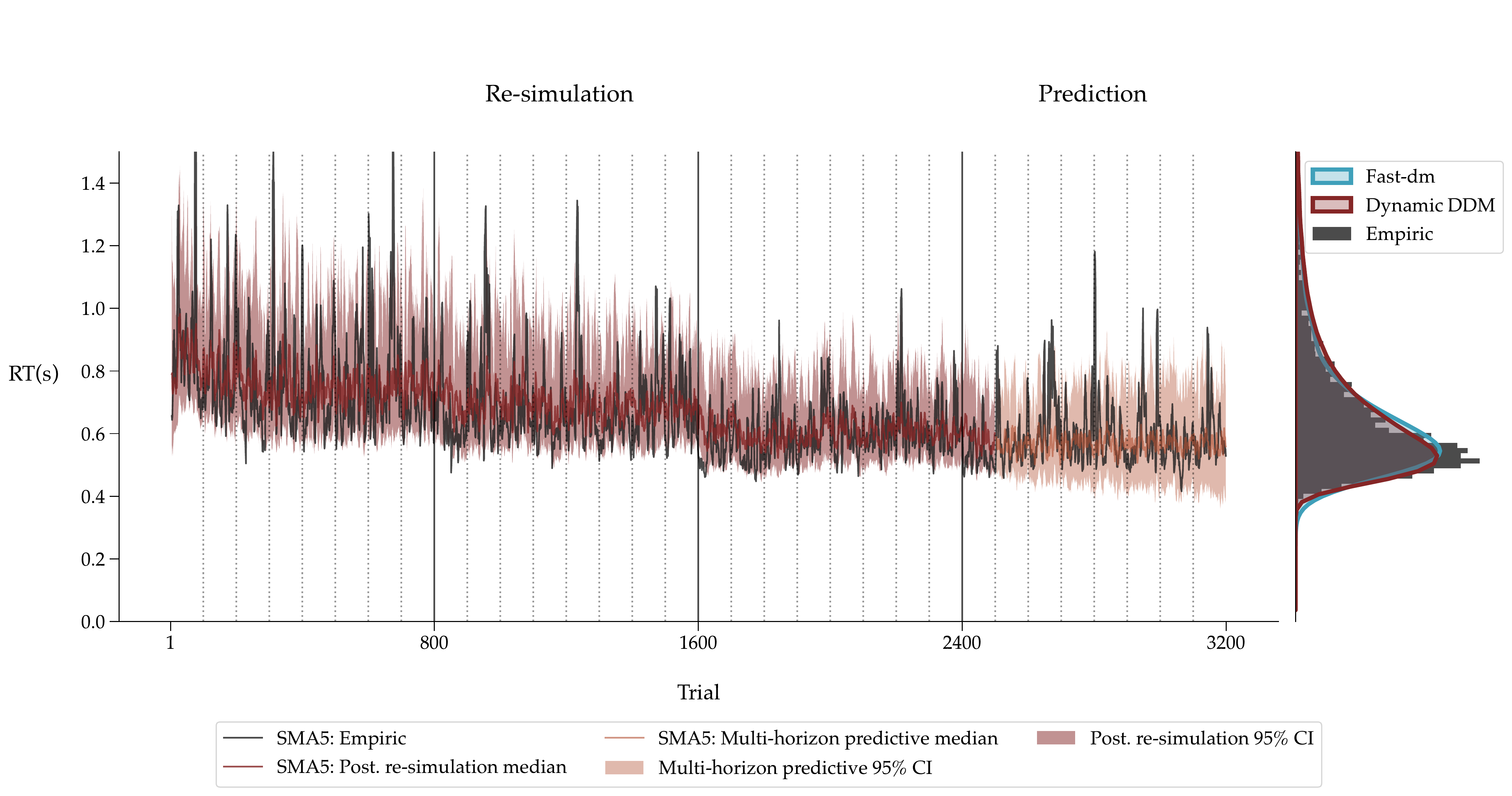}
\caption{\textbf{Left panel} The empirical RT time series of a single individual in black. From trial 1 to 2500, the median posterior re-simulation (aka \textit{retrodictive check}) using the dynamic DDM is shown in red. The models' multi-horizon prediction is depicted for the remaining trials in orange. The shaded areas for the posterior re-simulation and prediction correspond to the 95\% credibility interval. All the time series were smoothed via a simple moving average (SMA) with a period of 5. The dotted vertical lines indicate the end of an experimental block, and the solid vertical lines the end of an experimental session. \textbf{Right panel} The raw RT distribution is plotted as a histogram in black. The re-simulated RT distributions from the dynamic DDM and reference re-simulations from the static DDM using \texttt{fast-dm} are shown as kernel density estimates (KDEs) in red and blue, respectively.}
\end{figure}
\vfill
\hspace{0pt}
\newpage

\begin{figure}
\centering
\includegraphics[width=\textwidth]{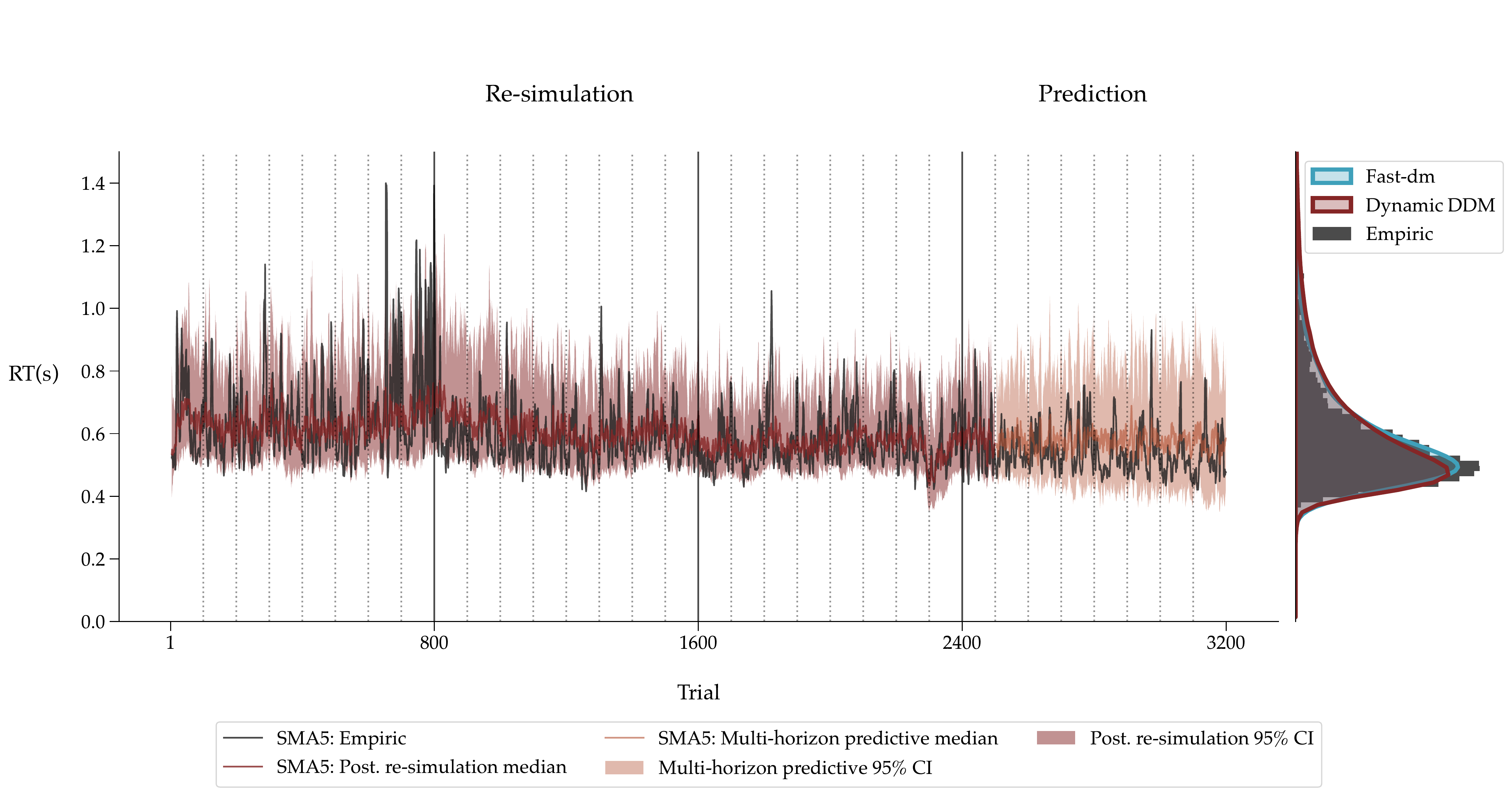}
\caption{\textbf{Left panel} The empirical RT time series of a single individual in black. From trial 1 to 2500, the median posterior re-simulation (aka \textit{retrodictive check}) using the dynamic DDM is shown in red. The models' multi-horizon prediction is depicted for the remaining trials in orange. The shaded areas for the posterior re-simulation and prediction correspond to the 95\% credibility interval. All the time series were smoothed via a simple moving average (SMA) with a period of 5. The dotted vertical lines indicate the end of an experimental block, and the solid vertical lines the end of an experimental session. \textbf{Right panel} The raw RT distribution is plotted as a histogram in black. The re-simulated RT distributions from the dynamic DDM and reference re-simulations from the static DDM using \texttt{fast-dm} are shown as kernel density estimates (KDEs) in red and blue, respectively.}
\end{figure}
\FloatBarrier

\begin{figure}
\centering
\includegraphics[width=\textwidth]{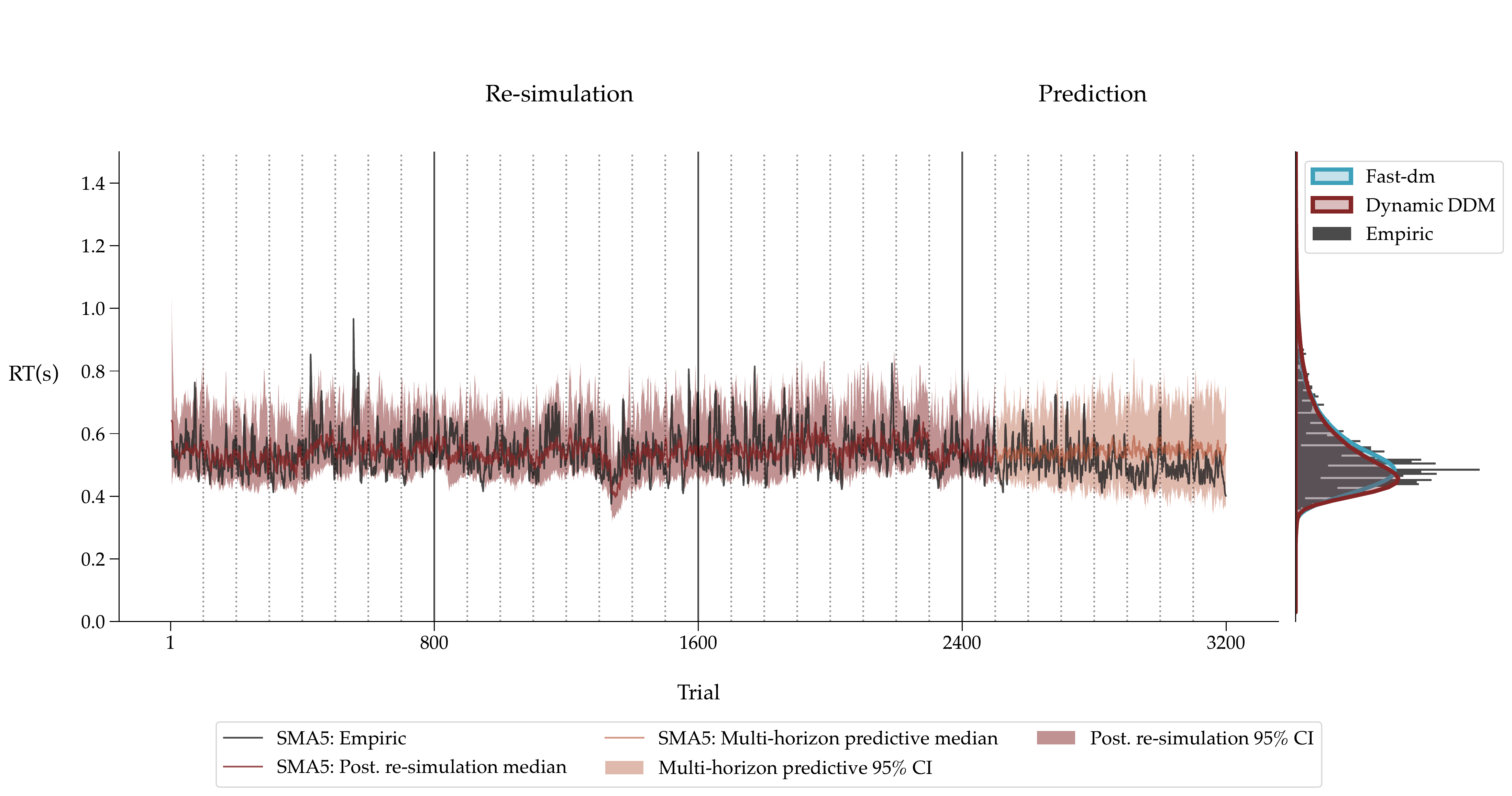}
\caption{\textbf{Left panel} The empirical RT time series of a single individual in black. From trial 1 to 2500, the median posterior re-simulation (aka \textit{retrodictive check}) using the dynamic DDM is shown in red. The models' multi-horizon prediction is depicted for the remaining trials in orange. The shaded areas for the posterior re-simulation and prediction correspond to the 95\% credibility interval. All the time series were smoothed via a simple moving average (SMA) with a period of 5. The dotted vertical lines indicate the end of an experimental block, and the solid vertical lines the end of an experimental session. \textbf{Right panel} The raw RT distribution is plotted as a histogram in black. The re-simulated RT distributions from the dynamic DDM and reference re-simulations from the static DDM using \texttt{fast-dm} are shown as kernel density estimates (KDEs) in red and blue, respectively.}
\end{figure}
\FloatBarrier

\begin{figure}
\centering
\includegraphics[width=\textwidth]{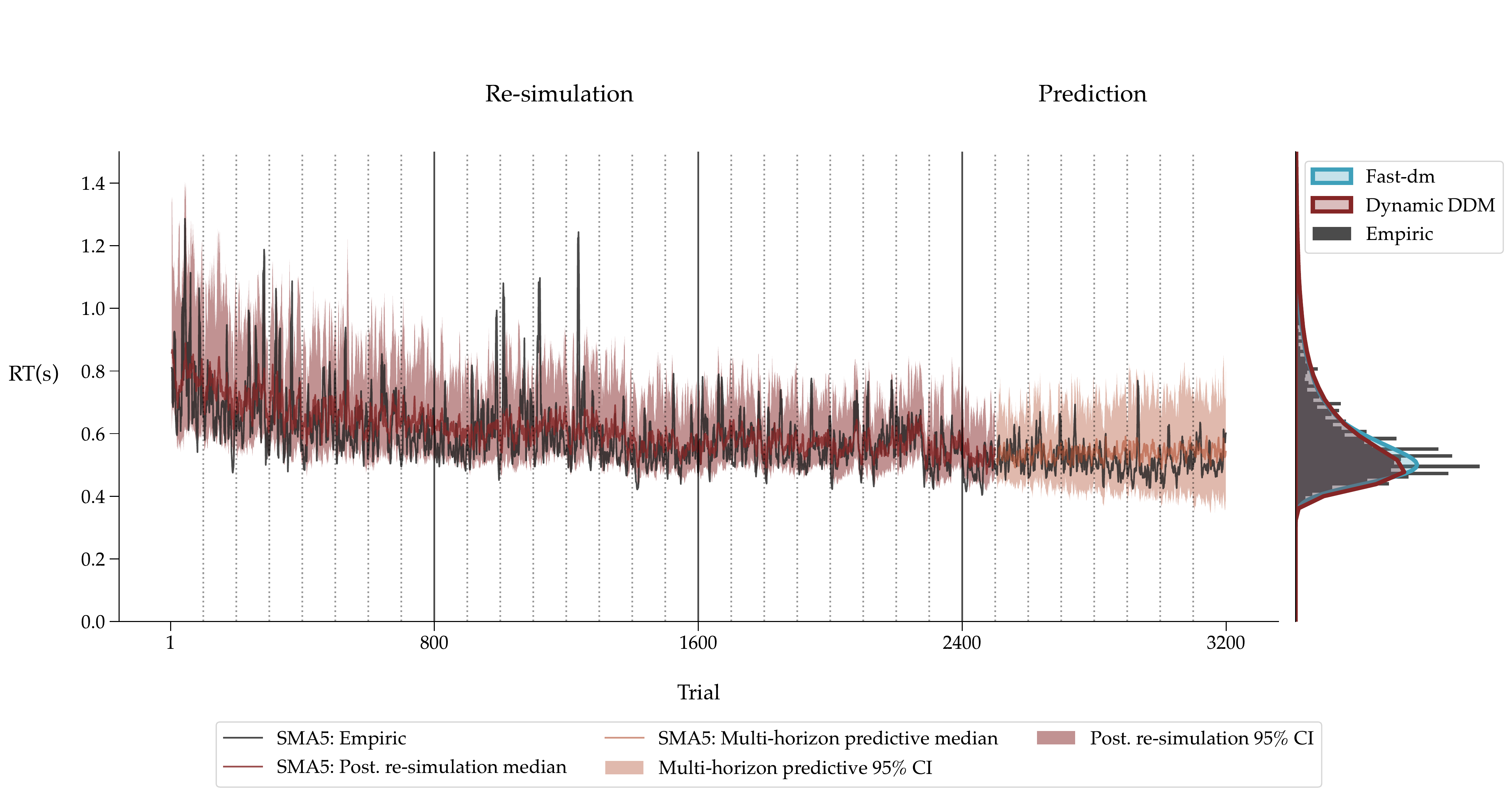}
\caption{\textbf{Left panel} The empirical RT time series of a single individual in black. From trial 1 to 2500, the median posterior re-simulation (aka \textit{retrodictive check}) using the dynamic DDM is shown in red. The models' multi-horizon prediction is depicted for the remaining trials in orange. The shaded areas for the posterior re-simulation and prediction correspond to the 95\% credibility interval. All the time series were smoothed via a simple moving average (SMA) with a period of 5. The dotted vertical lines indicate the end of an experimental block, and the solid vertical lines the end of an experimental session. \textbf{Right panel} The raw RT distribution is plotted as a histogram in black. The re-simulated RT distributions from the dynamic DDM and reference re-simulations from the static DDM using \texttt{fast-dm} are shown as kernel density estimates (KDEs) in red and blue, respectively.}
\end{figure}
\FloatBarrier

\begin{figure}
\centering
\includegraphics[width=\textwidth]{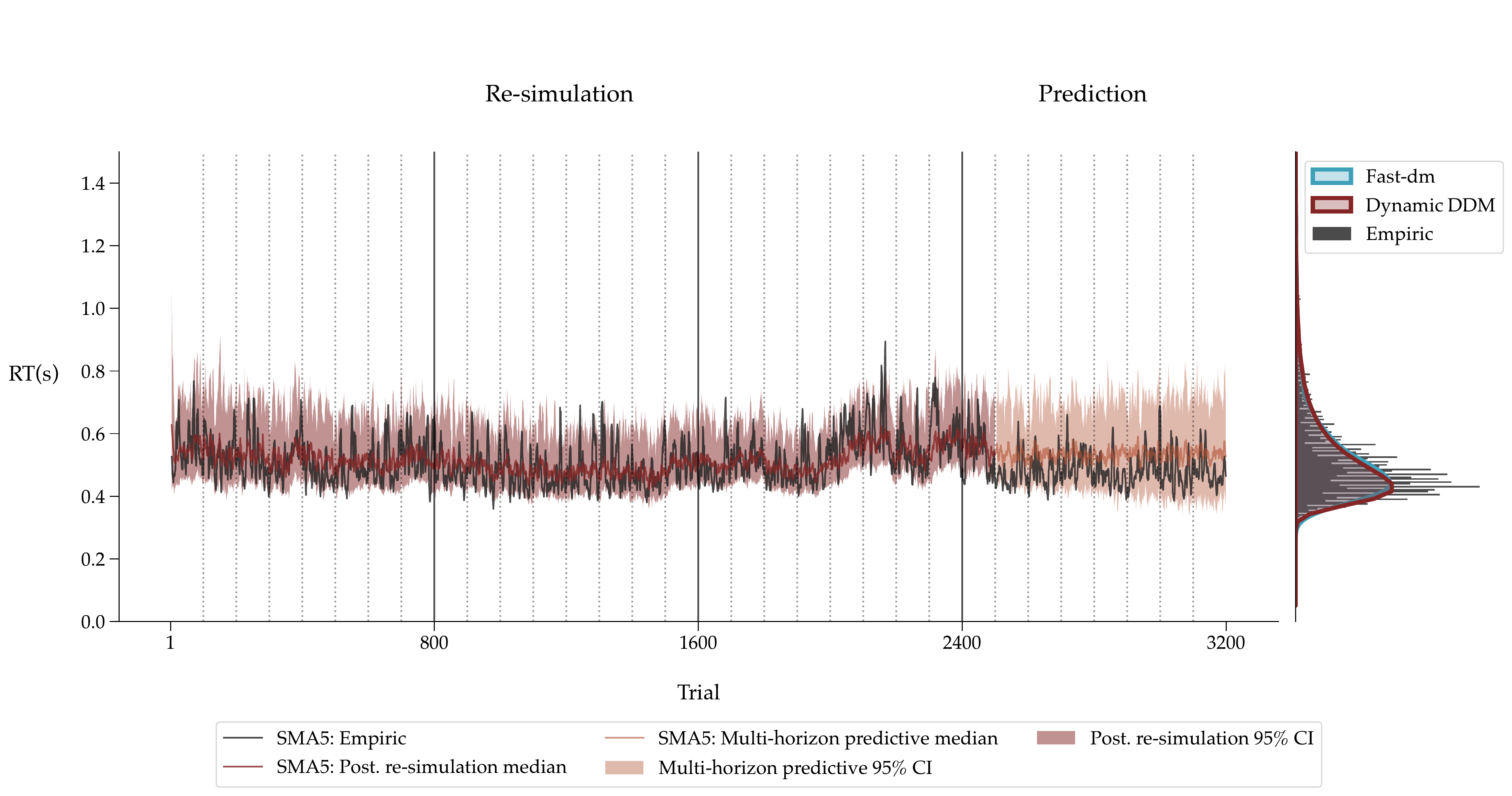}
\caption{\textbf{Left panel} The empirical RT time series of a single individual in black. From trial 1 to 2500, the median posterior re-simulation (aka \textit{retrodictive check}) using the dynamic DDM is shown in red. The models' multi-horizon prediction is depicted for the remaining trials in orange. The shaded areas for the posterior re-simulation and prediction correspond to the 95\% credibility interval. All the time series were smoothed via a simple moving average (SMA) with a period of 5. The dotted vertical lines indicate the end of an experimental block, and the solid vertical lines the end of an experimental session. \textbf{Right panel} The raw RT distribution is plotted as a histogram in black. The re-simulated RT distributions from the dynamic DDM and reference re-simulations from the static DDM using \texttt{fast-dm} are shown as kernel density estimates (KDEs) in red and blue, respectively.}
\end{figure}
\FloatBarrier

\begin{figure}
\centering
\includegraphics[width=\textwidth]{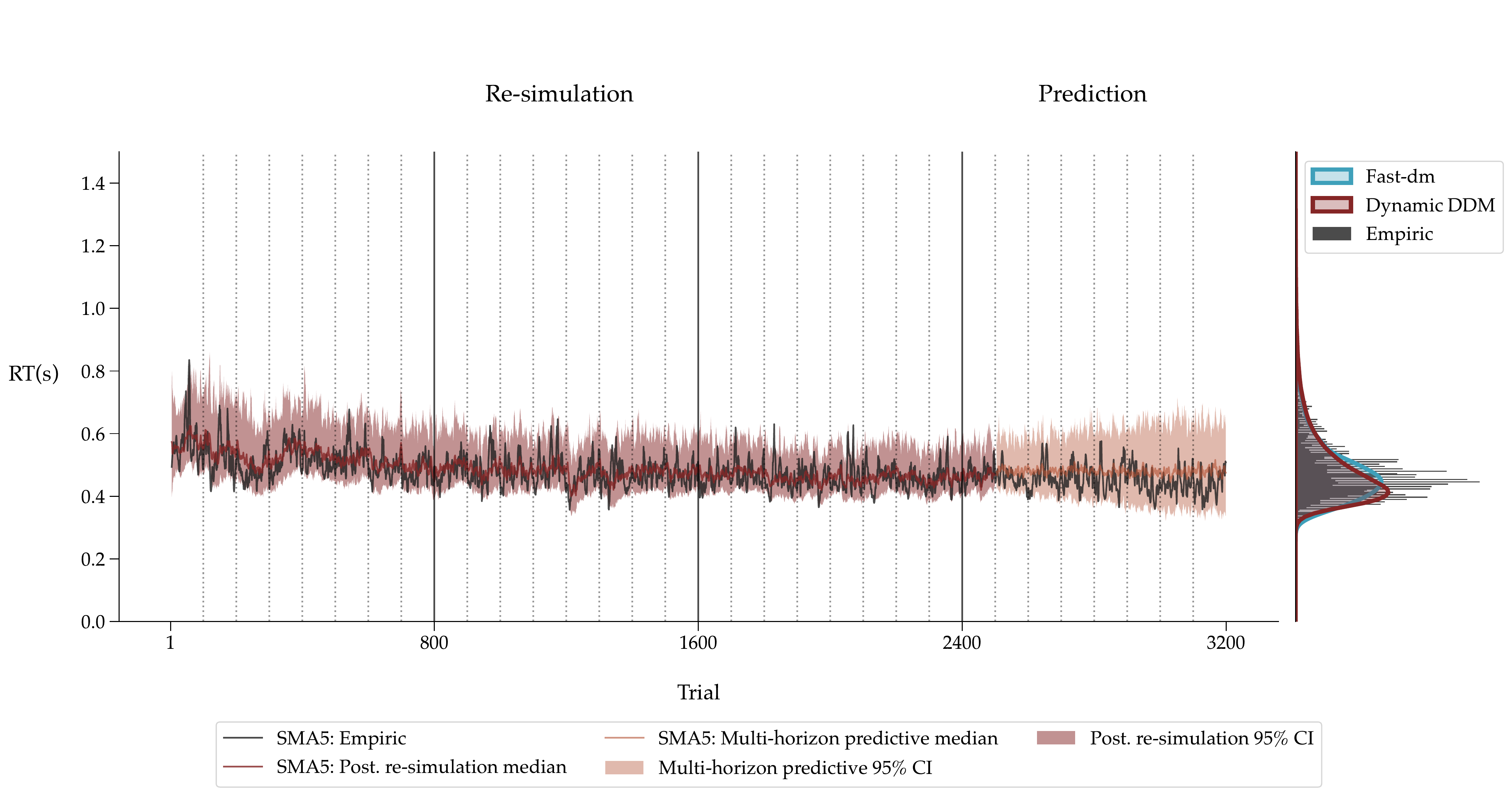}
\caption{\textbf{Left panel} The empirical RT time series of a single individual in black. From trial 1 to 2500, the median posterior re-simulation (aka \textit{retrodictive check}) using the dynamic DDM is shown in red. The models' multi-horizon prediction is depicted for the remaining trials in orange. The shaded areas for the posterior re-simulation and prediction correspond to the 95\% credibility interval. All the time series were smoothed via a simple moving average (SMA) with a period of 5. The dotted vertical lines indicate the end of an experimental block, and the solid vertical lines the end of an experimental session. \textbf{Right panel} The raw RT distribution is plotted as a histogram in black. The re-simulated RT distributions from the dynamic DDM and reference re-simulations from the static DDM using \texttt{fast-dm} are shown as kernel density estimates (KDEs) in red and blue, respectively.}
\end{figure}
\FloatBarrier

\begin{figure}
\centering
\includegraphics[width=\textwidth]{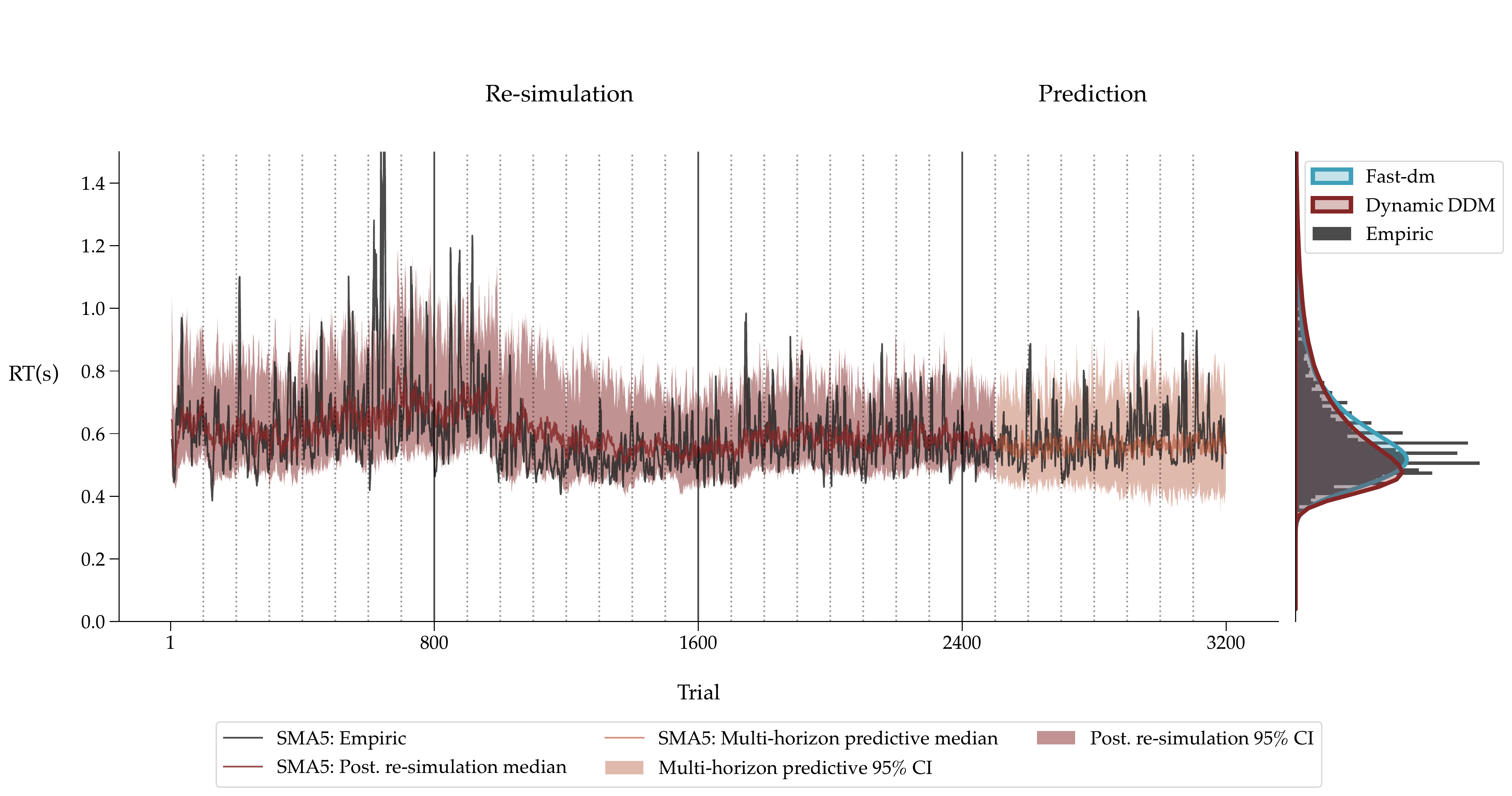}
\caption{\textbf{Left panel} The empirical RT time series of a single individual in black. From trial 1 to 2500, the median posterior re-simulation (aka \textit{retrodictive check}) using the dynamic DDM is shown in red. The models' multi-horizon prediction is depicted for the remaining trials in orange. The shaded areas for the posterior re-simulation and prediction correspond to the 95\% credibility interval. All the time series were smoothed via a simple moving average (SMA) with a period of 5. The dotted vertical lines indicate the end of an experimental block, and the solid vertical lines the end of an experimental session. \textbf{Right panel} The raw RT distribution is plotted as a histogram in black. The re-simulated RT distributions from the dynamic DDM and reference re-simulations from the static DDM using \texttt{fast-dm} are shown as kernel density estimates (KDEs) in red and blue, respectively.}
\end{figure}
\FloatBarrier

\begin{figure}
\centering
\includegraphics[width=\textwidth]{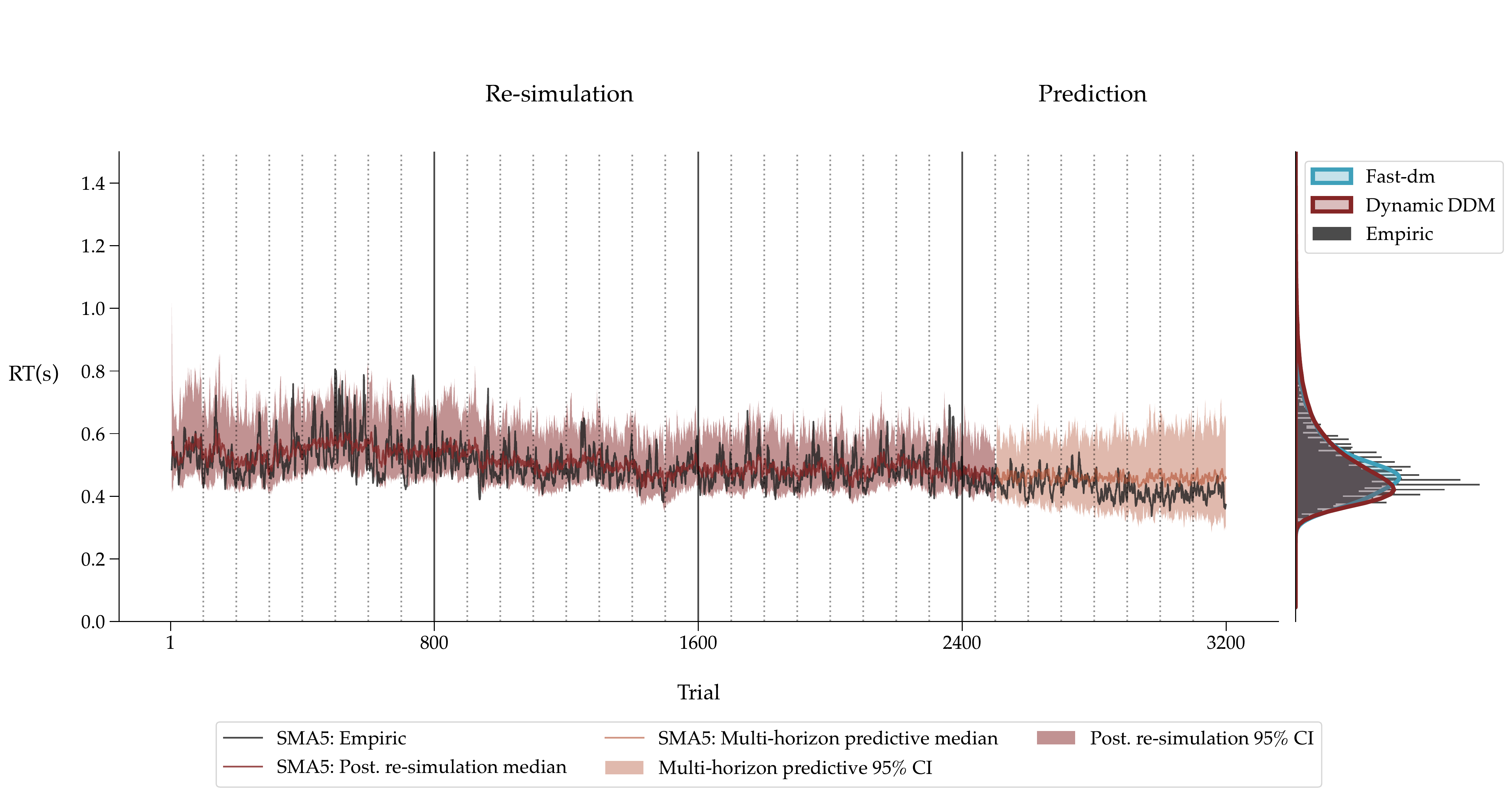}
\caption{\textbf{Left panel} The empirical RT time series of a single individual in black. From trial 1 to 2500, the median posterior re-simulation (aka \textit{retrodictive check}) using the dynamic DDM is shown in red. The models' multi-horizon prediction is depicted for the remaining trials in orange. The shaded areas for the posterior re-simulation and prediction correspond to the 95\% credibility interval. All the time series were smoothed via a simple moving average (SMA) with a period of 5. The dotted vertical lines indicate the end of an experimental block, and the solid vertical lines the end of an experimental session. \textbf{Right panel} The raw RT distribution is plotted as a histogram in black. The re-simulated RT distributions from the dynamic DDM and reference re-simulations from the static DDM using \texttt{fast-dm} are shown as kernel density estimates (KDEs) in red and blue, respectively.}
\end{figure}
\FloatBarrier

\begin{figure}
\centering
\includegraphics[width=\textwidth]{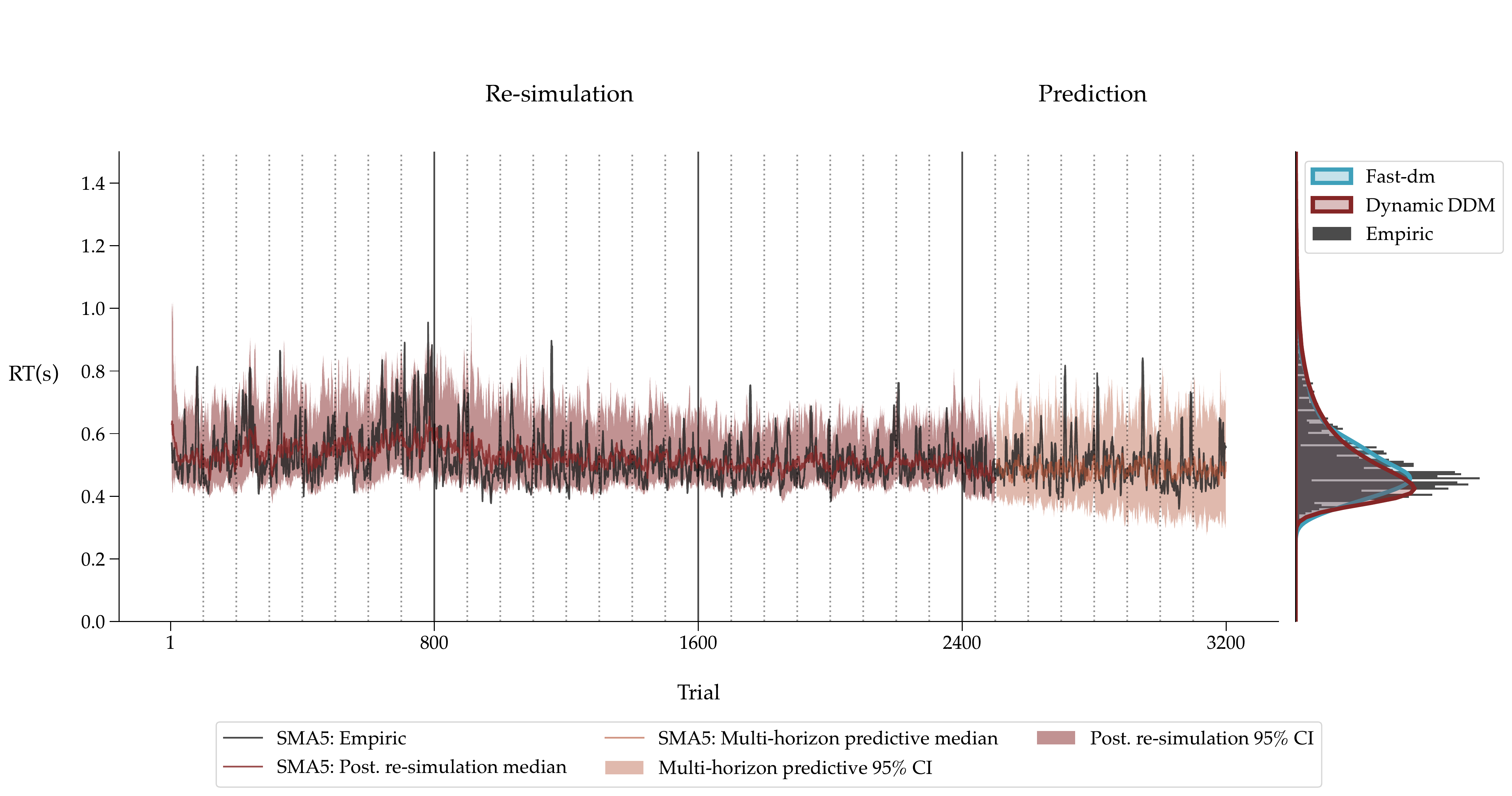}
\caption{\textbf{Left panel} The empirical RT time series of a single individual in black. From trial 1 to 2500, the median posterior re-simulation (aka \textit{retrodictive check}) using the dynamic DDM is shown in red. The models' multi-horizon prediction is depicted for the remaining trials in orange. The shaded areas for the posterior re-simulation and prediction correspond to the 95\% credibility interval. All the time series were smoothed via a simple moving average (SMA) with a period of 5. The dotted vertical lines indicate the end of an experimental block, and the solid vertical lines the end of an experimental session. \textbf{Right panel} The raw RT distribution is plotted as a histogram in black. The re-simulated RT distributions from the dynamic DDM and reference re-simulations from the static DDM using \texttt{fast-dm} are shown as kernel density estimates (KDEs) in red and blue, respectively.}
\end{figure}
\FloatBarrier

\begin{figure}
\centering
\includegraphics[width=\textwidth]{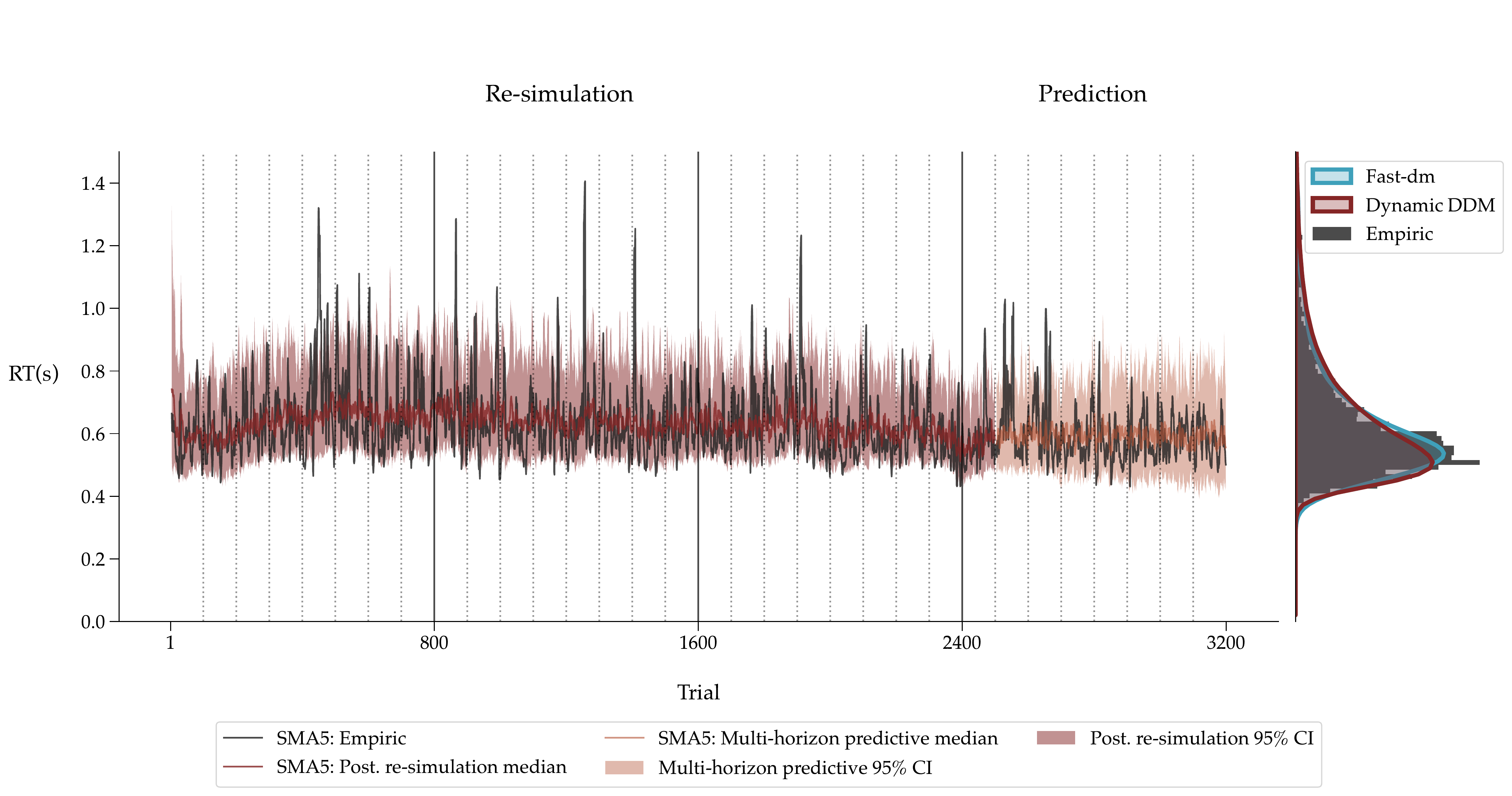}
\caption{\textbf{Left panel} The empirical RT time series of a single individual in black. From trial 1 to 2500, the median posterior re-simulation (aka \textit{retrodictive check}) using the dynamic DDM is shown in red. The models' multi-horizon prediction is depicted for the remaining trials in orange. The shaded areas for the posterior re-simulation and prediction correspond to the 95\% credibility interval. All the time series were smoothed via a simple moving average (SMA) with a period of 5. The dotted vertical lines indicate the end of an experimental block, and the solid vertical lines the end of an experimental session. \textbf{Right panel} The raw RT distribution is plotted as a histogram in black. The re-simulated RT distributions from the dynamic DDM and reference re-simulations from the static DDM using \texttt{fast-dm} are shown as kernel density estimates (KDEs) in red and blue, respectively.}
\end{figure}
\FloatBarrier

\subsection*{Individual Parameter Dynamics}
In the following, we show the inferred parameter dynamics of the remaining $10$ participants not shown in the main text.

\hspace{0pt}
\vfill
\begin{figure}[H]
\centering
\includegraphics[width=\textwidth]{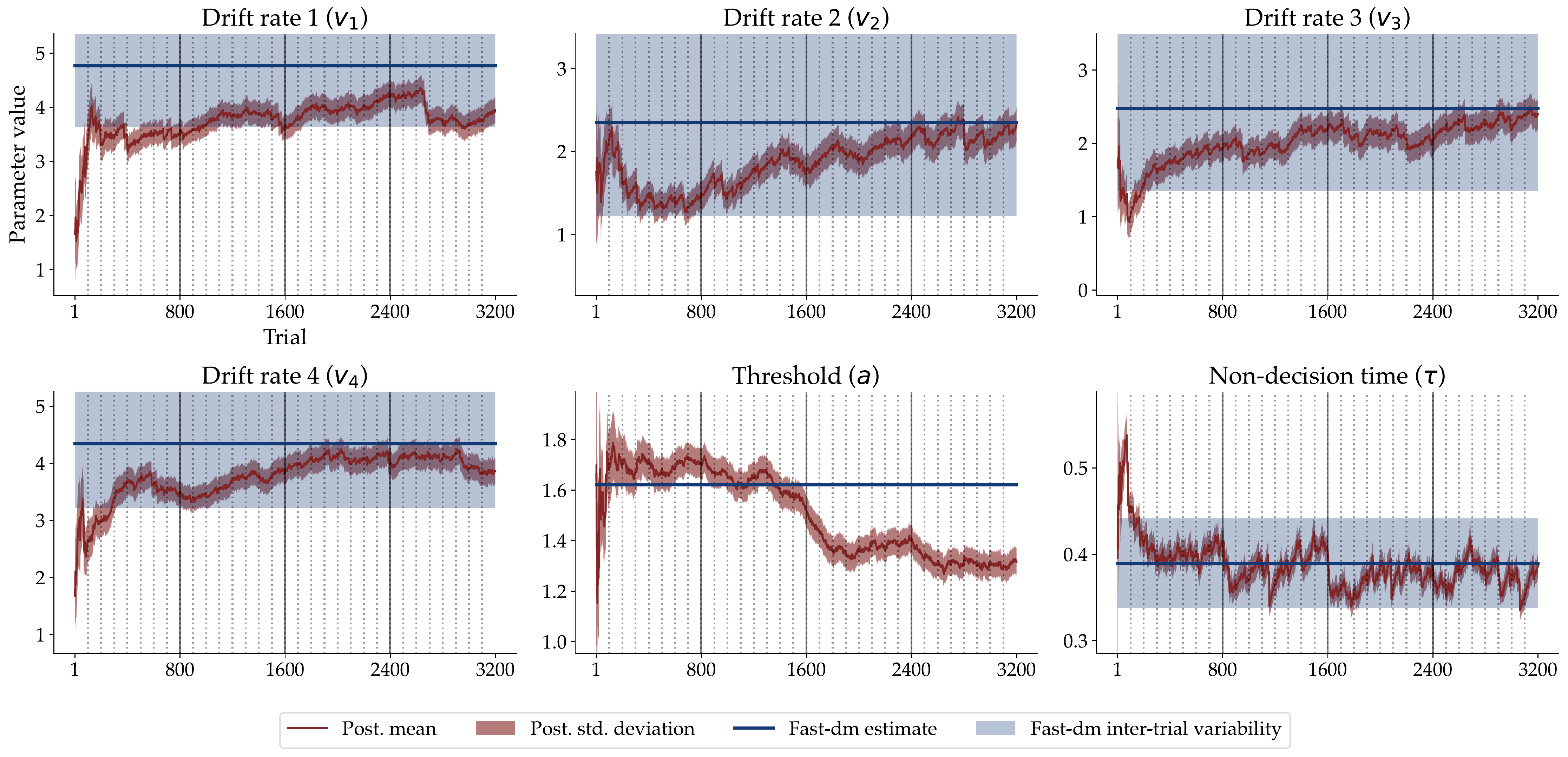}
\caption{The trial-wise posterior mean and $\pm1$ standard deviation for all six parameters, namely the four drift rates $v_1$ - $v_4$ (one for each experimental condition), the threshold $a$, and the non-decision time $\tau$ of an individual participant. The point estimates of the static DDM parameters and the corresponding inter-trial variabilities are shown in solid blue lines and blue shaded areas, respectively.}
\end{figure}
\vfill
\hspace{0pt}
\newpage

\hspace{0pt}
\vfill
\begin{figure}[H]
\centering
\includegraphics[width=\textwidth]{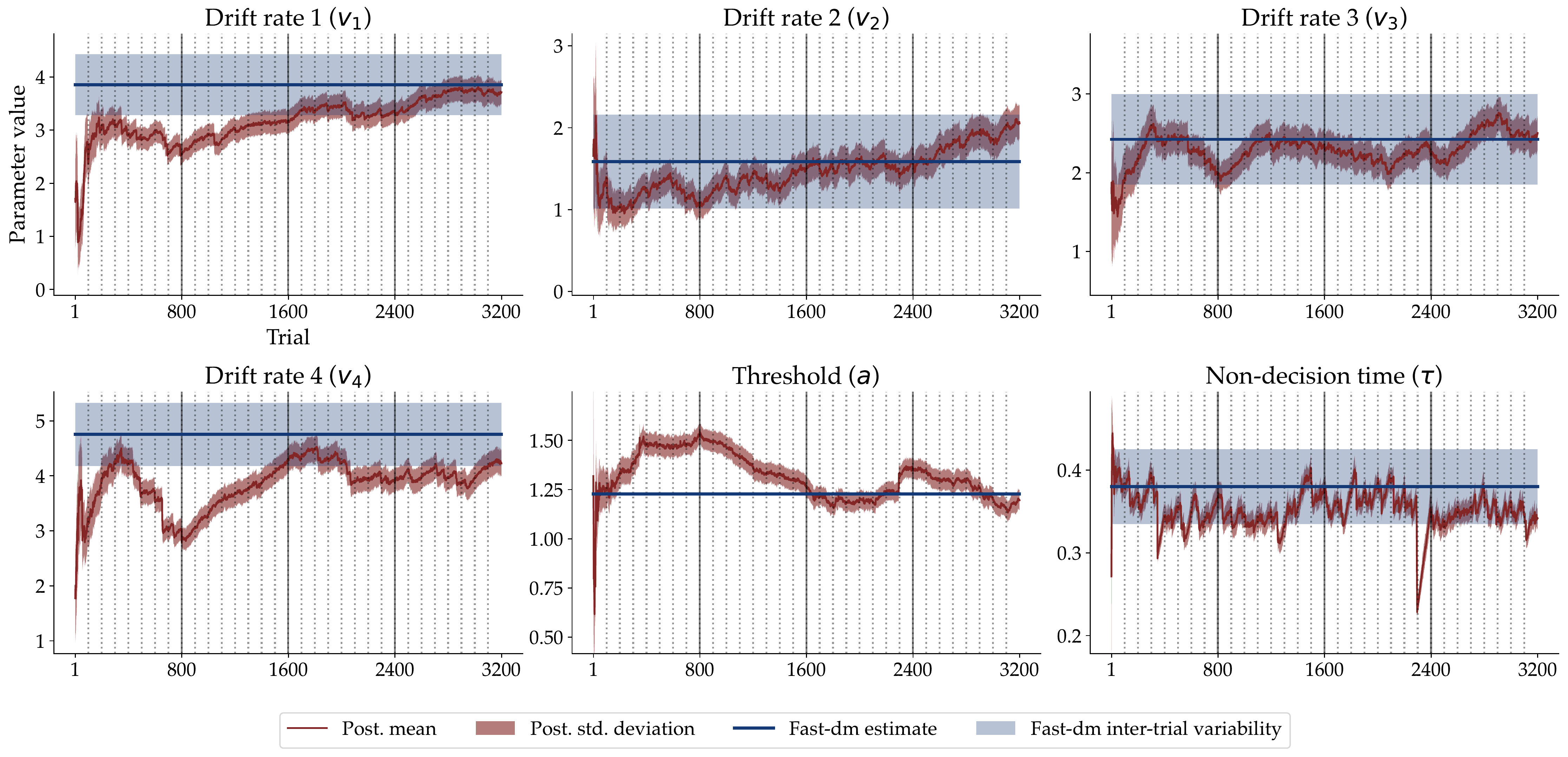}
\caption{The trial-wise posterior mean and $\pm1$ standard deviation for all six parameters, namely the four drift rates $v_1$ - $v_4$ (one for each experimental condition), the threshold $a$, and the non-decision time $\tau$ of an individual participant. The point estimates of the static DDM parameters and the corresponding inter-trial variabilities are shown in solid blue lines and blue shaded areas, respectively.}
\end{figure}
\vfill
\hspace{0pt}
\newpage

\hspace{0pt}
\vfill
\begin{figure}[H]
\centering
\includegraphics[width=\textwidth]{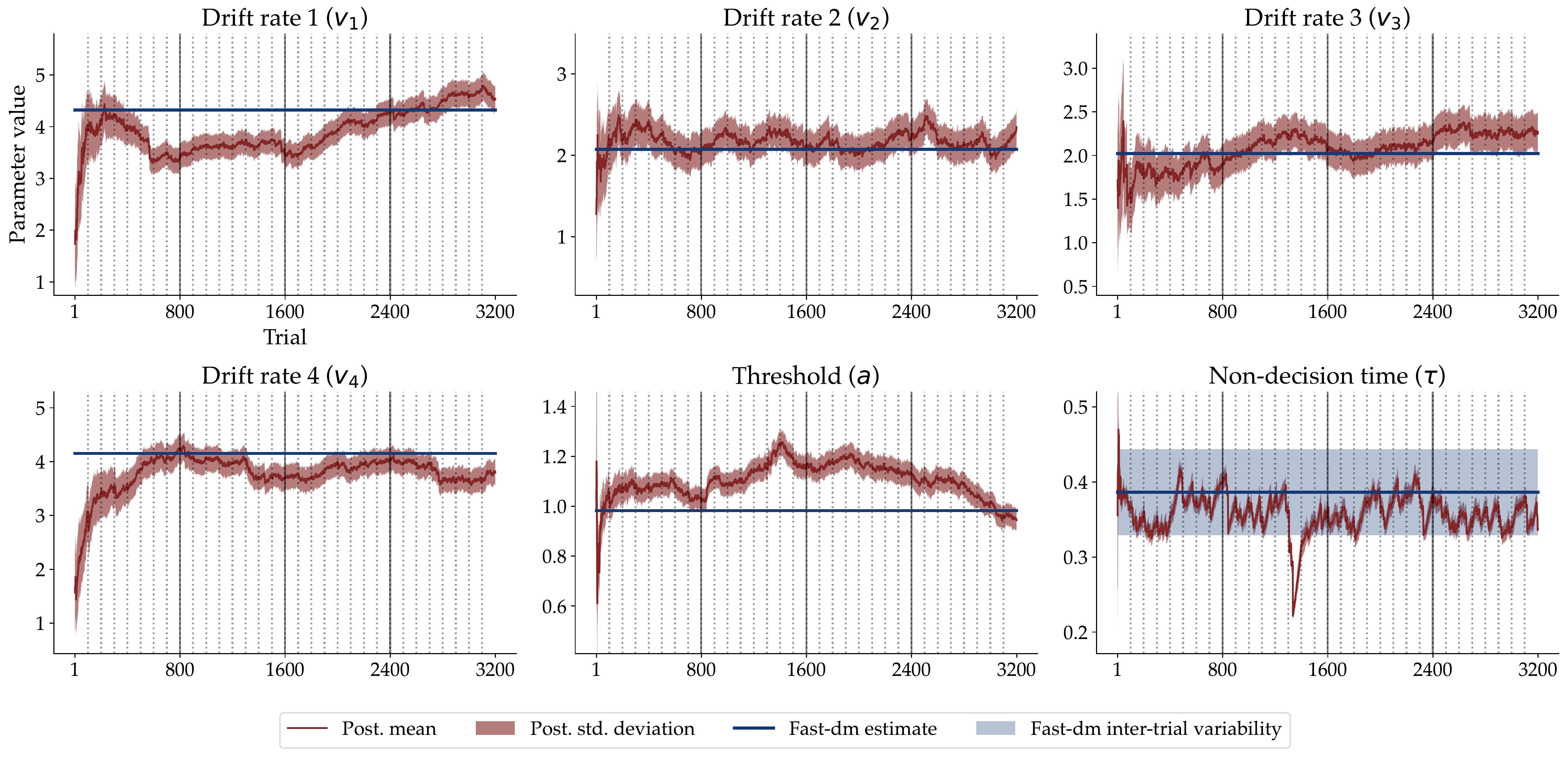}
\caption{The trial-wise posterior mean and $\pm1$ standard deviation for all six parameters, namely the four drift rates $v_1$ - $v_4$ (one for each experimental condition), the threshold $a$, and the non-decision time $\tau$ of an individual participant. The point estimates of the static DDM parameters and the corresponding inter-trial variabilities are shown in solid blue lines and blue shaded areas, respectively.}
\end{figure}
\vfill
\hspace{0pt}
\newpage

\hspace{0pt}
\vfill
\begin{figure}[H]
\centering
\includegraphics[width=\textwidth]{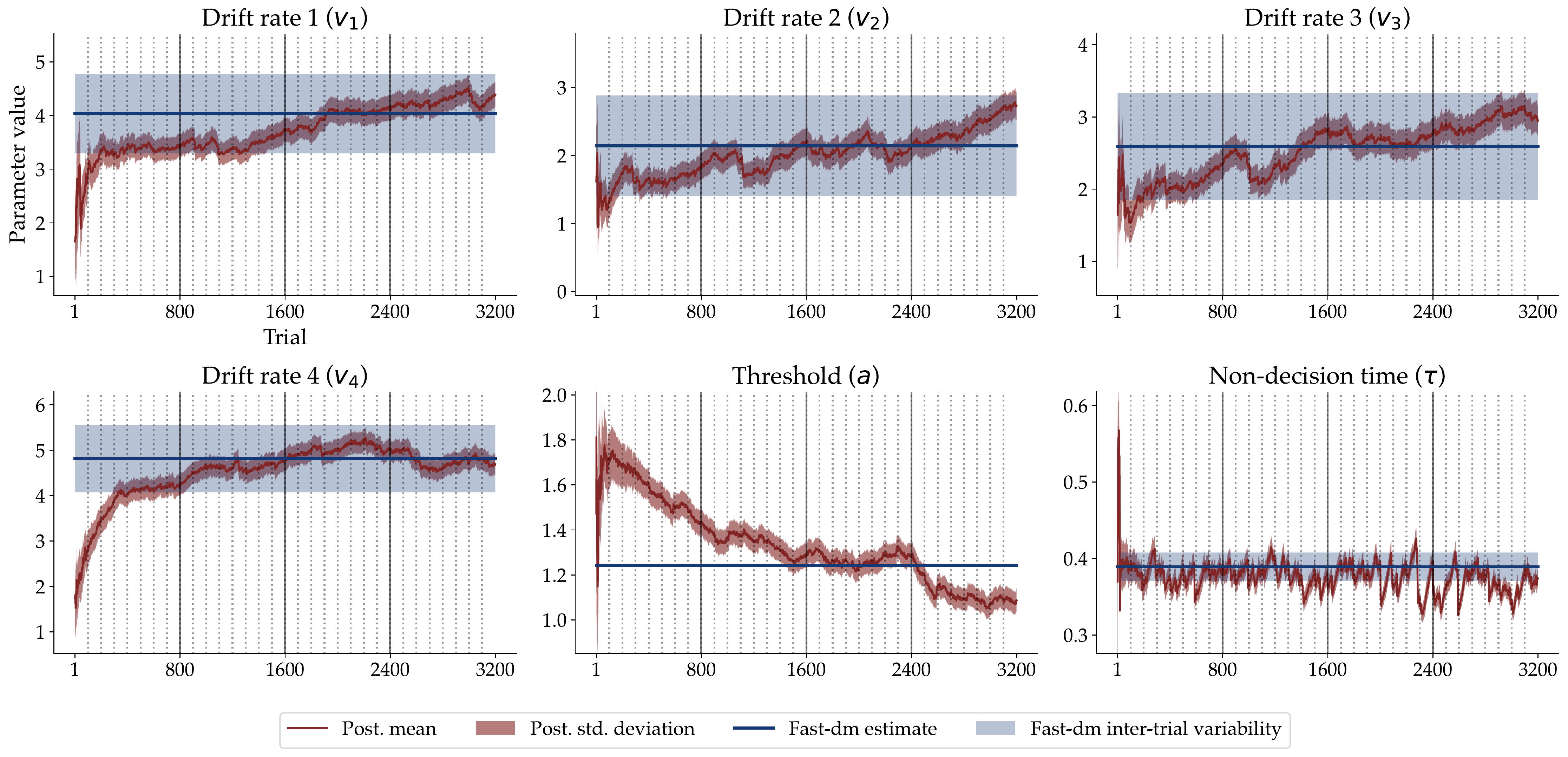}
\caption{The trial-wise posterior mean and $\pm1$ standard deviation for all six parameters, namely the four drift rates $v_1$ - $v_4$ (one for each experimental condition), the threshold $a$, and the non-decision time $\tau$ of an individual participant. The point estimates of the static DDM parameters and the corresponding inter-trial variabilities are shown in solid blue lines and blue shaded areas, respectively.}
\end{figure}
\vfill
\hspace{0pt}
\newpage

\hspace{0pt}
\vfill
\begin{figure}[H]
\centering
\includegraphics[width=\textwidth]{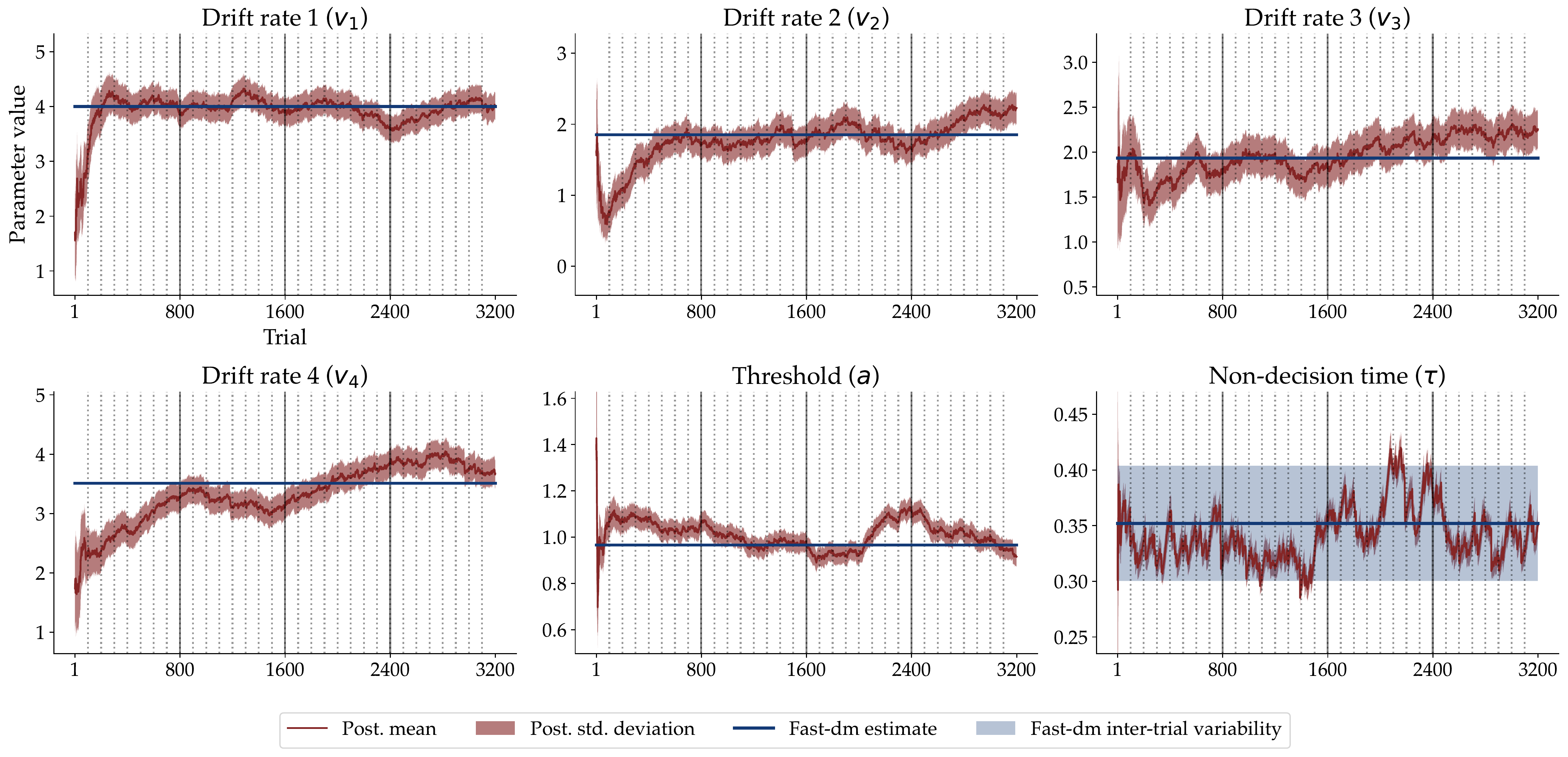}
\caption{The trial-wise posterior mean and $\pm1$ standard deviation for all six parameters, namely the four drift rates $v_1$ - $v_4$ (one for each experimental condition), the threshold $a$, and the non-decision time $\tau$ of an individual participant. The point estimates of the static DDM parameters and the corresponding inter-trial variabilities are shown in solid blue lines and blue shaded areas, respectively.}
\end{figure}
\vfill
\hspace{0pt}
\newpage

\hspace{0pt}
\vfill
\begin{figure}[H]
\centering
\includegraphics[width=\textwidth]{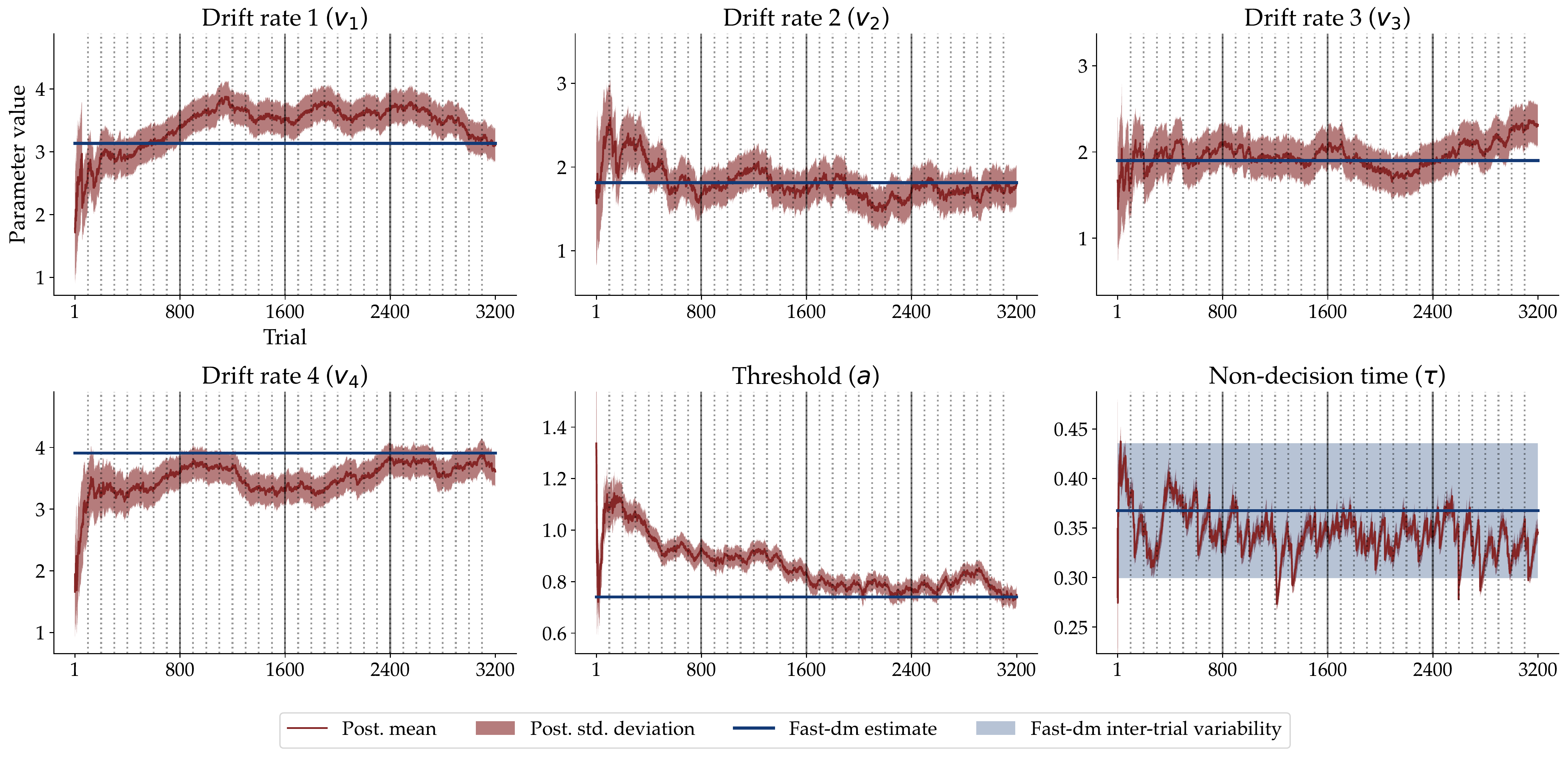}
\caption{The trial-wise posterior mean and $\pm1$ standard deviation for all six parameters, namely the four drift rates $v_1$ - $v_4$ (one for each experimental condition), the threshold $a$, and the non-decision time $\tau$ of an individual participant. The point estimates of the static DDM parameters and the corresponding inter-trial variabilities are shown in solid blue lines and blue shaded areas, respectively.}
\end{figure}
\vfill
\hspace{0pt}
\newpage

\hspace{0pt}
\vfill
\begin{figure}[H]
\centering
\includegraphics[width=\textwidth]{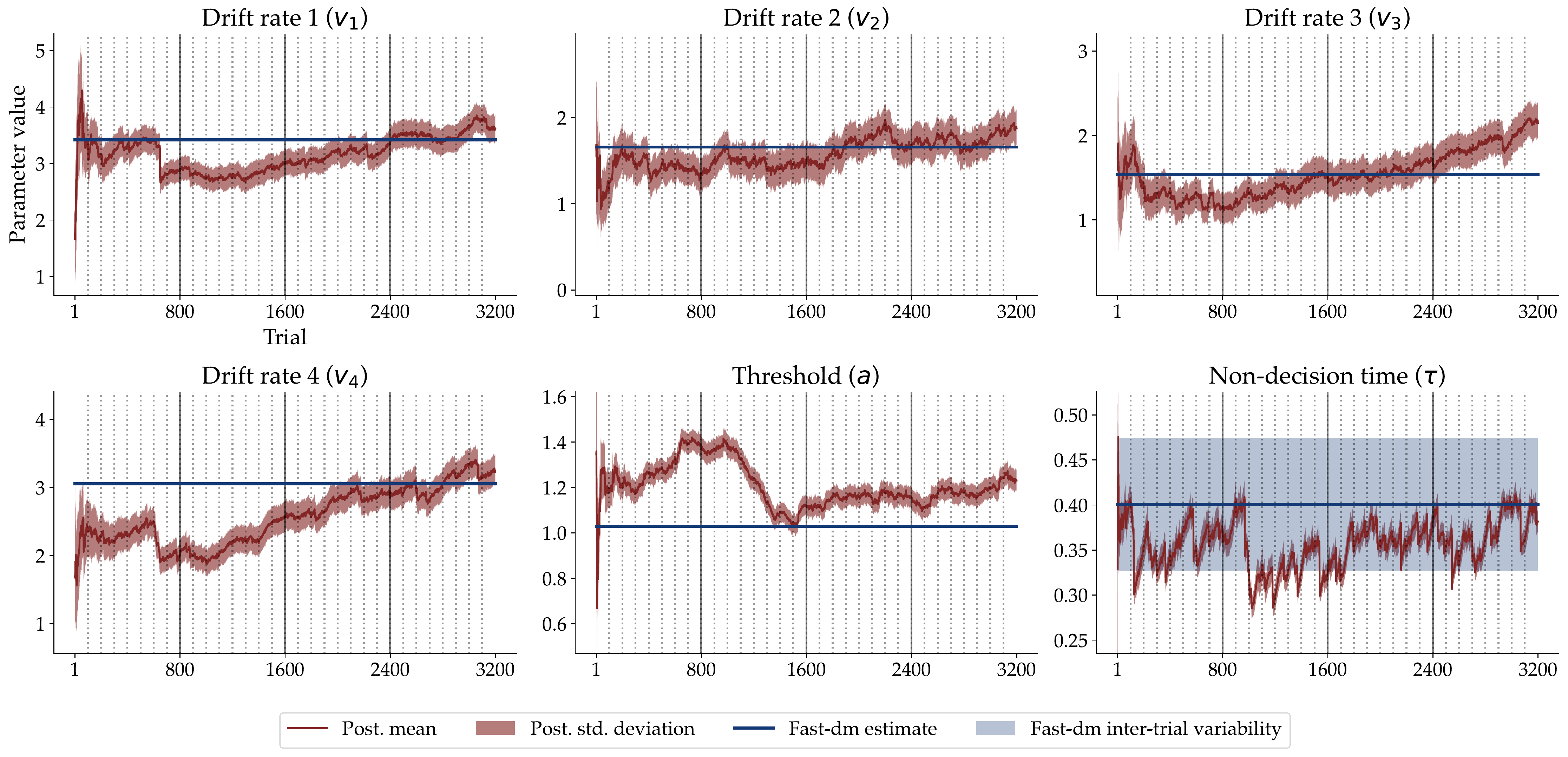}
\caption{The trial-wise posterior mean and $\pm1$ standard deviation for all six parameters, namely the four drift rates $v_1$ - $v_4$ (one for each experimental condition), the threshold $a$, and the non-decision time $\tau$ of an individual participant. The point estimates of the static DDM parameters and the corresponding inter-trial variabilities are shown in solid blue lines and blue shaded areas, respectively.}
\end{figure}
\vfill
\hspace{0pt}
\newpage

\hspace{0pt}
\vfill
\begin{figure}[H]
\centering
\includegraphics[width=\textwidth]{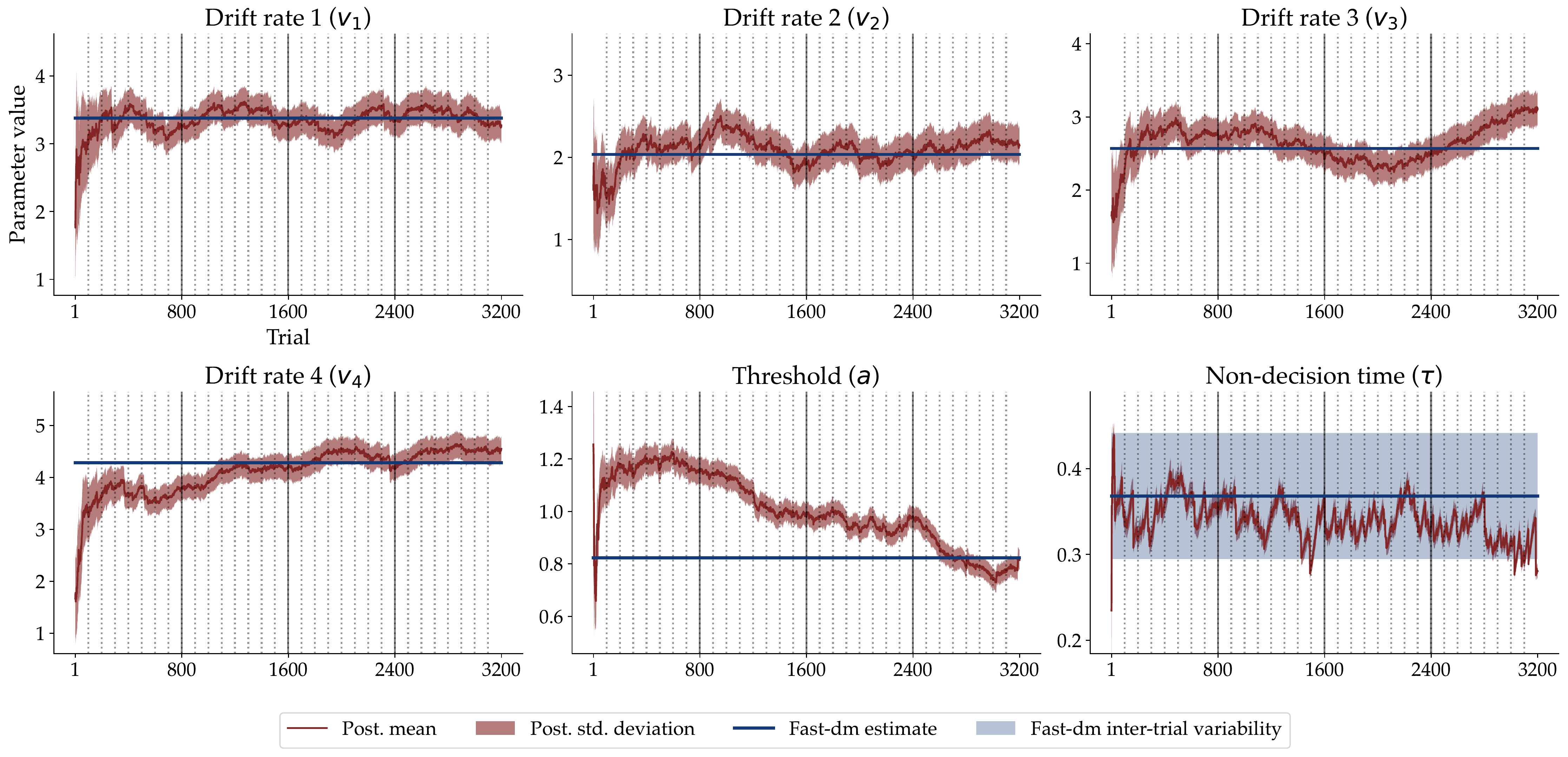}
\caption{The trial-wise posterior mean and $\pm1$ standard deviation for all six parameters, namely the four drift rates $v_1$ - $v_4$ (one for each experimental condition), the threshold $a$, and the non-decision time $\tau$ of an individual participant. The point estimates of the static DDM parameters and the corresponding inter-trial variabilities are shown in solid blue lines and blue shaded areas, respectively.}
\end{figure}
\vfill
\hspace{0pt}
\newpage

\hspace{0pt}
\vfill
\begin{figure}[H]
\centering
\includegraphics[width=\textwidth]{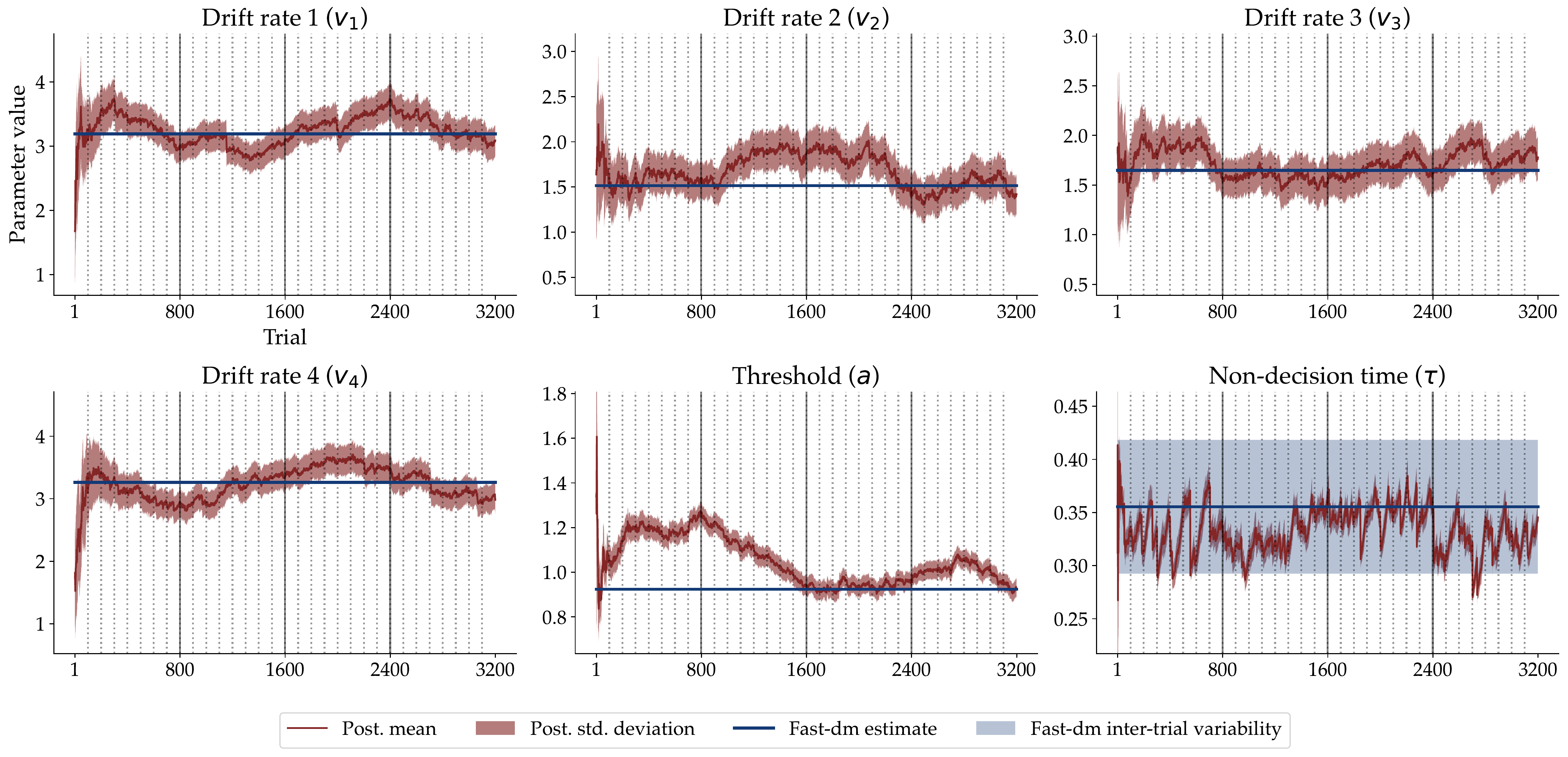}
\caption{The trial-wise posterior mean and $\pm1$ standard deviation for all six parameters, namely the four drift rates $v_1$ - $v_4$ (one for each experimental condition), the threshold $a$, and the non-decision time $\tau$ of an individual participant. The point estimates of the static DDM parameters and the corresponding inter-trial variabilities are shown in solid blue lines and blue shaded areas, respectively.}
\end{figure}
\vfill
\hspace{0pt}
\newpage

\hspace{0pt}
\vfill
\begin{figure}[H]
\centering
\includegraphics[width=\textwidth]{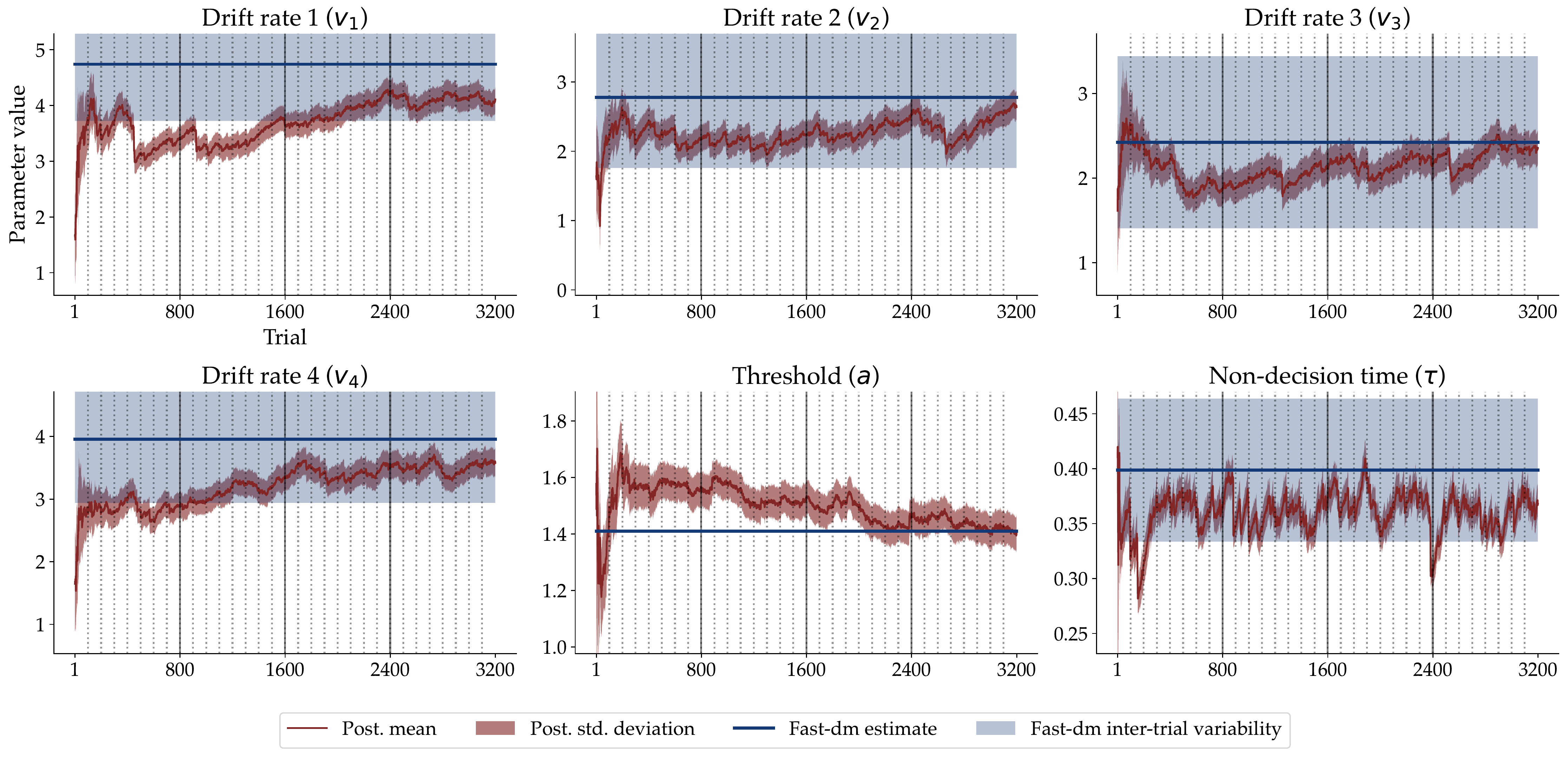}
\caption{The trial-wise posterior mean and $\pm1$ standard deviation for all six parameters, namely the four drift rates $v_1$ - $v_4$ (one for each experimental condition), the threshold $a$, and the non-decision time $\tau$ of an individual participant. The point estimates of the static DDM parameters and the corresponding inter-trial variabilities are shown in solid blue lines and blue shaded areas, respectively.}
\end{figure}
\vfill
\hspace{0pt}
\newpage

\subsection*{Average Parameter Dynamics}
\begin{figure}[H]
\centering
\includegraphics[width=\textwidth]{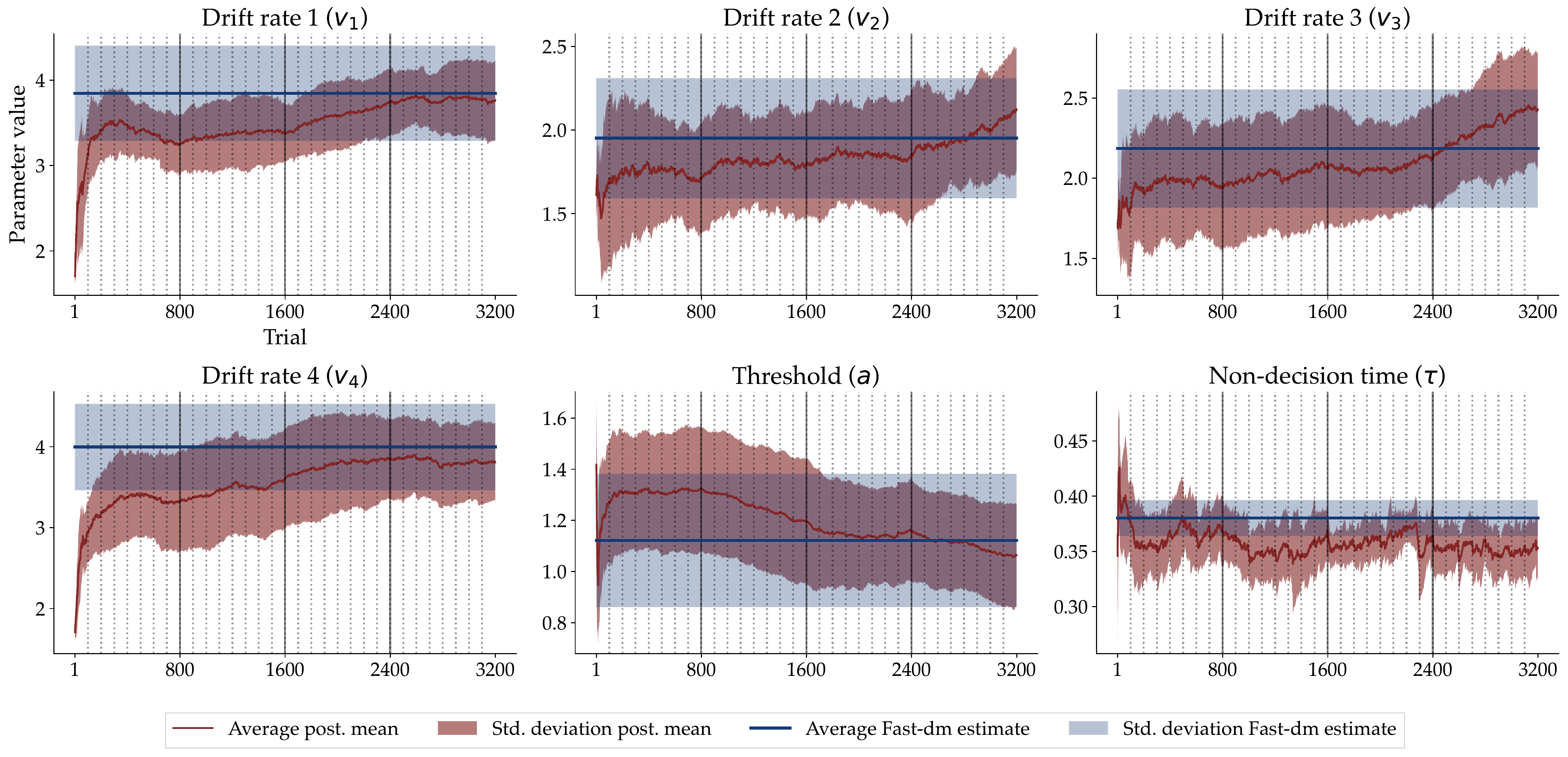}
\caption{The trial-wise posterior mean and $\pm1$ standard deviation for all six parameters, namely the four drift rates $v_1$ - $v_4$ (one for each experimental condition), the threshold $a$, and the non-decision time $\tau$ of averaged across all participant in solid red lines. The shaded red areas correspond to the $\pm1$ standard deviation of the posterior means of all individuals.
The point estimates of the static DDM parameters averaged across all participants and the corresponding standard deviations are shown in solid blue lines and shaded blue areas, respectively.}
\label{average_param_dynamic}
\end{figure}

\autoref{average_param_dynamic} shows the parameter dynamic averaged across all participants.

\hspace{0pt}
\vfill
\vfill
\hspace{0pt}
\newpage

\subsection*{Gaussian Random Walk Transition Model}
We wanted to test if our neural estimation method can also estimate dynamic models with a simpler high-level transition model than a Gaussian process (GP).
To this end, we fit a dynamic DDM with a Gaussian random walk as a transition model to the empirical data set described in the \textbf{Human data application} section:

\begin{align*}
    \theta_t = T(\theta_{t-1}, \eta, z_t) = \theta_{t-1} + \eta\,z_t \quad\text{with}\quad
    z_t \sim\mathcal{N}(0, 1)
\end{align*}

We use a Beta prior distribution parameterized with $\alpha$ and $\beta$ for the standard deviations $\eta_{j}$ of the Gaussian random walk transition model. 
The same prior distribution is used for all $j=6$ low-level parameter transitions:

\begin{align*}
    \eta_{j} &\sim \text{Beta}(1, 25)
\end{align*}

We trained the same neural network architecture as described in the main text for $50$ epochs, $1000$ batches per epoch, and a batch size of $8$.
The following figures show the results from simulation-based calibration (SBC), the model fit and inferred parameter dynamics for the same exemplar participant shown in the main text.
Additionally, we depict the estimated parameter dynamics averaged across all individuals for comparison.
These results are very similar to those obtained with the GP-DDM, which uses a Gaussian process as a transition model. However, the model with the Gaussian process transition model produces sharper predictions on unseen data.
Note, that the dynamics implied by the random walk transition model are less sharper (i.e., contain more uncertainty) than those implied by the GP transition model.

\newpage

\begin{figure}
\centering
\includegraphics[width=\textwidth]{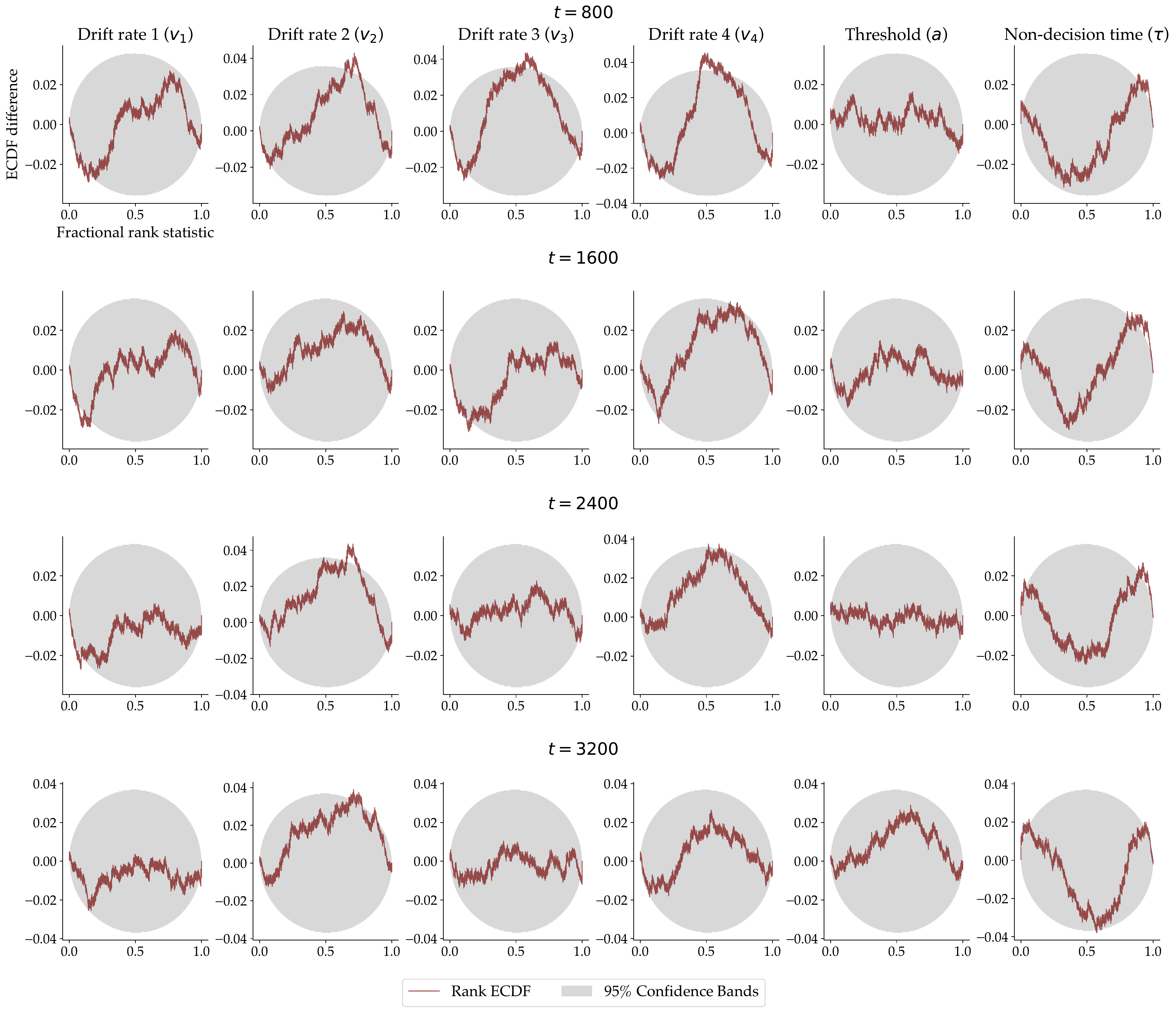}
\caption{\textbf{ECDF difference plot} 95\% simultaneous confidence bands (gray) for the empirical cumulative distribution function (ECDF; red) for all $6$ parameters at four selected time points (800, 1600, 2500, 3200) separately. We used the same settings as for the GP-DDM analysis.}
\end{figure}
\newpage

\begin{figure}
\centering
\includegraphics[width=\textwidth]{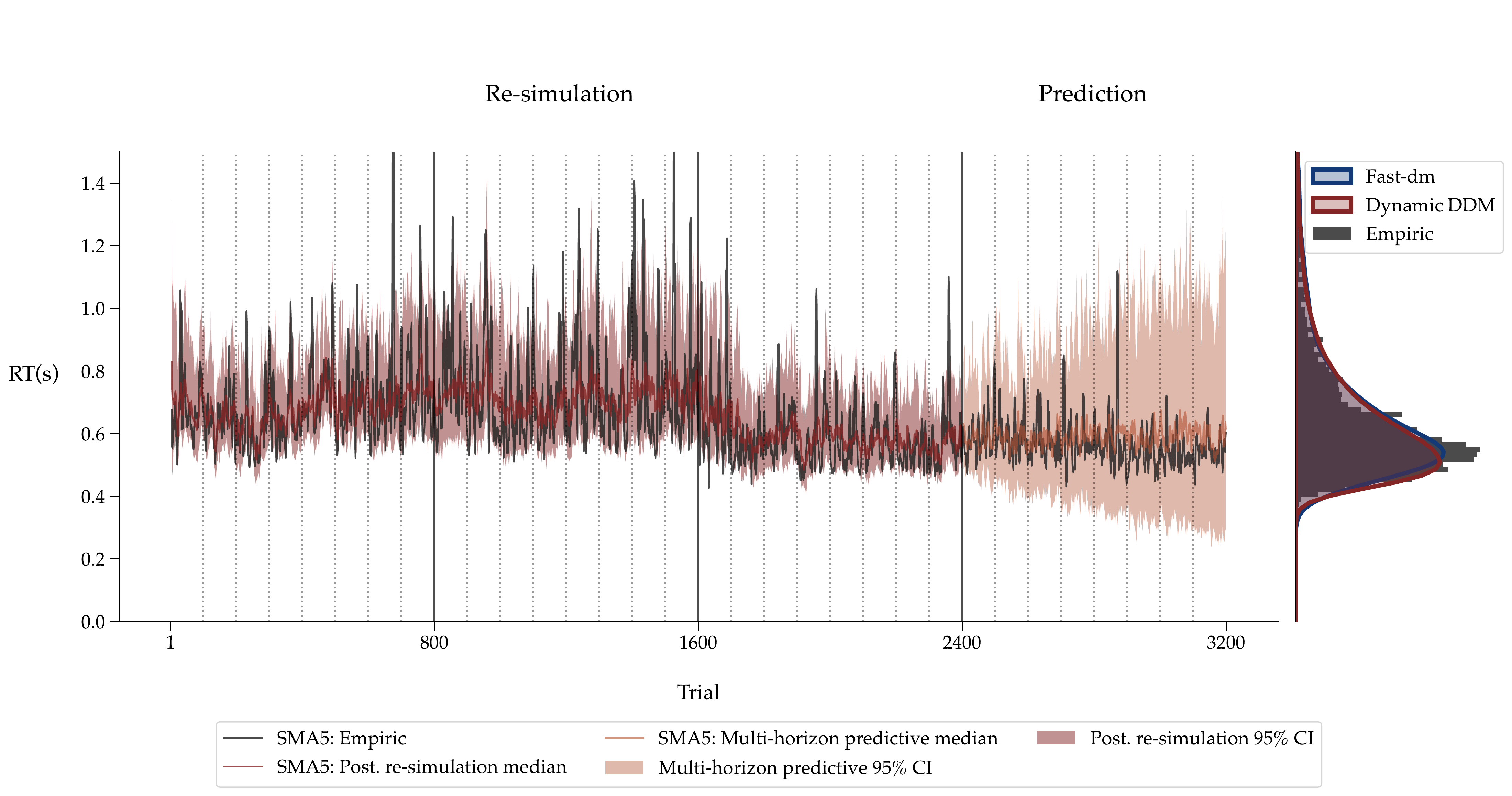}
\caption{\textbf{Left panel} The empirical RT time series of a single individual in black. From trial 1 to 2500, the median posterior re-simulation (aka \textit{retrodictive check}) using the dynamic DDM is shown in red. The models' multi-horizon prediction is depicted for the remaining trials in orange. The shaded areas for the posterior re-simulation and prediction correspond to the 95\% credibility interval. All the time series were smoothed via a simple moving average (SMA) with a period of 5. The dotted vertical lines indicate the end of an experimental block, and the solid vertical lines the end of an experimental session. \textbf{Right panel} The raw RT distribution is plotted as a histogram in black. The re-simulated RT distributions from the dynamic DDM and reference re-simulations from the static DDM using \texttt{Fast-dm} are shown as kernel density estimates (KDEs) in red and blue, respectively.}
\end{figure}

\clearpage

\begin{figure}
\centering
\includegraphics[width=\textwidth]{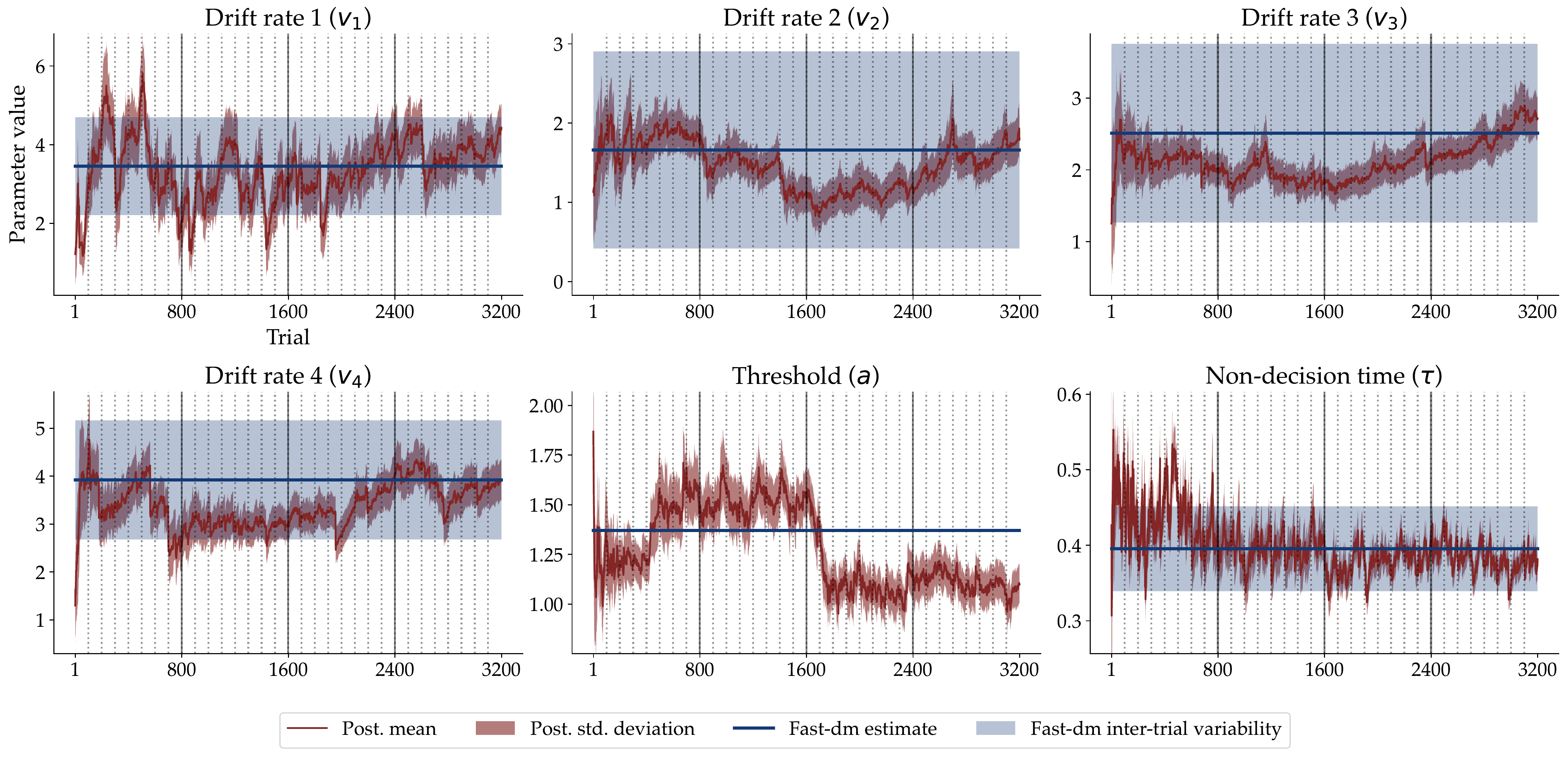}
\caption{The trial-wise posterior mean and $\pm1$ standard deviation for all six parameters, namely the four drift rates $v_1$ - $v_4$ (one for each experimental condition), the threshold $a$, and the non-decision time $\tau$ of an individual participant. The point estimates of the static DDM parameters and the corresponding inter-trial variabilities are shown in solid blue lines and blue shaded areas, respectively.}
\end{figure}

\clearpage

\begin{figure}
\centering
\includegraphics[width=\textwidth]{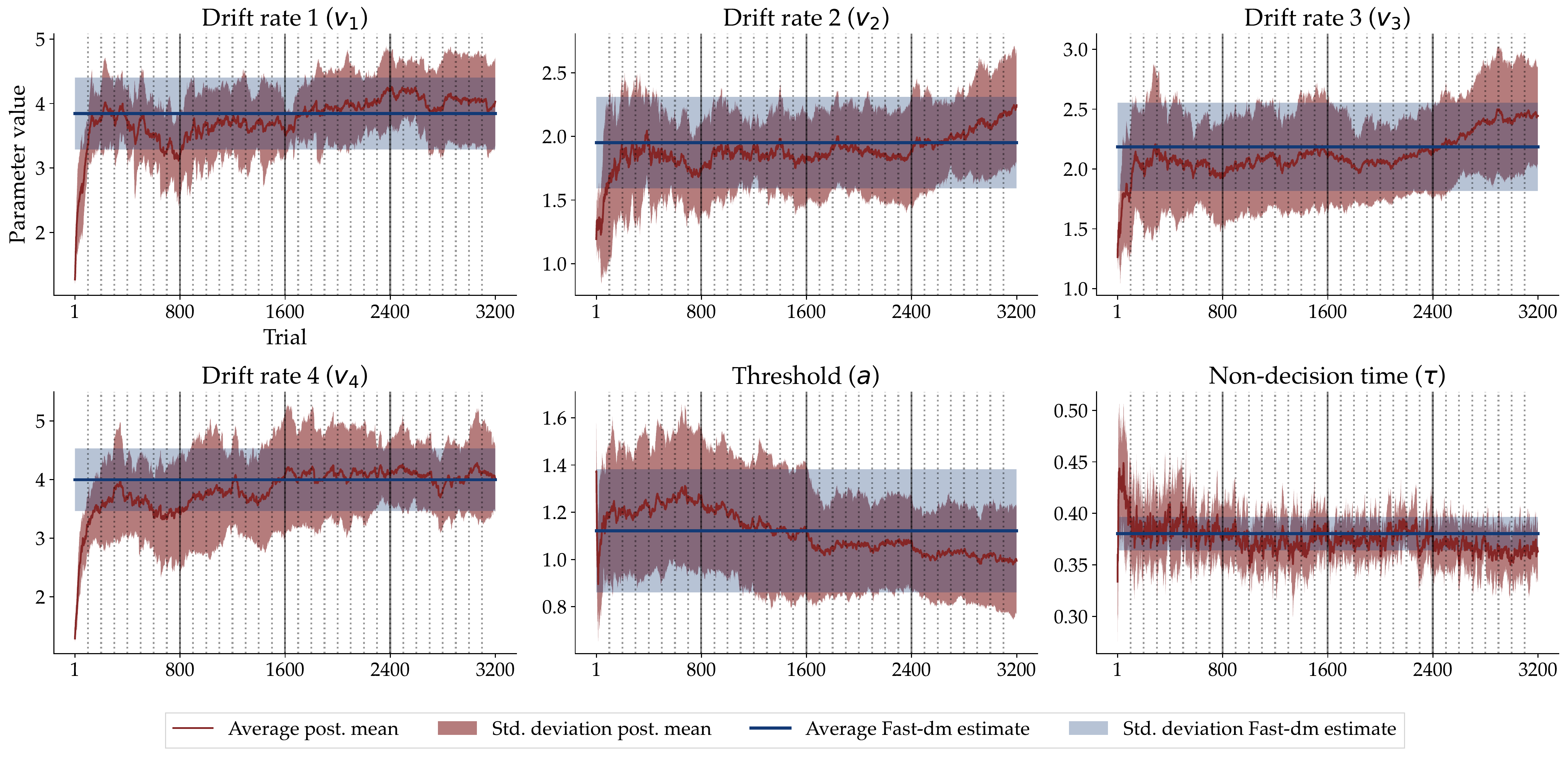}
\caption{The trial-wise posterior mean and $\pm1$ standard deviation for all six parameters, namely the four drift rates $v_1$ - $v_4$ (one for each experimental condition), the threshold $a$, and the non-decision time $\tau$ of averaged across all participant in solid red lines. The shaded red areas correspond to the $\pm1$ standard deviation of the posterior means of all individuals.
The point estimates of the static DDM parameters averaged across all participants and the corresponding standard deviations are shown in solid blue lines and shaded blue areas, respectively.}
\end{figure}

\end{document}